Structure and dynamics of the magnetopause and its boundary layers

Hiroshi Hasegawa

Institute of Space and Astronautical Science, Japan Aerospace Exploration Agency (JAXA), 3-1-1 Yoshinodai, Chuo-ku, Sagamihara, Kanagawa 252-5210, Japan.
E-mail: hase@stp.isas.jaxa.jp

**Abstract**
The magnetopause is the key region in space for the transfer of solar wind mass, momentum, and energy into the magnetosphere. During the last decade, our understanding of the structure and dynamics of Earth's magnetopause and its boundary layers has advanced considerably; thanks largely to the advent of multi-spacecraft missions such as Cluster and THEMIS. Moreover, various types of physics-based techniques have been developed for visualizing two- or three-dimensional plasma and field structures from data taken by one or more spacecraft, providing a new approach to the analysis of the spatiotemporal properties of magnetopause processes, such as magnetic reconnection and the Kelvin-Helmholtz instability (KHI). Information on the size, shape, orientation, and evolution of magnetic flux ropes or flow vortices generated by those processes can be extracted from *in situ* measurements. Observations show that magnetopause reconnection can be globally continuous for both southward and northward interplanetary magnetic field (IMF) conditions, but even under such circumstances, more than one X-line may exist within a certain (low-latitude or high-latitude) portion on the magnetopause and some X-lines may retreat anti-sunward. The potential global effects of such behavior are discussed. An overview is also given of the identification, excitation, evolution, and possible consequences of the magnetopause KHI: there is evidence for nonlinear KHI growth and associated vortex-induced reconnection under northward IMF. Observation-based estimates indicate that reconnection tailward of both polar cusps can be the dominant mechanism for solar wind plasma entry into the dayside magnetosphere under northward IMF. However, the mechanism by which the transferred plasma is transported into the central portion of the magnetotail, and the role of magnetopause processes in this transport remain unclear. Future prospects of magnetopause and other relevant studies are also discussed.

**Keywords**

Magnetopause, Low-latitude boundary layer, Magnetic reconnection, Kelvin-Helmholtz instability, Solar wind-magnetosphere interaction, Plasma transport, Collisionless plasma, Multi-spacecraft observations

**Contents**

**Glossary**

| | |
|---|---|
| CDPS | Cold and dense plasma sheet |
| C# | The Cluster-# spacecraft |
| E-t | Energy-versus-time |
| FTE | Flux transfer event |
| GS | Grad-Shafranov |
| GSE | Geocentric solar ecliptic |
| GSM | Geocentric solar magnetospheric |
| HT | deHoffmann-Teller |
| IMF | Interplanetary magnetic field |
| KAW | Kinetic Alfvén wave |
| KH | Kelvin-Helmholtz |
| KHI | Kelvin-Helmholtz instability |
| LLBL | Low-latitude boundary layer |
| MHD | Magneto-hydrodynamics |
| MLT | Magnetic local time |
| MMS | The Magnetospheric Multi-Scale mission |
| MP | Magnetopause |
| MSBL | Magnetosheath boundary layer |
| MSO | Mercury solar orbital |
| MVAB | Minimum variance analysis of the magnetic field |
| Pc | Continuous pulsations |
| PDL | Plasma depletion layer |
| RD | Rotational discontinuity |
| $R_E$ | Earth radius |
| SC | Spacecraft |
| SCOPE | The cross-Scale COupling in the Plasma universE mission |
| SEF | Slow-mode expansion fan |
| SMXR | Sequential multiple X-line reconnection |
| TD | Tangential discontinuity |
| TH# | The THEMIS-# spacecraft |
| ULF | Ultra-low frequency |
| UT | Universal time |
| VIR | Vortex-induced reconnection |

## 1. Introduction

A magnetosphere is formed around a celestial object, such as a star, a planet, and a moon, which has its own magnetic field of sufficient intensity. In contrast with atmospheres which do not have a well-defined outer boundary, magnetospheres generally have a sharp outer boundary called the magnetopause. This is because matter surrounding the magnetospheres are usually highly ionized and extremely weakly collisional, i.e., the diffusion of electrically-charged, non-collisional particles across the magnetic field is not efficient on a scale much larger than the associated Larmor radius and inertia length. The region governed by the field of a magnetized body and the surrounding medium are relatively well separated by the magnetopause. For example, the Earth's low-latitude magnetopause, observations of which were reported first by Cahill and Amazeen (1963), has a typical thickness (~800 km) of a few to about ten times the proton Larmor radius or inertia length in the region of the shocked solar wind, known as the magnetosheath (Berchem and Russell, 1982; Phan and Paschmann, 1996; Paschmann et al., 2005) (see also Panov et al. (2008) who showed that the high-latitude magnetopause has a thickness of one to about a hundred times the Larmor radius). This is much shorter than the size of the Earth's magnetosphere of the order of ten Earth radii (1 $R_E$ ~6371 km), although the physics underlying such thicknesses is not fully understood.

Essentially all matter and energy entering the magnetosphere cross its magnetopause. High-frequency electromagnetic radiation, electrically-neutral matters, highly-energetic cosmic rays, etc. are not, or only weakly, affected by the magnetopause. On the other hand, low-frequency electromagnetic waves/fields and low-energy charged particles, constituting space plasmas, strongly interact with the magnetopause and are modified when crossing, or reflected off, the magnetopause. In the case of the Earth, indeed, the magnetopause is the first line of defense against the flow of the shocked solar wind. Therefore, although it occupies only a small volume of the near-Earth space, the magnetopause is the vital region in the flow of mass and energy into or out of the magnetosphere; all magnetospheric structures and phenomena are directly or indirectly influenced by magnetopause processes.

Because of its importance in the interaction between the interstellar, stellar, or planetary wind and the magnetosphere, the magnetopause has been a subject of intensive research for decades. Nevertheless, our knowledge of the magnetopause is still lacking. This is probably because the magnetopause is one of the most complex boundaries or transition regions in space, in that it involves electrical current, gradients in the plasma density and pressure, flow shear, and/or anisotropy in velocity distributions of particles or in temperature. This means that various forms of free energy are available in and around the magnetopause, so that a variety of instabilities such as current-driven (e.g., magnetic reconnection or tearing), density or pressure gradient-driven (e.g., Rayleigh-Taylor), flow shear-driven (e.g., Kelvin-Helmholtz), and/or anisotropy-driven (e.g., mirror) instabilities can be excited there. In addition, the magnetopause is curved on a large scale and, if the impinging upstream flow is super-magnetosonic as in the case of the Earth, it is exposed to highly time-varying and inhomogeneous plasmas of the magnetosheath (e.g., Schwartz et al., 1996)—the region of shocked interstellar or stellar wind situated between the magnetopause and the bow shock. When the magnetopause thickness is comparable to, or less than, the ion Larmor radius, possibly as in the case of Mercury, the boundary structure could also be affected by kinetic effects (Cai et al., 1990; Nakamura et al., 2010a). These factors complicate the issues but, at the same time, make the magnetopause an ideal natural laboratory for testing various theoretical predictions and a representative of the interfaces between plasma regions of distinct natures which are ubiquitous in our universe. This contrasts with the magnetotail current layer which is sandwiched between the more-or-less similar, relatively stable plasmas of the tail lobes. It should also be stressed that the magnetopause is a unique region in space, where collision-less plasmas exist, and direct and in-depth measurements at the level of, for example, particle velocity distributions are possible.

In this section, we introduce simplified or conventional views of the magnetopause by referring to some basic theoretical concepts and earlier observations, the purpose being to set the stage for the rest of this monograph. Unless otherwise stated, the Earth's magnetopause is addressed.

## *1.1. Steady state picture*

### *1.1.1. A magnetohydrodynamic picture*

In the framework of ideal magnetohydrodynamics (MHD), the magnetopause can, in most cases, take either of the two types of discontinuity: tangential discontinuity (TD) or rotational discontinuity (RD) (e.g., Hudson, 1970). If there is only negligible diffusion of particles and magnetic field, a TD-type magnetopause is a sharp transition layer across which the density, magnetic field intensity, etc. can change drastically. There exists no magnetic field component normal to the TD-type magnetopause and, when it is more or less planar and is in approximate equilibrium, the total (magnetic plus plasma) pressure balances across the boundary. In principle, no particles can cross and enter into the other side of such a magnetopause, although, in reality, spacecraft often encounter a layer immediately earthward of the magnetopause, in which plasmas of solar wind origin are present (e.g., Eastman et al., 1976).

One of the physical processes that can transfer plasma across the magnetopause is magnetic reconnection, caused by a local breakdown in a current sheet of the frozen-in-flux condition. As a result of reconnection, the two sides of the current sheet become linked to each other via reconnected field lines that thread the outflow jets, namely, exhaust regions, situated on both downstream sides of the reconnection site (X-point or X-line). This process exerts profound influences on magnetospheric structures and dynamics, because (1) the change in magnetic topology allows for the plasma transfer along open field lines, whose one end is anchored to the Earth and the other end extends into interplanetary space (e.g., Paschmann, 1997) (mass transport); (2) part of the solar wind electric field penetrates into the magnetosphere, thereby driving large-scale plasma convection in the ionosphere as well as in the magnetosphere (Dungey, 1961) (momentum transport); (3) solar wind energy is transferred to the magnetosphere via the dynamo action or the Poynting flux on the open field lines and via entry of solar wind plasmas with substantial kinetic energy (e.g., Lee and Roederer, 1982) (energy transport).

Figure 1 shows an MHD, two-dimensional (2D) view of the exhaust region, resulting from steady reconnection in a current sheet across which there is a substantial density jump, as at the magnetopause (Levy et al., 1964). The magnetosheath plasma has a higher density and hence larger inertia than the magnetospheric plasma, so that a major (inertia) current flows in the upstream (magnetosheath-side) RD: a large-amplitude Alfvén wave whose front surface is standing in the X-line rest frame. This RD is usually identified as the magnetopause of an open magnetosphere. (Note, however, that in a rare case when the tangential field components on the two sides of the magnetopause are nearly parallel but have different intensities, no identifiable RD may form and the upstream edge of the slow-mode expansion fan (SEF) may be identified as the magnetopause. Such a situation may be achieved at the distant tail magnetopause far from the reconnection site. See Lin and Lee (1994) for the structure of reconnection layers under various boundary conditions. ) As at the TD, the total pressures on the two sides of a planar RD are balanced. However, the RD has a non-zero magnetic field component normal to it, whose polarity depends on the location of an observer relative to the reconnection site (Sonnerup et al., 1981). At such an RD-type magnetopause, the Walén relation is satisfied (e.g., Sonnerup et al., 1987):

$$\mathbf{v} - \mathbf{V}_{HT} = \pm \mathbf{V}_A, \qquad (1)$$

where **v** is the plasma flow velocity measured in or around the RD, $\mathbf{V}_{HT}$ is the velocity, relative to the frame of velocity measurement, of the deHoffmann-Teller (HT) frame (see Appendix B) in which the electric field is zero so that the flow is field-aligned, and $\mathbf{V}_A = \mathbf{B}/\sqrt{\mu_0 \rho}$ is the Alfvén velocity at the point of velocity measurement. The relation is derived from the stress balance tangential to the time-independent and planar RD, i.e., the balance between the magnetic tension (Maxwell stress) and the inertia force (which is, in the HT frame, the centrifugal force operating on the plasmas streaming along the kinked field lines). The RD is in a dynamical equilibrium. The sign of the right-hand side (RHS) of Eq. (1) is determined by whether the flow in the HT frame is parallel, or anti-parallel, to the magnetic field. Since the Alfvén wave emitted from the reconnection site and propagating toward the magnetosheath in the plasma-rest frame is identified as the RD-type magnetopause, this sign is plus (minus) when the field component $B_N$ normal to the boundary is negative (positive), i.e., when the field points into (out of) the magnetosphere. An isotropic temperature is assumed here, because to date there is no satisfactory observational confirmation of the Walén relation with temperature anisotropy taken into account (see Hudson (1970) for a complete form of the Walén relation). Numerous examples of the RD-type magnetopause, as evidence for magnetopause reconnection, have been reported (e.g., Sonnerup et al., 1995), and the structure corresponding to the SEF, formed downstream of the RD, has also been encountered by spacecraft in the transition region between the magnetosheath and the tail lobes (Hasegawa et al., 2002, and references therein).

*1.1.2. A kinetic picture*

While the MHD picture can be applied to large-scale properties of the field and bulk plasma parameters, a kinetic treatment is necessary to describe the behavior of particle transport along field lines. Figure 2 illustrates a 2D structure of the dayside low-latitude magnetopause region, generated by steady reconnection between southward interplanetary magnetic field (IMF) and geomagnetic field lines, at ~10 $R_E$ from the Earth's center (e.g., Fairfield, 1971). Once reconnection occurs, particles can spatially mix along the interconnected field lines. Consequently, a boundary layer is created on both sides of the RD-type magnetopause: the magnetosheath boundary layer (MSBL) and the low-latitude boundary layer (LLBL), the latter of which is the region immediately earthward of the magnetopause characterized basically by the coexistence of magnetosheath-like, cool and magnetospheric, hot plasmas. Observed in the MSBL are magnetosheath particles convected toward the magnetopause, and those of magnetosheath origin reflected off the magnetopause, and/or those of magnetosphere origin leaked out from the magnetosphere. (Although it is not fully understood exactly how particles are reflected at the magnetopause, a polarization electric field generated by ions with a larger Larmor radii than electrons (Willis, 1971) or a Hall electric field (e.g., Sonnerup, 1979), both of which can, on average, point toward the magnetosheath, may act to reflect the incoming ions. ) On the other hand, observed in the LLBL (of open magnetosphere models) are magnetospheric particles convected toward the magnetopause, and those of magnetosphere origin reflected off the magnetopause, and/or those of magnetosheath origin that crossed the magnetopause.

Multiple ion populations which originated in the distinct regions, as expected from Fig. 2, have indeed been observed (Fuselier et al., 1991). Another key feature of the reconnected boundary layers is separate ion and electron edges, shown as I1 and E1 (I2 and E2) in Fig. 2, of the MSBL (LLBL) on its magnetosheath (magnetospheric) side (Gosling et al., 1990). Such a separation is caused by time-of-flight effects: the speed of electrons, which cross or reflect off the magnetopause and can be energized by reconnection or in the RD, is much higher than that of ions, so that the electrons can reach closer to the separatrix (S1 and S2)—the topological boundary between the not-yet-reconnected and reconnected field lines. In addition, ion velocity distributions in both the MSBL and LLBL are often anisotropic (Gosling et al., 1990; Fuselier, 1995). This is because, unless there are multiple reconnection sites, the MSBL would contain no energetic ions streaming toward the magnetopause (in the HT frame), and the LLBL would contain no magnetosheath ions streaming toward the magnetopause, provided that there are no magnetosheath ions mirrored at low altitudes and streaming backward.

Kinetic ion signatures of magnetopause reconnection can be used to estimate the distance from an observing platform to the X-line at the magnetopause, information concerning which has been used to address a long-standing problem of anti-parallel versus component reconnection (e.g., Trattner et al., 2007; Trenchi et al., 2008; Fuselier et al., 2011): whether or not magnetopause reconnection occurs preferably along a line or lines where the magnetic shear angle, the rotation angle across the magnetopause of the field direction in the boundary plane, is nearly 180°. This issue has not been resolved yet, but the location as well as the rate of magnetopause reconnection may well depend on various conditions (Sections 3 and 6.2).

The LLBL is commonly characterized by an approximate balance between the fluxes of magnetic field-aligned and anti-field-aligned electrons at energies (50-400 eV) comparable to, or somewhat higher than, those of magnetosheath electrons (e.g., Ogilvie et al., 1984). One interpretation of this is that magnetosheath electrons transferred along reconnected field lines are fast enough to be mirrored at low altitudes and then bounce back to the observation site within a short period of time; the bidirectional feature may be seen in an open field line portion of the LLBL (e.g., Fuselier et al., 1995). Another interpretation is that the LLBL is on closed field lines; magnetosheath electrons are captured onto closed field lines through reconnection tailward of both polar cusps (cf. Fig. 3; Song and Russell, 1992) or are transferred via some cross-field diffusion process(es) (e.g., Phan et al., 1997).

*1.2. Identification of the magnetopause*

The magnetopause can usually be identified as a current layer when the magnetic field intensity changes perceivably across the boundary and/or when the magnetic shear is sufficiently large. However, this way of identification may not work when the IMF is strongly northward with its direction nearly parallel to geomagnetic field lines at low latitudes (Paschmann et al., 1993). For such conditions, reconnection seldom occurs at the dayside low-latitude magnetopause, so that the IMF field lines in the noon region are draped around the dayside magnetosphere (line 1 in Fig. 3). The subsolar part of such field lines is compressed by the magnetosheath flow, giving rise to a pressure gradient oriented toward the subsolar region within those flux tubes. This gradient force squeezes part of the magnetosheath plasma out of the subsolar region along the flux tubes. Formed as a result is a magnetosheath region immediately sunward of the magnetopause, called the plasma depletion layer (PDL), where the density and field intensity are substantially lower and higher, respectively, than in the outer portion (closer to the bow shock) of the magnetosheath (Zwan and Wolf, 1976). The field intensity in the PDL can indeed be comparable to that on the magnetospheric side of the magnetopause, with no, or only a weak, current flowing in the low-latitude magnetopause (see panels (f-i) in Fig. 4). In a kinetic view, particles with higher field-aligned speeds (in the Earth's rest frame) quickly stream out of the subsolar region, whereas those with lower field-aligned speeds stay there longer (velocity filter effects). As a result, and in combination with betatron acceleration associated with the field compression (across the bow shock and/or in the PDL), the PDL develops a substantial temperature anisotropy, with proton temperature transverse to the magnetic field higher than the parallel temperature ($T_{p\perp} > T_{p\|}$) (e.g., Phan et al., 1994), which can drive instabilities in the magnetosheath/PDL (e.g., Gary et al., 1993). Electrons in the PDL are nearly isotropic, but often show a weak anisotropy with $T_{e\perp} > T_{e\|}$ (Paschmann et al., 1993).

If the magnetic field information alone is not sufficient for identifying the magnetopause, what other information could be used? In the case of a TD-type magnetopause, changes of plasma properties—such as in the density and temperature—from their magnetosheath to magnetospheric values, or vice versa, can be used. However, even in the case of a strongly northward IMF, reconnection is known to occur at the magnetopause poleward of the cusp (e.g., Kessel et al., 1996), sometimes in both the northern and southern hemispheres (Onsager et al., 2001; Lavraud et al., 2006a). In such circumstances, the identification of the low-latitude magnetopause is not a trivial issue because a variety of boundary layers can be generated. This is the case especially when, as shown in Fig. 3, poleward-of-the-cusp reconnection occurs in both hemispheres but not simultaneously (Bogdanova et al., 2008).

Figure 3 illustrates various types of field lines that could be encountered as one traverses the dayside magnetopause under a northward IMF. The first type of field line encountered earthward of a PDL region threaded by IMF-type field lines (line 1 in Fig. 3) is open field lines, newly formed by reconnection in either of the two hemispheres (in Fig. 3, line 2 resulting from reconnection poleward of the northern cusp). On the magnetosheath-side part of these field lines, unidirectional, field-aligned fluxes of higher-energy electrons and/or ions, characteristic of the MSBL, are often observed (Region 2 in Fig. 4(l)). The polarity of these fluxes is controlled by the location of the reconnection site relative to an observer (anti-parallel to the field in the case of Figs. 3 and 4). Once an MHD (Alfvén or slow-mode) wave or plasma bulk flow, emanated from the reconnection site and propagating along the reconnected field lines, travels past the observer, it now becomes embedded earthward of the magnetopause (line 3 in Fig. 3). The magnetopause can then be identified by an increase in ion temperature from the MSBL to the reconnection exhaust region that contains heated magnetosheath ions (at ~18:45 UT in Fig. 4(b)). Field line 3 may also be signified by unidirectional field-aligned fluxes of heated magnetosheath electrons (Region 3 in Fig. 4(k)), and/or by appreciable southward (or northward, when reconnection occurs first in the southern hemisphere) ion flows originating from the reconnection site (~18:46:45 UT in Fig. 4(c)) (Hasegawa et al., 2009a). However, if reconnection in both hemispheres is almost simultaneous and the observing probe is near the equator roughly equidistant from the two reconnection sites, no such bulk flows but a temperature increase would be observed. Observations reported by Øieroset et al. (2008) are basically consistent with this picture.

Line 4 in Fig. 3 represents field lines newly closed by reconnection in the southern hemisphere. These field lines are characterized by bidirectional field-aligned fluxes of heated magnetosheath electrons (Region 4 in Fig. 4(k, l)) (see also Lavraud et al., 2006a). A part of the field lines is still on the magnetosheath side because the Alfvén wave launched from the southern reconnection site has not yet reached their lower-latitude portion. Therefore, ions energized in the southern reconnection region and streaming northward are not observed at low latitudes; significant southward ion flows can still be seen ($V_L \sim -100$ km s$^{-1}$ for interval ~18:35–18:44 UT in Fig. 4(c)). However, these field lines eventually become fully embedded in the magnetosphere (line 5 in Fig. 3), possibly with a weaker bulk flow along the field lines. (For the event shown in Fig. 4, the reconnection site in the southern hemisphere would have been much farther from the subsolar point than the northern reconnection site, under large geomagnetic dipole tilt conditions (Hasegawa et al., 2009a; Li et al., 2009), leading to the magnetosheath flow speed to be higher and the Alfvén speed to be possibly lower around the southern reconnection site. Consequently, only a weak equatorward flow was probably generated through reconnection tailward of the southern cusp. This explains why the significant southward flows are still seen in the region marked as "4 or 5", i.e., for the interval ~18:33–18:44 UT, in Fig. 4, part of which appears to be threaded by type 5 field lines in Fig. 3. ) Since these newly closed flux tubes contain plasmas of magnetosheath origin, poleward-of-the-cusp reconnection in both hemispheres, often called dual lobe reconnection or double high-latitude reconnection, is suggested to be, and does operate as, a mechanism by which the closed field-line portion of the dayside LLBL is created under northward IMF (Song and Russell, 1992; Lavraud et al., 2006a). Indeed, the density in Region 4 can be comparable to that in the magnetosheath (compare an LLBL interval ~18:33–18:44 UT and PDL interval ~18:47–18:50 UT in Fig. 4(a)), which indicates a rather efficient plasma transfer across the magnetopause and makes it difficult to identify the magnetopause by a density jump. Line 6 in Fig. 3 represents closed field lines that have not experienced reconnection but may contain a mixture of magnetosheath-like and magnetospheric plasmas. Such a region of the LLBL may be generated by diffusive processes that could work at the magnetopause or in the LLBL (Treumann and Sckopke, 1999), but may be difficult to discern from the portion of the LLBL threaded by type 5 field lines.

In summary, the most general definition (in our view) of the magnetopause is the outermost MHD-type discontinuity between the magnetosheath and the magnetosphere. By "MHD-type" is meant that the magnetopause is identifiable in the magnetic field and/or ion bulk parameter data, rather than in the electron data. The magnetopause should be somewhere earthward of the inner boundary of a magnetosheath region threaded by IMF-type field lines that have never crossed the magnetopause. One may be tempted to define the magnetopause as the boundary between field lines of distinct magnetic topologies, e.g., the transition region from open to closed field lines. This definition does not work however; open field lines 2 and 3 in Fig. 3 are, at low altitudes, undoubtedly within the magnetosphere, and a high-latitude part of closed field line 4 is on the magnetosheath side of the magnetopause current layer. Note also that bidirectional field-aligned fluxes of heated magnetosheath electrons, often used as the signature of closed field lines, can be observed outside the magnetopause current layer for a northward IMF (Onsager et al., 2001; Lavraud et al., 2005a).

### *1.3. Layout of this monograph*

This monograph presents recent results from studies, mostly of our own, on the structure and dynamics of the magnetopause and its boundary layers, and discusses future perspectives on magnetopause studies. Our earlier understanding of the magnetopause was more or less limited to locally 1D and/or time-stationary aspects, largely because in most observational studies only single-point measurements were available. Over the last decade, however, much progress has been made regarding multi-dimensional and/or time-dependent aspects, thanks especially to the advent of multi-spacecraft missions for *in situ* observations of the magnetosphere and surrounding regions, such as Cluster (Escoubet et al., 2001) and THEMIS (Angelopoulos, 2008) (observation), to the development of new analysis methods for single- and multi-spacecraft data (Paschmann and Daly, 1998, 2008) (method), and to insights from theoretical or computer simulation studies on fundamental plasma processes and the solar wind-magnetosphere interaction (theory).

The rest of the monograph is organized as follows: In Section 2, several data analysis methods for studies of magnetopause and space plasma structures are summarized, in particular with the aim of making clear the differences among, and the capabilities and limitations of, relatively new techniques for recovering 2D or 3D maps of magnetic field/plasma structures in space from *in situ* measurements. In Section 3, recent results on the temporal aspects of magnetopause reconnection are discussed. In Section 4, a review is presented on observations, identification, evolution, and consequences of the Kelvin-Helmholtz instability developed at the magnetopause or in the LLBL. Section 5 discusses the effects of magnetopause and boundary layer processes on large-scale structures and phenomena in the magnetosphere. Section 6 gives a summary and discussion of open issues that should be addressed in the future. For earlier reviews of magnetopause studies, the readers are referred to, e.g., Lundin (1988), Song et al. (1995), Sibeck et al. (1999), De Keyser et al. (2005), Phan et al. (2005), Paschmann (2008), and Lavraud et al. (2011).

## 2. Analysis Methods for Magnetopause Study

In contrast with the magnetotail current sheet, which can macroscopically be assumed to be parallel to the GSM *x-y* plane, the magnetopause surface is 3D and the local orientation of the boundary depends on its location. Analysis of data taken in and around the magnetopause is thus performed often in the LMN boundary coordinate system (Russell and Elphic, 1978), which is basically defined by use of some magnetopause model (see Appendix A for details).

When the magnetopause is in a non-steady state, its local normal direction can deviate substantially from the average or model normal direction, and it may also involve inward (earthward) or outward motion. Determining the local normal and motion of the magnetopause is then of fundamental importance, e.g., for the accurate estimation of mass or energy flow across the boundary, and thus a number of techniques have been developed for analyzing a planar boundary/discontinuity. See Sonnerup and Scheible (1998) and Sonnerup et al. (2006b) for some single-spacecraft methods, and Schwartz (1998) and Haaland et al. (2004) for multi-spacecraft methods. Note also a recent attempt to estimate the orientation and motion of a 2D, non-planar boundary (Blagau et al., 2010).

In this section, we briefly summarize and discuss several methods for the identification, analysis, and understanding of physical processes operating in and around the magnetopause. A special focus is given to relatively recently developed techniques for visualizing 2D or 3D magnetic field/plasma structures using one or more spacecraft data.

*2.1. Walén test*

Since the Walén relation (1) is briefly explained in Subsection 1.1.1 and its use is detailed by Paschmann and Sonnerup (2008), here we only comment on a common feature of the Walén test: even for the magnetopause that appears to be of RD-type, the ion velocity $v_i$ in the HT frame is lower than the Alfvén speed $V_A$, i.e., the magnitude of the velocity changes across the open magnetopause is considerably smaller than that of the Alfvén velocity changes (e.g., Paschmann et al., 2005), or the absolute value of the Walén slope, defined as the slope of a linear regression line fitted to the data points in a scatter plot of the three components of $\mathbf{v}_i$ and of $\mathbf{V}_A$, is smaller than unity (see Figure 5.2 in Sibeck et al. (1999) for examples of the Walén plot). Several possible explanations, other than temperature anisotropy, have been given for such subunity Walén slopes, as discussed by, e.g., Phan et al. (1996) and Paschmann and Sonnerup (2008). They include (1) underestimation of mass density, (2) the presence of multi-dimensional structures, (3) the presence of time-evolving structures, and (4) breakdown of ideal MHD in the interior of the RD.

Cold particle populations or heavier ions, e.g., of ionospheric origin, may be missed in the plasma measurements, which leads to an overestimation of the Alfvén speed used in the test. It has been shown that in their usual concentration (~5%), alpha particles ($He^{++}$) in the solar wind and magnetosheath are probably not the primary cause of the subunity Walén slope (Puhl-Quinn and Scudder, 2000). On the other hand, oxygen ions ($O^+$) of magnetospheric origin, if abundant, may significantly affect the actual Alfvén speed (e.g., Kasahara et al., 2008).

Plasma or field inhomogeneity tangential to the discontinuity or in the direction along the X-line may also affect the tangential stress balance so that the underlying assumption of a strictly 1D structure is invalid. For example, part of the magnetic tension at the RD may balance the force from tangential gradients of the plasma and/or magnetic pressure (in Subsection 2.2.1, it is shown that this appears to be the case in a magnetopause where reconnection is developing). Also, if the X-line is so short that the resulting reconnection jets are narrow, the plasma in the jets may be subject to magnetic or viscous stress from the neighboring regions where no reconnection activity is present. Moreover, if the observing platform is not very far from the reconnection site so that the SEF is not sufficiently separated from the RD (Fig. 1), a data interval used in the Walén test may contain not only the RD but also a part of the SEF. Indeed, such cases are shown to lead to a subunity Walén slope (Hasegawa et al., 2002).

Temporal evolution of the structures may also affect the tangential stress balance. If the observations are made close to an X-line which is becoming active or in regions where the jet is interacting with the plasma preexisting in front of it so that the plasma experiences substantial acceleration or deceleration, the inertia force ($\rho\, \partial \mathbf{v}/\partial t$) arising from a temporal change in the velocity cannot be neglected. Another possibility is that within thin RDs, gyro-viscous or ion inertia effects may become important (Hau and Sonnerup, 1991), so that the Walén relation, derived under the assumption of the frozen-in condition, may not be well fulfilled for the ion but for electron velocity measurements (Scudder et al., 1999) (remember that it is an inertia current that flows within RDs; in the field line-rest frame, the current in RDs is carried mostly by less-well magnetized ions).

## 2.2. Reconstruction methods

Over the past 15 years, various types of techniques for reconstructing 2D structures in a space plasma from *in situ* measurements, i.e., from plasma and field data taken as the structures are swept past an observing platform, have been developed. These techniques can be categorized into three types: (1) reconstruction of 2D structures based on the Grad-Shafranov (GS) or GS-like equation (Sonnerup et al., 2006a), (2) reconstruction of 2D structures based on a set of the MHD or Hall-MHD equations (Sonnerup and Teh, 2008; Sonnerup and Teh, 2009), and (3) reconstruction of slow time evolution of quasi-static 2D structures (Sonnerup and Hasegawa, 2010). Most recently, it has also been shown that reconstruction of 3D structures is in principle possible if the spatial gradient in certain direction(s) can be estimated from closely separated two or more spacecraft measurements (Sonnerup and Hasegawa, 2011). Here, we summarize those data analysis methods, give some caveats and comments on their capabilities and limitations, and show examples of their application. For a quick survey, see Table 1. In the following, it is assumed, for the reconstructions of 2D structures, that physical quantities have no, or only a weak, gradient along the *z*-direction ($\partial/\partial z \sim 0$), which we call the invariant axis.

*2.2.1. Frame velocity determination*

All the reconstructions summarized in Subsections 2.2.3-2.2.9 need to be conducted in a proper frame of reference in which the structure, or at least its central part, such as the reconnection X-line and the center of a magnetic island/flux rope or a flow vortex, looks stationary. Determination of such a frame velocity allows for a time series of data, taken during an interval, to be converted into spatial information at points along the path of the observing probe. If the configuration of individual magnetic field lines does not vary significantly in time, the deHoffmann-Teller (HT) velocity (Appendix B) can be used as the velocity, $\mathbf{V}_0$, of the co-moving frame. In the case of steady reconnection, however, the configuration of each field line evolves with time, although the field structure when seen in the X-line rest frame is time independent as a whole (Fig. 1). The velocity of the X-line or O-line rest frame can be estimated by such methods as described by Shi et al. (2006), Sonnerup and Hasegawa (2005), and Zhou et al. (2006b). For a test of some of these methods, see Denton et al. (2010) who concluded that a slightly modified version of the Shi et al. method gives a reasonable estimate of the X-line velocity.

Some of the above multi-spacecraft techniques (Shi et al., 2006; Zhou et al., 2006b) could also be used to estimate the motion of structures with flow transverse to the magnetic field, such as flow vortices, based on ion or electron velocity (rather than magnetic field) measurements. However, since no high-quality four-point measurements of the velocity are available at present, this has not been attempted yet. In their studies of Kelvin-Helmholtz (KH) vortices observed in the dusk-flank LLBL, Hasegawa et al. (2007a, 2009b) estimated the vortices velocity based on a linear theory (e.g., Chandrasekhar, 1961) or simulations of the KH instability.

*2.2.2. Coordinate axis determination*

For successful reconstruction of 2D structures, not only the frame velocity, but also proper coordinate axes, need to be determined. The key orientation is that of the invariant (*z*) axis along which gradients are assumed to be negligible. This axis can be estimated by the methods developed by Hasegawa et al. (2004c), Shi et al. (2005), and Zhou et al. (2006a), when data are available from multi-spacecraft with a sufficiently close separation. For single-point measurements, one may use the method developed by Hu and Sonnerup (2002), which is applicable to 2D magneto-hydrostatic structures (namely, Grad-Shafranov equilibria) only, or that developed by Sonnerup and Hasegawa (2005), which assumes time independence of the structure in the co-moving frame. Once the invariant axis orientation is chosen, the x-axis is defined to be anti-parallel to the projection of the frame velocity, $\mathbf{V}_0$, onto the plane perpendicular to the z-axis; the y-axis completes the right-handed orthogonal system. The x-axis (or y = 0) thus represents the path of the observing platform in the x-y frame. The dimension of the reconstruction domain in the x-direction is then $L_x = |V_{0x}|(T_{en} - T_{st})$, where $T_{st}$ and $T_{en}$ are the start and end times of the data interval used for the reconstruction.

*2.2.3. Magneto-hydrostatic structures*

When inertia terms can be neglected in the MHD equation of motion, the structure is magneto-hydrostatic so that the governing equation is $\mathbf{j} \times \mathbf{B} = \nabla p$, which represents the balance between magnetic tension and the force from the total pressure gradient. Under the assumption of a 2D structure ($\partial/\partial z \sim 0$), this force balance equation can be reduced to the Grad-Shafranov equation in a Cartesian coordinate system (e.g., Grad and Rubin, 1958; Shafranov, 1958; Sonnerup et al., 2006a):

$$\nabla^2 A = \frac{\partial^2 A}{\partial x^2} + \frac{\partial^2 A}{\partial y^2} = -\mu_0 \frac{dP_t(A)}{dA}, \tag{2}$$

where A is the partial vector potential (z-component of the magnetic vector potential) so that the magnetic field is $\mathbf{B} = \nabla A \times \hat{\mathbf{z}} + B_z \hat{\mathbf{z}}$ or $(B_x, B_y, B_z) = [\partial A/\partial y, -\partial A/\partial x, B_z(A)]$. The transverse pressure $P_t(A) = p(A) + B_z^2(A)/2\mu_0$, plasma pressure $p(A)$, and axial field $B_z(A)$ are all constant along the transverse field lines; this property can be used to estimate the orientation of the invariant (z) axis, when the same field lines, characterized by equal A values, are encountered more than once along the spacecraft path during an interval under consideration (Hu and Sonnerup, 2002). See Subsection 2.2.2 for other methods to estimate the z-direction from spacecraft data.

For the reconstruction of the magnetic field in a rectangular domain in the x-y plane,

the GS equation is solved numerically by using the magnetic field measurements as spatial initial values (Sonnerup and Guo, 1996; Hau and Sonnerup, 1999). This is usually performed in the HT frame. The integration of $A$ and $B_x$ in the $\pm y$-directions uses the following (second and first order) Taylor series

$$A(x, y \pm \Delta y) = A(x,y) \pm \Delta y \frac{\partial A(x,y)}{\partial y} + \frac{1}{2}(\Delta y)^2 \frac{\partial^2 A(x,y)}{\partial y^2}, \qquad (3)$$

and

$$B_x(x, y \pm \Delta y) = B_x(x,y) \pm \Delta y \frac{\partial B_x(x,y)}{\partial y}, \qquad (4)$$

respectively. Here $\partial A(x,y)/\partial y = B_x$ is known and $\partial^2 A(x,y)/\partial y^2 = \partial B_x(x,y)/\partial y = -\partial^2 A(x,y)/\partial x^2 - \mu_0\, dP_t/dA$ is taken from the GS equation (2). The integration is stopped at certain $y$-points, which define the $y$-boundaries of the reconstruction domain. The transverse field, $\mathbf{B}_t = \nabla A \times \hat{\mathbf{z}}$, in the $x$-$y$ plane is then represented by equi-$A$ contour lines ($A$ is constant along the transverse field lines). The axial field and pressure in the maps are computed with the use of the functions $B_z(A)$ and $p(A)$, which become available by determining a polynomial (or exponential) functional relationship between the actually measured $B_z$ and $A(x,0)$ and between $p$ and $A(x,0)$, respectively, for the analyzed interval (Hau and Sonnerup, 1999). Here, $A(x,0)$ is calculated from the spatial integration of measured $B_y$ $(= -\partial A/\partial x)$ along the spacecraft path ($x$-axis). For improvements to reduce numerical errors inherent in the GS solver and to take into account the acceleration of the HT frame (Appendix B), see Hu and Sonnerup (2003); their benchmark test using an exact analytical solution of the GS equation demonstrates that within the range of $|y| < 0.1 L_x$ (see Subsection 2.2.2 for the definition of $L_x$), integration errors are sufficiently small (less than ~2 %).

Figure 5 shows an example of the application of the magneto-hydrostatic GS reconstruction to Cluster observations on 5 July 2001 of a dawn-flank magnetopause (Hasegawa et al., 2004c). Figure 5(a), recovered from Cluster 1 (C1) data, shows a TD-type current layer with a gentle curvature. The Walén slope for a C1 interval (0623:23–0623:40 UT) of the current sheet is 0.567. This suggests that the field lines around C1 were, at least partly, reconnected with a negative field component ($B_N < 0$) normal to the magnetopause, but inertia effects were not so large. Part of the magnetosheath field lines indeed appears to cross the boundary near C1 at ($x$, $y$) ~ (11000, 0) km into the magnetotail. In the map, however, magnetic tension of those reconnected field lines in principle balances the force from the total pressure gradient oriented toward the lower-right in the current sheet (note that non-zero inertia effects reflected in the appreciable Walén slope are neglected in the magneto-hydrostatic GS reconstruction).

Figure 5(b), recovered from C3 data, shows distinctly different features from the C1 map. The current layer is thicker (about 1000 km) along the C3 path (Haaland et al., 2004) and contains a larger amount of reconnected field lines with larger negative $B_N$ components. These reconnected field lines are wedge-shaped and the plasma appears to be flowing along the field lines from the magnetosheath into the magnetotail. The Walén slope for a C3 interval (0623:52–0624:37 UT) is 1.03, consistent with full-blown reconnection and the field-aligned entry of the magnetosheath plasma. Comparison with Fig. 5(a) indicates that in the course of the development of reconnection near Cluster, the magnetopause structure evolved in a dramatic way in about a 30-s interval between the C1 and C3 crossings. This example demonstrates that the application of the method to multi-spacecraft data can reveal not only the 2D structure of the magnetopause but also its temporal evolution.

It must be borne in mind that this type of reconstruction method for 2D structures cannot recover the topology of magnetic field lines, if some (even a weak) 3D effect is present. Hasegawa et al. (2007b) indeed showed, based on an experiment using synthetic data from a 3D MHD simulation of localized (namely 3D) reconnection, that the GS method may lead to closed magnetic field loops that seemingly suggest the presence of more than one X-line, even when there is only a single X-line in the simulation. This kind of test demonstrates that data including 3D effects must be analyzed and interpreted with great care (see also Dorelli and Bhattacharjee, 2009).

*2.2.4. Structures with field-aligned flow*

When the structure in a proper frame is time-stationary and involves significant flows along curved magnetic field lines, an inertia term needs to be included in the force balance equation. The equation to consider is then

$$\nabla \cdot (\rho \mathbf{v}\mathbf{v} - \mathbf{B}\mathbf{B}/\mu_0) = -\nabla(p + B^2/2\mu_0), \tag{5}$$

which represents the balance between the total pressure gradient force (RHS) and the sum of magnetic tension and the inertia force exerted by the field-aligned flow (left-hand side) (note that for planar RDs with no tangential gradients, the RHS of Eq. (5) vanishes so that the Walén relation (1) can be derived). Sonnerup et al. (2006a) showed that, assuming that the flow is isentropic in the co-moving frame, Eq. (5) can be written as a GS-like equation,

$$\nabla \cdot [(1 - M_A^2)\nabla A] = \mu_0 \rho \left[T \frac{dS}{dA} - \frac{dH}{dA}\right] - B_z \frac{dC_z}{dA} - \frac{B^2}{2\mu_0 \rho} \frac{dG^2}{dA}, \qquad (6)$$

where $M_A = v\sqrt{\mu_0\rho}/B$ is the Alfvén Mach number based on the flow speed $v = (v_x^2 + v_y^2 + v_z^2)^{1/2}$. The four field-line (and equivalently streamline) invariants are: the entropy $S(A) = c_v \ln(T/\rho^{(\gamma-1)})$; the total enthalpy $H(A) = c_p T + v^2/2$; the mass flux invariant $G(A) = \mu_0 \rho v/B = M_A \sqrt{\mu_0 \rho}$; and the axial momentum invariant $C_z(A) = (1 - M_A^2)B_z$, where the ratio of specific heats is $\gamma = c_p/c_v$.

Equation (6) can be used to reconstruct structures such as reconnection layers where the flow speed in the HT frame is comparable to the Alfvén speed. Difficulties involved in this type of reconstruction are that a functional form must be defined for all four field-line invariants ($S(A)$, $H(A)$, $G(A)$, and $C_z(A)$) using measured values of these invariants and calculated $A(x, 0)$, and that a 7×7 matrix must be inverted to solve a set of simultaneous equations for the integration of the fundamental quantities ($A$, density, velocity, temperature, etc.) in the ±$y$-directions. The inversion becomes impossible when the field-aligned flow speed is equal to either of the three MHD wave speeds so that the matrix is singular. Such a situation may occur in the analysis of RDs where Alfvénic, or nearly Alfvénic, field-aligned flow in the HT frame is expected.

Teh et al. (2007) developed a numerical code for this type of reconstruction and applied it to the same Cluster event as shown in Fig. 5. Figure 6 shows the 2D maps of the magnetic field and the Alfvén Mach number recovered from the C3 data. The configuration of the transverse field lines is qualitatively similar to that for the magneto-hydrostatic reconstruction (Fig. 5(b)). A remarkable difference is that the axial field $B_z$ is no longer preserved along the field lines. The plasma streaming along a flux tube is compressed or rarefied, and is accelerated or decelerated, in the course of change in the cross-sectional area of the flux tube (which is proportional to $1/B$). It was also shown that within some flux tube (marked by the red line in Fig. 6(b)), the variations of the density, sonic Mach number, etc. all show the behavior expected for supersonic flow. Assuming that the interconnected field lines within the current layer were all reconnected during the 30-s interval between the C1 and C3 crossings of the magnetopause, an estimate of the reconnection electric field, $E_z$ = 0.47 mV/m, was given (Teh et al., 2007); this is equivalent to the non-dimensional reconnection rate of ~0.025. Here, the non-dimensional rate is defined to be the reconnection electric field normalized to $E_0 = V_{A0}B_0$, where $B_0$ is the intensity of the background field component transverse to the X-line in question, and $V_{A0}$ is the background Alfvén speed based on $B_0$.

*2.2.5. Structures with flow transverse to the magnetic field*

When the magnetic field is unidirectional ($\mathbf{B} = B(x,y)\hat{\mathbf{z}}$), so that there is no magnetic tension, but the structure involves flow transverse to the field, the force balance equation can be written as

$$(\mathbf{v} \cdot \nabla)\rho\mathbf{v} = -\nabla(p + B^2/2\mu_0). \tag{7}$$

This equation represents the balance between the total pressure gradient force and the inertia force operating on the plasma flowing along time-independent, but curved, streamlines. For such a steady structure, it is shown from the mass continuity equation ($\nabla \cdot (\rho\mathbf{v}) = 0$) that the velocity field can be expressed using the compressible stream function $\psi(x, y)$ as $\rho\mathbf{v}_t = \nabla\psi \times \hat{\mathbf{z}}$, i.e., $(v_x, v_y) = [1/\rho\,(\partial\psi/\partial y), -1/\rho\,(\partial\psi/\partial x)]$. Here, $\mathbf{v}_t = (v_x, v_y)$ is the velocity component transverse to the invariant ($z$) axis. A field-aligned velocity component $v_z = v_z(\psi)$ can be included as well, since it has no effect on the force balance. Assuming that the flow is isentropic, Eq. (7) can be written as a GS-like equation for the stream function (Sonnerup et al., 2006a),

$$\nabla \cdot \left[\frac{1}{\rho}\nabla\psi\right] = \rho\left[\frac{d\tilde{H}}{d\psi} - T\frac{dS}{d\psi}\right] - \frac{\rho^2}{2\mu_0}\frac{dF^2}{d\psi}, \tag{8}$$

where the three streamline invariants are: the entropy $S(\psi)$; the generalized total entropy $\tilde{H}(\psi) = c_p T + (v_x^2 + v_y^2)/2 + B^2/(\mu_0\rho)$; and the frozen-flux function $F(\psi) = B/\rho$.

Hasegawa et al. (2007a) developed a numerical code for this type of reconstruction, which requires that a functional form for the three streamline invariants, $S(\psi)$, $\widetilde{H}(\psi)$, and $F(\psi)$, is determined from the measurements. The matrix to be inverted for the spatial integration becomes singular when $v_y$ reaches the fast-mode magnetosonic speed $v_y = (c_s^2 + v_A^2)^{1/2}$; however such a situation rarely occurs in and around the magnetopause, except in the magnetosheath at large downtail distances. The code has been benchmarked by the use of both an exact numerical solution of the GS-like equation (8) and an MHD simulation of a time-evolving Kelvin-Helmholtz (KH) vortex (Hasegawa et al., 2007a). In the latter case, the streamline invariants are not even approximately constant along streamlines, because of the time dependence. Nevertheless, the method is able to well recover the velocity field in the vortex; the reconstructed streamlines are roughly parallel to the exact velocity vectors from the simulation.

The technique has been applied to flow vortices in the flank LLBL (Hasegawa et al., 2007a, 2009b; Eriksson et al., 2009; Nishino et al., 2011) and those observed in the tail plasma sheet (Tian et al., 2010). Figure 7 shows the streamlines reconstructed from C1 data when the spacecraft traversed rolled-up vortices in the dusk-flank LLBL. Based on the map showing that two vortices of roughly equal size are embedded within one dominant KH wavelength (which was estimated from a wavelet spectral analysis of the total pressure shown in Fig. 20), Hasegawa et al. (2009b) suggested that a dominant-mode vortex may have broken up into two vortices of about half the size. Such a feature suggests that a forward cascade of bulk flow energy was induced in the KH vortices.

*2.2.6. General MHD structures*

Sonnerup and Teh (2008) have shown that reconstruction from single-spacecraft data is, in principle, possible for any 2D, time-independent, ideal MHD structures. Such a reconstruction is based on the standard set of ideal MHD equations, so that it differs from those presented in the previous subsections that are governed by a GS-type equation. As before, however, the magnetic and velocity fields in the x-y plane are described by the partial vector potential and stream function as $\mathbf{B}_t = \nabla A \times \hat{\mathbf{z}}$ and $\rho \mathbf{v}_t = \nabla \psi \times \hat{\mathbf{z}}$, respectively. The velocity is now no longer necessarily aligned with the magnetic field (as it was in Subsection 2.2.4), or transverse to it (as it was in Subsection 2.2.5). The advantage is that the technique can be applied to structures for which no HT frame exists or the HT frame is not the frame in which the structure looks as stationary as possible. An example is the structure around the site of steady reconnection, viewed in the frame co-moving with the X-line (Fig. 1). The one and only quantity assumed to be conserved along streamlines is the entropy $S(\psi) = c_v \ln(T/\rho^{(\gamma-1)})$. The method can reveal large-scale structures of, for example, reconnection events, by reconstructing 2D maps of essentially all MHD parameters from measurement of the bulk plasma parameters (first three moments) and magnetic field.

A numerical code to perform the MHD reconstruction has been developed and successfully benchmarked, by the use of an exact, axially symmetric solution of the ideal MHD equations (Sonnerup and Teh, 2008). The method has been applied to observations of the magnetopause current layer (Teh and Sonnerup, 2008), a magnetic flux rope on an undulating flank magnetopause (Eriksson et al., 2009), and a large-scale reconnection exhaust and magnetic field directional discontinuities in the solar wind (Teh et al., 2009, 2011a). Figure 8 shows an example of the MHD reconstruction, applied to a flux rope observed on the sunward-facing edge of a magnetopause surface wave. The reconstruction of both magnetic field lines and streamlines demonstrates that a meso-scale flux rope of diameter ~1000 km was roughly collocated with a vortex of a similar size. Eriksson et al. (2009) suggested that the flux rope was generated by magnetic reconnection induced by the growth of large-scale vortices, which are not clearly seen in the maps but, as demonstrated by other analyses, were located to the left and right of the meso-scale vortex that is fully recovered in the maps.

More recently, it has also been shown that effects of electrical resistivity can be included in the MHD reconstruction, when at least two components of directly measured electric fields are available (Teh et al., 2010b). Such a reconstruction is applicable to structures in which, for example, a reconnection diffusion region is embedded. A benchmark test using a 3D resistive MHD simulation of reconnection indeed demonstrates that the resistive version can recover reasonably well the X-type magnetic field configuration in the non-zero resistivity region, whereas the ideal MHD reconstruction cannot.

*2.2.7. Hall-MHD structures*

It has recently been shown further that reconstruction from single-spacecraft data is possible for steady 2D Hall-MHD structures, when at least two components of the electric field can be directly measured (Sonnerup and Teh, 2009). This type of reconstruction recovers both magnetic and electric fields and streamlines for both ion and electron fluids. It can be a viable tool for the study of the ion diffusion region surrounding a reconnection site, as depicted in Fig. 9. Theory has also been developed to include the effects of the resistivity and non-zero, but isotropic, electron pressure. However, the Hall-MHD reconstruction cannot be applied to the electron diffusion region (marked by the black rectangle in Fig. 9) and works only within the ion diffusion and nearby regions where the Hall effects are significant. Therefore, a combined use of the MHD and Hall-MHD reconstructions and some other method that can reconstruct the electron diffusion region may be necessary to fully accommodate an entire MHD-scale structure surrounding a reconnection site. Despite these complications, the Hall-MHD reconstruction has recently been applied successfully to a reconnection event observed in the Earth's magnetotail (Teh et al., 2011b).

*2.2.8. Evolution of quasi-magnetohydrostatic structures*

Sonnerup and Hasegawa (2010) have developed a theory for reconstructing the slow evolution of quasi-magnetohydrostatic 2D structures from essentially single-spacecraft measurements. Such a reconstruction is based on the classical GS equation for magneto-hydrostatic equilibria, but advances the quantities measured at points along the spacecraft path backward and forward in time by using a set of equations. As a result, those quantities are recovered on the plane $y=0$ in a 3D space-time ($x$, $y$, $t$). The recovered values on ($y$, $t$) = (0, $t_i$) are used as the spatial initial values for the standard GS reconstruction at a certain time step $t_i$, the product being multiple 2D field maps, or a movie, for a sufficiently short period of time before and after the actual measurements. The essentials underlying the time integration can be described as follows. The Maxwell equations indicate that the electric field is expressed by use of the electric scalar potential $\phi$ and magnetic vector potential $\mathbf{A}$,

$$\mathbf{E} = -\nabla\phi - \frac{\partial \mathbf{A}}{\partial t}. \tag{9}$$

Under the 2D assumption ($\partial/\partial z \sim 0$), this equation can be used to derive the time derivative of the partial vector potential $A$, viz., of the $z$-component of $\mathbf{A}$,

$$\frac{\partial A}{\partial t} = -E_z = v_x B_y - v_y B_x, \tag{10}$$

where the frozen-in condition is used to express the electric field component $E_z$ by measured velocity and magnetic field components. For Eq. (10) to be used in each time integration step, the velocity components, $v_x$ and $v_y$, as well as $A$ and $B_x$, must be advanced in time. This can be done using the equation:

$$\frac{\partial \mathbf{v}}{\partial t} = -(\mathbf{v} \cdot \nabla)\mathbf{v}, \tag{11}$$

which results from the MHD equation of motion under the condition of strictly magneto-hydrostatic force balance ($\mathbf{j} \times \mathbf{B} = \nabla p$), and of other equations assuming a incompressible plasma, i.e., $\nabla \cdot \mathbf{v} = 0$ (see Sonnerup and Hasegawa (2010) for details). The advantage of the latter assumption is that the $y$-gradient of $v_y$, which is not directly known from single-point measurements, can be converted to the $x$-gradient of $v_x$, which can be calculated from the measurements along the spacecraft path ($x$-axis): $\partial v_y/\partial y = -\partial v_x/\partial x$ (likewise, $\partial B_y/\partial y = -\partial B_x/\partial x$ from $\nabla \cdot \mathbf{B} = 0$).

A numerical code for the time evolution method has been developed and benchmarked by the use of a 2D resistive MHD simulation of magnetic island formation (Hasegawa et al., 2010b). The benchmark test demonstrated that convective motion of the transverse field lines can be recovered reasonably well in regions, located well away from the reconnection site, where the magneto-hydrostatic force balance is approximately satisfied and both the compressibility and resistive electric field are negligible. Figure 10 shows the first application of the method to a flux rope traversed by Cluster at the high-latitude magnetopause equatorward of the northern cusp, which has been studied by Sonnerup et al. (2004), Sonnerup and Hasegawa (2005), and Hasegawa et al. (2006a). The result suggests that the field lines within the flux rope were moving toward its center. Since the amount of magnetic flux reconnected during a specified time interval can be calculated from the maps, the method could be used as a tool for estimating the electric field, or the rate, of 2D reconnection.

One caveat is that the present version of the method assumes that the input data were taken instantaneously at a certain time, thereby allowing for the data to be projected onto a line in the 3D space-time characterized by $t$ = const. and $y$ = 0. However, the data are, in fact, time-aliased because it takes a finite time for the observing spacecraft to traverse the structure that may be evolving significantly; the actual spacecraft path in the space-time (i.e., the world line) has a non-zero angle with the plane $t$ = const. An improvement is thus necessary to eliminate the effects of time aliasing, which will be addressed in a forthcoming paper.

*2.2.9. Three-dimensional magneto-hydrostatic structures*

Most recently, Sonnerup and Hasegawa (2011) have shown that reconstruction of steady, 3D, magneto-hydrostatic structures is in principle possible, when plasma and magnetic field data from two closely separated spacecraft are available. A numerical code developed for such a reconstruction solves a set of four equations: the magneto-hydrostatic equation in the form $\nabla P = (\mathbf{B} \cdot \nabla)\mathbf{B}/\mu_0$ (which consists of three scalar equations) and $\nabla \cdot \mathbf{B} = 0$ (one scalar equation), where $P$ is the total pressure, $P = p + B^2/(2\mu_0)$. Since no invariant axis exists for 3D structures, there is no need to predetermine a coordinate axis for this type of reconstruction; its coordinate system is automatically defined once the moving direction of the structure and the direction in which the two spacecraft are separated are known.

The basics of the 3D reconstruction can be summarized as follows. Suppose that the spacecraft traverse a time-independent structure at a constant velocity. Then, spatial gradient $\partial U/\partial x$ along the spacecraft path, which is defined to be the $x$-axis, can be computed from a time series of data for the four quantities, $U \equiv \{B_x; B_y; B_z; P\}$. When the vector representing the spacecraft separation has a component transverse to the $x$-axis, say, a $y$-component, gradients along $y$, $\partial U/\partial y$, at the mid-point between the two spacecraft, where we set $(y, z)$ = (0, 0), can also be calculated from the two point measurements. It now turns out that we have the four equations for the four unknowns $\partial U/\partial z$, so that $\partial U/\partial z$ can eventually be computed. The four quantities $U$ can then be integrated in the $\pm z$-directions from the line $(y, z)$ = (0, 0) which is halfway between the two spacecraft paths; the result is the integrated $U$ values on the plane $y$ = 0 (see Sonnerup and Hasegawa (2011) for actual integration procedures). This information is then used, along with the four equations, to compute $\partial U/\partial y$ at each point on $y = 0$ ($x$-$z$ plane); the integration along $\pm y$ is also possible.  The product is a 3D distribution of the magnetic field and pressure reconstructed in a narrow rectangular parallelepiped surrounding the spacecraft paths.

Figure 11 shows an example of the benchmark test using a spheromak-type, axi-symmetric solution of the magneto-hydrostatic equations (Sonnerup and Hasegawa, 2011).  Outside the spheromak, the magnetic field is of a scalar potential and the plasma pressure is set at zero; the boundary of the sphere can be identified by red arcs on the back and bottom faces of the reconstruction box.  Overall, the reconstructed pressure and fields are similar to the exact solution, demonstrating that reasonably accurate reconstruction of 3D, magneto-hydrostatic structures is possible.  Successful application to actual events would require a proper spacecraft separation and well-calibrated multipoint data for the accurate estimation of spatial gradients characterizing the structure.  Nonetheless, the method could be a useful tool for studies of various structures in space, such as magnetic clouds in the solar wind (e.g., Lepping et al., 1990), flux ropes at the magnetopause (e.g., Hasegawa et al., 2006a), and plasmoids in the magnetotail (e.g., Slavin, 1998).  Experiments using synthetic data from 3D MHD simulations and application to a flux rope event observed near the magnetopause are currently under way.

Results from the 3D reconstruction can be used to estimate various parameters in 3D space that characterize the structures, such as current, pressure/magnetic field gradient, field line curvature (e.g., Shen et al., 2003), and the dimensionality of, and direction of spatial variation in, the structures (Shi et al., 2005).  Such information would be useful to assess, for example, whether the conditions for some instabilities are fulfilled there. Furthermore, if the method is extended to incorporate reconnection diffusion regions, it could enable us to identify magnetic nulls and separators (e.g., Cai et al., 2001; Xiao et al., 2006; Dorelli et al., 2007), which may exist not only within the tetrahedron constructed by four satellites such as Cluster but possibly also in the adjacent region.

*2.2.10. Discussion of the reconstruction methods*

The various types of reconstruction techniques reviewed in Subsections 2.2.3-2.2.9 have the following common features: (1) they solve an initial value problem, i.e., a Cauchy problem, so that integration errors accumulate as the distance in space-time from the path(s) of the spacecraft, namely the number of integration steps, increases; (2) in any application to real events, the underlying assumptions are not strictly satisfied, so that the solution would always show some deviations from reality (even if there are no numerical errors). For example, most of the reconstructions require that there exists one or more field-line, or streamline, invariants, and that an appropriate functional form is determined for those invariants, even though actual data points do not exactly fall onto a single, or double-branched, curve (e.g., Hasegawa et al., 2007a); (3) the integration itself requires data from only a small number of spacecraft (one spacecraft for the reconstruction of 2D structures, and two for the reconstruction of steady 3D structures), so that, when data from more than one (or two) spacecraft are available, the validity of the model assumptions can be checked and/or the solution can be improved or optimized in some way. For example, Hasegawa et al. (2004c) used multi-spacecraft information to check the consistency between the 2D maps recovered independently for two spacecraft, and to optimize the orientation of the invariant axis. In this sense, the reconstruction methods presented here are expected to perform better when they are applied to multi-spacecraft data such as taken by the Cluster, THEMIS, and STEREO spacecraft, and would become a useful set of tools in future multi-satellite missions such as Magnetospheric Multi-Scale (MMS) (e.g., Burch and Drake, 2009) and cross-Scale COupling in the Plasma universE (SCOPE) (Fujimoto et al., 2009). We also stress that the above-mentioned consistency check and comparison of the results from different methods could allow us to infer which effects (2D/3D, force terms, time dependence, and/or possibly kinetic effects) are important in the structure in question.

Our experience suggests that the more sophisticated the reconstruction method is, the more rapidly the integration errors grow. This is because a more sophisticated method uses more equations, or a more complicated system matrix that can become singular more often. For example, the 6×6 matrix used in the ideal MHD reconstruction (Subsection 2.2.6) becomes singular whenever $B_y = 0$ and $v_y = 0$, or when $v_y$ equals either of the three MHD wave speeds along $y$. As for the Hall-MHD reconstruction (Subsection 2.2.7), singularities occur at points where the $y$-component of the ion or electron velocity changes sign. Also, two of the equations for the reconstruction of 3D magneto-hydrostatic structures (Subsection 2.2.9) have terms divided by $B_y$ or $B_z$, which are a source of numerical errors. From the numerical point of view, therefore, the magneto-hydrostatic GS reconstruction is the simplest and thus most robust.

Let us conclude the present section with a discussion of possible tests and improvements that should be implemented in future. These include: the development of the theory and numerical code for the reconstruction of more general and/or complex structures, such as governed by a two-fluid system of equations, electron diffusion regions embedded in the reconnection region, and 3D and time-dependent structures; more benchmark tests using computer simulations to understand how the accuracy of the various types of reconstruction decreases with the extent to which the model assumptions (2D, time independence, MHD, etc.) are violated; development of better integration schemes that suppress the growth of singularities and errors; incorporation into the reconstruction codes of a data assimilation technique (e.g., Nakano et al., 2008) that makes use of multipoint measurements that are not used as the initial values, to correct or improve the solution and thus to yield a larger window of view and more accurate field map. It is also noted that the various kinds of reconstruction can all produce a type of 2D or 3D equilibrium from a time series of data. An intriguing possibility is to use those equilibria as the initial condition for a proper type of 2D or 3D simulation such as MHD and Hall-MHD. We would then be able to analyze time evolution of actually encountered plasma structures in space, and study the relation to microphysical processes/phenomena, insights into which can be obtained from measurements of high-frequency electromagnetic waves and/or of particle velocity distributions.

## 3. Temporal Aspects of Magnetopause Reconnection

There are continued debates on whether magnetic reconnection is an intrinsically steady or non-steady process, and on whether it is a continuous or intermittent process (e.g., Frey et al., 2003; Trattner et al., 2008). Since the magnetopause is exposed to time-varying flows of the shocked solar wind, it is not easy to address these questions with observations of the magnetopause and surrounding regions. Also, there seems to be some confusion about terminology, arising from the fact that discussion is sometimes given without making clear exactly what is meant by continuous or intermittent reconnection. Thus, we first summarize our views on the temporal nature of magnetopause reconnection in Table 2 and Fig. 12.

In Table 2, the term "FTE" (flux transfer event) is used, from a phenomenological point of view, to represent a signature observed at or around the magnetopause, characterized by a bipolar variation of the magnetic field component $B_N$ normal to the nominal magnetopause and an enhancement of the field intensity. Such signatures were first reported, and were attributed to patchy and transient reconnection at the magnetopause, by Russell and Elphic (1978). A number of mechanisms for FTE generation have been proposed ever since, some invoking bursty, single X-line reconnection (Scholer, 1988; Southwood et al., 1988) and others requiring multiple X-lines (see, e.g., Scholer (1995) and Raeder (2006) for reviews of FTE models). Nowadays, it can be said that an FTE is a signature of some form of time-dependent reconnection. In Section 3.1, we discuss some global models of FTE generation and recent FTE observations, demonstrating that FTEs can be generated by time-dependent reconnection accompanying multiple X-lines or separator-lines. Hereafter, the term "X-line" is used to represent both an X-line in 2D reconnection and a separator in 3D reconnection (e.g., Dorelli et al., 2007).

Magnetopause reconnection can be quasi-steady only when the associated X-line is stationary in the rest frame of a celestial object in question (in this case, the Earth). This is because the boundary conditions external to the magnetopause current layer vary, depending on its location, so that the temporal behavior of reconnection would be modified as the X-line travels. To our knowledge, there is no report of observational evidence of quasi-steady reconnection. This implies that reconnection is intrinsically non-steady, that the external conditions are always too time-dependent for reconnection to remain quasi-steady, or that the reconnection site is moving along the associated current sheet. As demonstrated in Subsection 2.2.3 (Fig. 5), it is now certain that magnetopause reconnection often occurs in a time-dependent manner, although we still don't understand the fundamental physics underlying such an evolution. There are also observational indications of the X-line motion, as presented in Sections 3.1 and 3.2.

It should be emphasized that even if a particular reconnection site is transiently active, a collection of such locally transient reconnections may be taken as continuous reconnection from a global viewpoint (Fig. 12). We thus define reconnection as globally continuous if the reconnection rate is non-zero somewhere within the area or volume under consideration (Table 2(b)). Events consistent with such globally-continuous reconnection are discussed in some detail in Section 3.2. On the other hand, observations consistent with locally continuous reconnection (Table 2(a)) have been reported by Phan et al. (2004), who suggested that a reconnection site in the subsolar magnetopause region can be quasi-stationary and continuously active for more than two hours under steady southward IMF conditions. Phan et al. (2004) also suggested, based on the fact that the FTE signatures were observed in their event, that although reconnection was continuous, its rate was varying with time; FTEs can be generated by continuous but time-dependent reconnection. A simulation study suggests that such time-dependent (in this case, oscillatory) reconnection could be induced in the presence of significant pressure anisotropy ($p_\perp > p_\parallel$), through coupling with the mirror-mode instability (Chiou and Hau, 2003).

### *3.1. Multiple X-line reconnection*

*3.1.1. Models of multiple X-line reconnection*

Multiple X-line reconnection or tearing instabilities in the magnetopause have been put forward as one mechanism for FTE generation (Lee and Fu, 1985). This process is essentially time-dependent because an X-line exists in the direction of the outflow jet from another X-line such that the activity of the former X-line can be affected by that of the latter, and vice versa (e.g., Nakamura et al., 2010b). (Note, however, that if the X-line has a relatively short extent, as inferred from FTE observations reported by Fear et al. (2010), multiple X-lines and the resultant reconnection jets may exist more or less independently of each other. ) Lee and Fu (1985) postulated that multiple X-lines at different latitudes become active simultaneously. A drawback of this model is that it cannot explain in a simple way why FTE occurs repeatedly, with an average interval between two neighboring FTEs of ~8 min (e.g., Rijnbeek et al., 1984). Raeder (2006) suggested that a key factor to overcome this difficulty is the tilt of the geomagnetic dipole axis, which has an angle of ~11° with the Earth's spin axis, so that the dipole tilt varies diurnally and biannually in the GSM or GSE coordinate system.

Figure 13 shows the evolution of the dayside magnetopause current layer seen in a global MHD simulation of the solar wind-magnetosphere interaction under a large dipole tilt, reported by Raeder (2006). The dipole axis is tilted sunward (summer) in the northern hemisphere, so that the magnetic equator on the dayside is substantially southward of $z = 0$ in GSE coordinates. The IMF is set to be nearly due southward, allowing for magnetopause reconnection to occur in the subsolar region. Figure 13 clearly demonstrates that the magnetopause current layer evolves significantly with time and often accompanies more than one X-line.

In the stage of reconnection initiation, an X-line forms in the subsolar current layer somewhere between the flow stagnation point near $z = 0$ and the magnetic equator located to the south. This is because the current density is maximized there. Since the magnetosheath field intensity has a maximum near the stagnation point, around which the IMF is most compressed, and since the field intensity in the outer magnetosphere earthward of the dayside magnetopause has a local maximum near the magnetic equator (e.g., Mead, 1964; Roederer, 1970), the magnetic field gradient (and hence current density) in the magnetopause has a local maximum somewhere in between, where magnetic diffusivity is maximized in the presence of only numerical dissipations in the simulation (Raeder, 2006). Since there is a southward magnetosheath flow immediately outside the reconnecting current sheet, the X-line in the subsolar region cannot stand still but starts to move southward, i.e., into the winter hemisphere, along the magnetopause. It is indeed seen in Fig. 13 that the X-line at $(x, z) \sim (8.5, -3)$ $R_E$ in panel (a) moved to $(x, z) \sim (8, -5)$ $R_E$ in panel (b), and to $(x, z) \sim (7, -7)$ $R_E$ in panel (c). As it travels, this X-line becomes inactive, or at least less active, probably because the external conditions farther away from the initiation site are less favorable for reconnection. As a result, the subsolar magnetopause can become thin again and a new X-line is formed near the location of the previous X-line formation (at $(x, z) \sim (8.5, -3)$ $R_E$ in panel (d)). A magnetic island or flux rope can thus be created between the two X-lines. The main contention by Raeder (2006) is that this flux rope is identified as an FTE.

The advantage of Raeder's model for FTE generation is that it can explain the recurrence of FTEs in a natural way: the sequence of the motion and reformation of X-line and flux rope, as shown in Fig. 13, can repeat itself (see also Fig. 14(b)). This is why Raeder calls this process "sequential multiple X-line reconnection" (SMXR). With no dipole tilt, there may be only a single X-line in the subsolar region and this X-line may sit still, because the stagnation point and the magnetic equator can be collocated (Raeder, 2006) (Fig. 14(a)). If this is the case, no flux ropes would form, and no FTEs would occur unless the reconnection rate varies significantly in time (Phan et al., 2004). Possibly testable predictions of the present model are that most FTEs move into, and are observed in, the winter hemisphere, and that FTEs or, more specifically, flux rope-type FTEs (rather than FTEs emerging from a single X-line with no full turn of field line in the plane perpendicular to the X-line) occur more frequently in summer/winter, when the dipole tilt is larger, than in spring/autumn. Subsection 3.1.2 presents examples of FTEs that are essentially consistent with the SMXR process.

On the other hand, other global hybrid as well as MHD simulations show that FTEs can be created with no dipole tilt (Omidi and Sibeck, 2007; Dorelli and Bhattacharjee, 2009; Tan et al., 2011). A key difference from Raeder's simulations is the way electrical or numerical resistivity is imposed in the simulations: in the MHD simulation by Dorelli and Bhattacharjee (2009), as high a spatial resolution as possible was used; in the 2D hybrid simulations by Omidi and Sibeck (2007), a spatially uniform resistivity was imposed; whereas the 3D hybrid simulations by Tan et al. (2011) prescribed a resistivity that was proportional to the current density. How the reconnection electric field arises and is sustained, and how in reality it is related to the current density are poorly understood. Thus, it is necessary to reveal the magnetic topology and occurrence patterns (latitudinal distribution and motion, and their dependence on the dipole tilt angle) of FTEs on the basis of a statistical survey. Such studies could also help constrain the microphysics of reconnection.

*3.1.2. Observations of multiple X-line reconnection*

Boudouridis et al. (2001) reported DMSP satellite (at an altitude of ~800 km) observations in the high-latitude magnetosphere that are consistent with multiple X-line reconnection at the dayside magnetopause. Their observations showed an overlap of two distinct ion energy dispersions within the same flux tube, indicating that the injection of magnetosheath ions into that flux tube occurred twice, i.e., the corresponding field lines experienced magnetopause reconnection at two distinct locations and/or times. Until recently, however, there was no observational evidence that FTEs can be generated by multiple X-line reconnection, while evidence has been reported of a flux rope resulting from multiple X-line reconnection in the tail (e.g., Eastwood et al., 2005a). Here, we present key observations of the first report of an FTE that is precisely consistent with formation via multiple X-line reconnection in the subsolar magnetopause (Hasegawa et al., 2010a). The event occurred on 14 June, 2007, namely, near the summer solstice in the northern hemisphere.

Figure 15 shows observations by four (THB, THC, THD, and THE) of the five THEMIS spacecraft in the dayside magnetopause region under a weakly southward and dominantly duskward IMF condition. The four satellites were separated by about 1 $R_E$, and in the order of THE, THC, THD, and THB in the radial direction (Fig. 16). THC was at (10.2, 3.7, −2.3) $R_E$ in GSM coordinates, somewhat southward of the equator, at ~0400 UT when it encountered the FTE. THEMIS was generally traveling from the dayside magnetosphere into the magnetosheath, as inferred from the outbound magnetopause crossings by THC and THD. At ~0400 UT, all four satellites observed signatures of a southward-moving FTE, an enhancement of the field magnitude (Fig. 15(f)) and negative-to-positive perturbation of $B_x$, which is roughly the $B_N$ component near the subsolar magnetopause, (Fig. 15(g)). Consistently, THD, closer to the subsolar point than THC, observed the field reversal in the bipolar signature slightly earlier than THC (Fig. 15(g)), and the HT velocity determined for the FTE interval, which can be taken to represent the FTE motion, had a southward and weakly duskward component: $\mathbf{V}_{HT}$ = (−46, 11, −103) km s$^{-1}$ in GSM.

A conspicuous feature of this event is a flow reversal around the FTE, seen by THC and THD: significant northward and dawnward flows before the FTE, and southward and duskward flows after the FTE (Fig. 15(d, e)). Such flow reversals have often been interpreted as a signature of a northward moving X-line. However, this is not the case here, because the FTE between the two oppositely-directed jets was moving southward: an O-line was moving southward. The result indicates that the two jets originated from two different X-lines, respectively, i.e., there was more than one X-line. Other signatures consistent with multiple X-line models are: slow speed (52 km s$^{-1}$ in the plane transverse to the FTE axis) of the HT frame, and bidirectional, field-aligned fluxes of heated magnetosheath electrons (electron MSBL), seen by THB on the magnetosheath side of the FTE (at ~0400:10 UT in Fig. 15(l, m)). Our argument is that if the FTE arose from a single X-line, it would have been swept away by an Alfvénic reconnection jet, so that its speed should have been comparable to the magnetosheath Alfvén speed 244 km s$^{-1}$ (Hasegawa et al., 2010a). The latter electron signature demonstrates that the field lines encountered by THB were reconnected on both north-dawn (subsolar) and south-dusk sides of the FTE, as depicted in Fig. 16. Hasegawa et al. (2010a) also produced 2D maps of the FTE cross-section with the magneto-hydrostatic GS reconstruction (Section 2.2.3). The flux rope constituting the FTE was ~1 $R_E$ in diameter (Fig. 16) and, importantly, its core part was elongated in the magnetopause normal direction, suggesting that the FTE flux rope was squeezed by the two oppositely-directed jets, converging toward the FTE.

There are also features in agreement with the SMXR model: the FTE was encountered in, and was traveling into, the southern (i.e., winter) hemisphere, and only anti-field-aligned but no significant field-aligned fluxes of energetic ions were observed in the MSBL (see the highest energies at ~0400:10 UT in Fig. 15(j, k)).  The latter suggests that reconnection on the subsolar side of the FTE became active later than that on the south-dusk side; the field-aligned streaming ions from the subsolar X-line had not yet reached THB (Fig. 16).  We also note that another FTE was encountered by THB at ~0347 UT (not shown), about 13 min before the FTE in Fig. 15.  The preceding FTE was also moving southward, i.e., into the winter hemisphere, and accompanied a flow reversal from northward to southward.  All these features are essentially consistent with the SMXR, and the recurrence of similar FTEs suggests that the process of X-line (or equivalently FTE) motion and reformation, as illustrated in Fig. 14(b), repeated itself.  FTEs suggestive of generation by multiple X-line reconnection have also been reported by Zhang et al. (2008) and Trenchi et al. (2011), and FTEs analyzed by Hasegawa et al. (2006a) also had properties not inconsistent with the SMXR, in that they were encountered in the winter hemisphere and repeatedly.  Note also a statistical tendency for FTEs to occur in the winter hemisphere (Korotova et al., 2008).  Most recently, Øieroset et al. (2011) showed evidence for the 3D nature of an FTE flux rope generated by multiple X-line reconnection: its core part had a magnetically open topology.

The observations reported by Boudouridis et al. (2001) are now revisited.  We argue that their event is also basically consistent with the SMXR model.  Their event occurred on 10 January, 1990, when the DMSP satellites were in the southern hemisphere, namely, in the summer hemisphere.  We note that in the simple 2D picture of the SMXR, as depicted in Fig. 14(b), the overlap of two ion energy dispersions would occur only in the summer hemisphere (for example, on field line **a** in the lower middle panel).  This is because in the winter hemisphere, the field lines that have experienced reconnection more than once (field lines **a''** and **b''** in Fig. 14(b)) are fully embedded in a magnetic island/flux rope and are not connected to the ionosphere. (Note, however, that in the actual 3D space, the ends of the field lines constituting the flux ropes may extend into interplanetary space or be connected to the ionosphere in either hemisphere. ) Thus, the observation in the summer hemisphere of the overlapped energy dispersions (Boudouridis et al., 2001) is not inconsistent with the SMXR picture, although the absence of the overlap in the winter hemisphere has not been confirmed.  Our speculation is then that, if the SMXR model is correct, the dispersion overlaps would be found more frequently in the summer than in the winter hemisphere. This could be tested by investigating the occurrence patterns of the dispersion overlaps, based on ion precipitation data such as from DMSP and Cluster satellites (Trattner et al., 2012).

### *3.2. Globally continuous reconnection*

#### *3.2.1. Remote observations of globally continuous reconnection*

While *in situ* observations during a steady southward IMF suggest that magnetopause reconnection in the subsolar region can persist with a quasi-stationary single X-line (Phan et al., 2004), reconnection tailward of the cusp may exhibit a distinct behavior, because of the fast anti-sunward flow present in the high-latitude or flank magnetosheath. In particular, Cowley and Owen (1989) theoretically showed that in the presence of a super-Alfvénic magnetosheath flow (in the Earth's rest frame), no steady reconnection is possible, or the reconnection site cannot be stationary. Even under sub-Alfvénic flow conditions, an X-line may move as Raeder's simulations and the observations presented in Section 3.1 show. The question is thus whether tailward-of-the-cusp reconnection can be continuous and, if so, in a local sense or only in a global sense (Table 2).

Frey et al. (2003) demonstrated that tailward-of-the-cusp reconnection can persist, at least in a global sense, for hours under northward IMF. Figure 17 shows the conjunctive observations by the Cluster and IMAGE spacecraft, studied also by Phan et al. (2003). For a prolonged northward IMF, Cluster traversed the high-latitude magnetopause poleward of the northern cusp and detected high-speed proton jets, first directed tailward (~1455 UT in the top left panel) and then sunward (~1457 UT), from a high-latitude reconnection site. At the same time (fourth image in the bottom panels), IMAGE observed a proton auroral spot at latitude ~80° poleward of the dayside auroral oval, which can be linked to the reconnection jets via the Tsyganenko model field lines (Tsyganenko, 2002, and references therein). The energy fluxes of the precipitating protons observed by Cluster can account for the intensity of the aurora emission, establishing the one-to-one correspondence between the aurora spot and the reconnection jets: the spot is a remote signature of, and hence can be used to monitor, high-latitude reconnection (Phan et al., 2003).

Striking features are that the bright proton auroral spot existed uninterruptedly for ~4 hours (Fig. 17), with roughly constant emission intensity and at similar latitudes (Frey et al., 2003). The images were taken on a 2-min cadence, thus the results imply that there was no cessation of reconnection for any time period longer than ~4 minutes over the 4-hour period; reconnection was essentially continuously active on the field lines that map to the aurora spot.

The magnetic local time (MLT) location of the spot responded rather quickly to changes in the east-west ($y$) component of the IMF (whose orientation is indicated by the arrows in the green inserts in Fig. 17, with positive $z$ pointing up and positive $y$ to the left). The spot moved to post-noon for positive $B_y$ and to pre-noon for negative $B_y$, following the locations favorable for anti-parallel reconnection. This correlation suggests that magnetopause reconnection is a process driven by the impinging solar wind and IMF with variable directions, and that the anti-parallel reconnection site, which can accelerate plasma more strongly than component reconnection, produces the most intense proton aurora.

The present observations demonstrate that tailward-of-the-cusp (as well as subsolar) reconnection can be continuous, at least when viewed globally. The next question is whether the persistence of the proton spot is due to a single reconnection site that is more or less stationary and active continuously in the same portion of the high-latitude magnetopause (continuous in a local sense as well), or whether it is a combined effect of multiple X-lines, each of which may be transiently active and/or be moving (continuous only in a global sense). Since the field lines in the magnetopause boundary region may not be well represented by the model field, it is hard to tell exactly to which part of the magnetopause the aurora spot is connected. *In situ* observations are thus indispensable for revealing the actual behavior of tailward-of-the-cusp reconnection.

3.2.2. *In situ observations of globally continuous reconnection*

Hasegawa et al. (2008) reported Cluster observations on 19-20 November 2006 of the high-latitude magnetopause region, which suggest continuous tailward-of-the-cusp reconnection only in a global sense. For a prolonged northward and dawnward IMF interval, Cluster was skimming the dusk-flank magnetopause in the southern hemisphere (from $x \sim -5$ to $+1$ $R_E$ in GSM), the portion of the magnetopause where anti-parallel reconnection is expected for the observed IMF orientation.  Reconnection jets in the form of Alfvénic acceleration of magnetosheath ions were identified almost always when either of the Cluster spacecraft traversed the magnetopause during a 16-hour interval, implying that tailward-of-the-cusp reconnection was active quasi-continuously.  Retinò et al. (2005) also reached the same conclusion based on similar Cluster observations on another day.

For the majority of the 16-hour interval, the reconnection jets were directed tailward and had a speed higher than the magnetosheath flow, indicating that a dominant X-line was on the sunward side of Cluster but was tailward of the southern cusp.  However, Cluster occasionally observed flows consistent with acceleration on the sunward side of an X-line, and ion velocity distributions that consisted of two distinct populations of magnetosheath origin, separated in the field-aligned direction: one with a speed lower than the magnetosheath flow and the other with a higher speed.  The former indicates that an X-line was occasionally on the tailward side of Cluster, and the latter suggests that more than one X-line existed, at least for part of the interval.  The HT frame velocity determined for a jet sunward of an X-line had a tailward component ($\mathbf{V}_{HT} = (-26, 53, 107)$ km s$^{-1}$ in GSM) (Hasegawa et al., 2008).  Thus, the associated X-line must have been retreating tailward at an even higher tailward velocity, because in the rest frame of the field lines sunward of the X-line, the X-line should be moving tailward at a velocity comparable to the magnetosheath Alfvén speed.  This non-stationary behavior of the X-line suggests that tailward-of-the-cusp reconnection was quasi-continuous, not locally but only globally.

It is also noted that Cluster observed a tailward jet soon (~5 min) after encountering the jet on the sunward side of the retreating X-line, implying that another X-line existed sunward of the retreating X-line.  The tailward jet had the HT velocity $\mathbf{V}_{HT} = (-327, 201, -150)$ km s$^{-1}$.  Assuming that the O-line between the two X-lines was moving at about the mean of the two HT velocities (one for the sunward jet and the other for the tailward jet) and that the X-line velocity was similar to the O-line velocity, the velocity is estimated to be about $(-177, 127, -21)$ km s$^{-1}$, comparable to the magnetosheath flow velocity.  The X-line was possibly being swept anti-sunward by the magnetosheath plasma.

Hasegawa et al. (2008) interpreted these observations with a scenario as illustrated in Fig. 18, which has some similarity to the SMXR on the dayside magnetopause discussed in Subsection 3.1.1 (Raeder, 2006; 2009). For most of the interval under consideration, a dominant X-line was probably near, but tailward of, the southern cusp (top sketch). At some point, this X-line began traveling tailward, eventually down to some location tailward of Cluster, thereby allowing for the identification of the jets accelerated sunward and the reconnected field lines sunward of an X-line being convected tailward. The two counter-streaming (in the field-line rest frame) ion populations can be observed on magnetic island-type or flux rope-type field lines, created through the formation of a new X-line sunward of Cluster, possibly near the location of the initial X-line formation (middle sketch). This process of retreat and reformation of X-line in the tailward-of-the-cusp magnetopause can recur, as in the SMXR process, and thus may lead to a repetitive encounter of flux rope-type FTEs in the high-latitude magnetopause region. This remains to be confirmed observationally.

The present results suggest that the persistence of the dayside proton spot, as reported by Frey et al. (2003), may only indicate that an active reconnection site exists continuously somewhere within the area on the magnetopause linked to the spot, and may be a collective effect of multiple reconnection sites, each of which may be only transiently active and/or retreating. We also note that the reversal from a tailward to a sunward jet, in the observation reported by Phan et al. (2003) (Fig. 17, top left panel), can in fact be explained by a tailward retreat of the X-line. Therefore, the *in situ* observations discussed in this section (Phan et al., 2003; Hasegawa et al., 2008, 2010a) all seem to suggest that magnetopause X-lines travel in the presence of significant magnetosheath flow. We point out, however, that the X-line motion may be affected also by magnetospheric plasmas, especially when they have a large inertia and stream at a substantial velocity, as at Jupiter.

4. **Kelvin-Helmholtz Instability in the Magnetopause Region**

The Kelvin-Helmholtz instability (KHI) is a widely-known fluid instability excited on an interface between two media flowing at different velocities relative to each other. It has been suggested to occur along the magnetopause (Dungey, 1955) and the inner edge of the LLBL (Sonnerup, 1980; Sckopke et al., 1981), across which a significant velocity shear exists between the roughly stagnant magnetospheric plasma and the anti-sunward streaming magnetosheath or LLBL plasma. The KHI in these regions has been studied extensively over the past few decades, because it could be a key ingredient for mass, momentum, and energy input from the solar wind into the magnetosphere (see, e.g., Fujimoto and Terasawa (1994) for theoretical work on mass transport, Miura (1984) for momentum transfer, and Atkinson and Watanabe (1966), and Southwood (1968), for a form of energy transfer). In this section, we first describe possible roles of the KHI in order to emphasize the importance of understanding the KHI in the magnetopause region (Section 4.1; see also a brief summary in Table 3). This is followed by a discussion of how plasma vortices generated by the KHI can be identified from *in situ* data (Section 4.2). Key observations are then reviewed, regarding the excitation (Section 4.3), evolution (Section 4.4), and consequences (Section 4.5) of the magnetopause KHI.

*4.1. Possible roles of the Kelvin-Helmholtz instability*

The magnetopause KHI, as an important ingredient of magnetospheric physics, was discussed perhaps first by Atkinson and Watanabe (1966), who suggested that the KHI could lead to the generation of magnetospheric ultra-low frequency (ULF) waves in the Pc5 (2-7 mHz) band. Theoretical works on the ULF wave generation, directly by the KHI, or via a coupling between resonant field lines and magnetopause surface waves that can be excited by the KHI, were carried out by Southwood (1968, 1974) and Chen and Hasegawa (1974). Magnetospheric ULF waves may further interact with and accelerate electrons in the outer radiation belt (e.g., Elkington, 2006). Also, the KHI excited on the LLBL inner edge or in the plasma sheet may drive field-aligned currents that connect to the ionosphere (e.g., Sonnerup and Siebert, 2003), and thus may have some relation to dayside aurorae with spatially-periodic patterns (e.g., Lui et al., 1989; Yamamoto, 2008). Accordingly, the KHI could play a key role in the solar wind energy transfer to the magnetosphere and also to the ionosphere.

The KHI as an agent for momentum transfer between the shocked solar wind and the magnetosphere has been studied intensively by, e.g., Miura (1984). It could drive large-scale convection within the magnetosphere through some kind of viscous interaction associated with Maxwell and/or Reynolds stress. However, the relevant studies are mostly limited to an MHD framework. Without a full understanding of the microphysical process underlying the Reynolds-like stress, the significance of a KHI-induced momentum transfer remains unclear. While the viscous interaction in the magnetopause or LLBL, possibly associated with the KHI, appears to make some contribution to magnetospheric convection under northward IMF (Sundberg et al., 2009), magnetopause reconnection is likely to play the primary role.

The most recent advances in KHI research concern the mass, viz., plasma, transfer across the magnetopause. A pioneering work using hybrid (particle ions and fluid electron) simulations was done by Fujimoto and Terasawa (1994), who showed that ion mixing across the velocity shear layer can be enhanced by the nonlinear growth of the KHI, which leads to an overturning of the surface waves, namely, to the generation of rolled-up plasma vortices. Most recent simulations have been conducted using fully kinetic particle codes (Matsumoto and Hoshino, 2006; Nakamura et al., 2011), which demonstrate that particle transport through vortex-induced reconnection or turbulent mixing can occur within and/or around a developed KH vortex. It has been suggested that the plasma transport enhanced by the KHI is the key to the formation of the LLBL and the cold-dense plasma sheet (CDPS) that become prominent under northward IMF conditions (Mitchell et al., 1987; Terasawa et al., 1997; Wing and Newell, 2002; Hasegawa et al., 2004b) when reconnection is unlikely to occur at the low-latitude magnetopause.

The KHI, in general, is also suggested to initiate an eddy turbulence, which is ubiquitous in fluids and plasmas (e.g., Roberts et al., 1992; Matsumoto and Hoshino, 2004). Turbulence is known to transport energy from large to small spatial scales, and vice versa (e.g., Miura, 1999), and to play a role in structure formation/decay. When the energy in turbulence is dissipated, the ambient media is heated, or particles are energized (e.g., Zimbardo et al., 2010). An understanding of the magnetopause KHI could thus have a broader impact which would not be limited to magnetospheric physics.

### *4.2. Identification of rolled-up Kelvin-Helmholtz vortices*

Cluster multipoint measurements have allowed us to identify complex, multi-dimensional structures such as KH vortices, and to clarify the signatures of overturning, or its absence, of KH surface waves that can be identified from single-spacecraft measurements. In this section, we discuss in some detail the basic structure of KH vortices and methods for the detection of the rolled-up vortices, i.e., overturned waves, from *in situ* data. First, it must be kept in mind that not all magnetopause surface waves are of KH origin, because there are other mechanisms which lead to the excitation of surface waves, such as dynamic pressure variations in the solar wind or magnetosheath (e.g., Sibeck et al., 1989; Sibeck, 1990), non-steady magnetopause reconnection that can generate bulges in the magnetopause or FTEs, and possible Rayleigh-Taylor instabilities.

*4.2.1. Basic structure of the rolled-up vortex*

As the KHI grows to a nonlinear phase, rolled-up vortices are generated as depicted in Fig. 19, which shows the density, streamline, and total pressure patterns. Although these vortices never reach a strictly steady state in such an open system as the magnetosphere, the force balance there can be approximated by Eq. (7), which indicates that in the vortex rest frame, the centrifugal (and other minor inertia) forces acting on the plasmas flowing along the curved streamlines balances the total pressure gradient force. The centrifugal forces push out plasmas from the central part of the vortices, thereby generating a local minimum in the total pressure at the center ("L" in Fig. 19) and a maximum at the hyperbolic point ("H" in Fig. 19) between the vortices (e.g., Miura, 1999).

Since the vortices growing along the magnetopause generally travel tailward, a spacecraft, which is almost stationary in the Earth's rest frame at geocentric distances of the magnetopause, would traverse the vortices from their tailward to sunward side, as shown by the white arrow in Fig. 19. Accordingly, it is expected that the spacecraft observes quasi-periodic oscillations in the density, total pressure, and *N* and *M* components of the velocity. Such variations were indeed seen in the Cluster observations on 20 November, 2001 of rolled-up vortices at the dusk-flank magnetopause (Hasegawa et al., 2004a), as shown in Fig. 20, and also in Geotail spacecraft observations reported by Fairfield et al. (2007). The wavelet analysis of the total pressure (Fig. 20(d)) indicates that the surface waves had a period of ~200 s, which can be converted to a wavelength of ~40,000 km via an estimate of the wave phase speed ~200 km s$^{-1}$ tangential to the nominal magnetopause (Hasegawa et al., 2009b) (this estimate of the phase speed is based on the mean ion bulk velocity during the magnetopause traversal; see Hasegawa (2009) for a discussion of how to estimate the phase velocity of surface waves).

Another salient feature expected when the spacecraft path passes through the central part of the vortices is the approximate coincidence of total pressure maxima and density jumps from the magnetosphere to the magnetosheath, which occurs around the hyperbolic points (marked by "H" in Fig. 19). This is also confirmed in the Cluster observations, which show that rapid density increases at the outbound magnetopause crossings mostly have corresponding total pressure maxima (red dashed lines in Fig. 20). This result thus suggests that whether or not an observing platform has traversed the vicinity of a vortex center can be inferred from single-point measurements alone.

*4.2.2. Semi-quantitative assessment of total pressure variation*

In the initial stage of the KHI, the magnetopause can be approximated by a planar surface, so that the total pressure should roughly be balanced, i.e., there would be no significant total pressure variations across and around the magnetopause. Let us discuss how large the total pressure difference $\Delta P$ between the hyperbolic point and vortex center should be if the KHI is in the nonlinear stage characterized by rolled-up vortices.

Suppose that $r_c$ is the curvature radius for a streamline at the magnetosheath- or magnetosphere-side edge of the vortex, where the flow speed in the vortex rest frame is $v_0 \sim \Delta V/2$. Here, $\Delta V$ is the total velocity jump across the magnetopause. Before the KHI growth, the streamlines are nearly straight with $r_c \gg \lambda_{KH}$, where $\lambda_{KH}$ is the wavelength of the KHI. In the phase of a rolled-up vortex, however, the curvature radius would become comparable to, or smaller than, a quarter of the wavelength, $r_c \lesssim \lambda_{KH}/4$, where it is assumed that a roughly circular vortex with radius $\lambda_{KH}/4$ is embedded within one KHI wavelength (cf. Fig. 19). We note that the force balance at the edges of the vortex can be written as

$$\frac{\rho v_0^2}{r_c} = \frac{\partial P}{\partial r} \sim \frac{\Delta P/2}{\lambda_{KH}/4}, \tag{12}$$

where $r$ is the distance from the vortex center and it is assumed that the magnitude of the total pressure gradient is constant in and around the vortex on the scale of one KHI wavelength. The curvature radius is then

$$r_c \sim \frac{\rho v_0^2}{\Delta P} \frac{\lambda_{KH}}{2} \sim \frac{\rho \Delta V^2}{2\Delta P} \frac{\lambda_{KH}}{4}. \tag{13}$$

It is seen that $r_c$ decreases as $\Delta P$ increases during the course of the KHI growth. With Eq. (13), the above condition, $r_c \leq \lambda_{KH}/4$, for the roll-up of the KH vortex, can be rewritten as

$$\frac{r_c}{\lambda_{KH}/4} \sim \frac{\rho \Delta V^2}{2\Delta P} \leq 1. \tag{14}$$

Note that this can be checked without information on the actual wavelength of the KHI and even with single-spacecraft data.

With the observed density $\rho \sim 3$ cm$^{-3}$ and velocity jump $\Delta V \sim 200$ km s$^{-1}$ (Fig. 20(a, b)), condition (14) becomes $\Delta P \geq \sim 0.1$ nPa. Figure 20(c) shows that the total pressure difference observed was about 0.1 nPa, consistent with the view that at least some of the vortices encountered were rolled up. As reviewed in the next subsection, the formation-flight observations by Cluster indeed demonstrated the presence of overturned, viz., rolled-up, vortices for the event shown in Fig. 20.

*4.2.3. Multi-spacecraft detection of rolled-up vortices*

A striking feature of rolled-up vortices is that at certain $M$ locations corresponding to some specific wave phases, the density is greater closer to, rather than farther from, the tail center (Fig. 19), because of the intrusion of the magnetosheath plasma into the magnetosphere during the course of roll-up of the vortices. However, it is difficult, if not impossible, to identify such density distributions from single-point measurements, because time series of data can be interpreted in terms of either spatial structures or temporal variations. Evidence for the rolled-up vortices was thus found for the first time by Cluster multi-spacecraft observations, made at the dusk-flank magnetopause during an extended period of northward IMF (Hasegawa et al., 2004a).

The Cluster ion measurements focusing on the interval 2026–2042 UT on 20 November, 2001 are shown in Fig. 21, in which time progresses from right to left for an easier comparison with Fig. 19. The $x$-position of the $i$th spacecraft is defined by $x_i = |\mathbf{V}_{mean}|t + \Delta x_{i1}$, where $\mathbf{V}_{mean}$ is a rough estimate of the vortex velocity, determined by averaging the measured velocities over the interval, $t$ is the time elapsed from the start of the interval, and $\Delta x_{i1}$ is the distance along $x$ (which is parallel to $\mathbf{V}_{mean}$) of the $i$th spacecraft relative to Cluster 1. The separation $\Delta y_{i1}$ along $y$ is doubled in Fig. 21. Ion density variations observed by three of the four Cluster spacecraft, for which ion data are available, agree well with a virtual spacecraft observation of a simulated rolled-up vortex (panel b). Figure 21(c) shows the densities projected onto the *x-y* plane along the spacecraft paths, with the magnetosheath at the bottom and subsolar region to the left. Consistent with the roll-up of KH vortices, the density is greater on the magnetospheric side (at C1), than on the magnetosheath side (at C3), at some phases (marked by the red bars in panel b) near the vortex centers, whose locations can be inferred from the pattern of vortical velocity perturbations shown in Fig. 21(d) and are denoted by red vertical dashed lines.

This kind of direct construction of the 2D distributions using multi-spacecraft data can unambiguously verify the overturning of KH waves. Note, however, that the present method basically assumes time independence of the structures over an interval comparable to the KHI wave period (~200 s for the case in Fig. 20). Also, the spacecraft separation along the nominal magnetopause normal *N* must be modestly, but not extremely, smaller than the width of the vortices, in order for the overturned density structure to be detected. Likewise, the separation along *L* or *M* should probably be not larger than the KHI wavelength, for the construction of 2D distributions not to be contaminated by the effects of possible three-dimensionality or time evolution of the vortex structure. In the Cluster event, the separation was appropriately about 2000 km for the inferred vortex width of order 1 $R_E$ (Fig. 7).

*4.2.4. Single-spacecraft detection of rolled-up vortices*

In the presence of a significant density jump across the initial velocity shear layer, secondary shear layers are created as the KHI enters the nonlinear phase (Nakamura et al., 2004). This is because, at a certain distance from the vortex center, the centrifugal force $\rho v_\phi^2/r_c$ would be nearly equal for both denser and less dense media, where $v_\phi$ is the flow speed in the vortex rest frame; the less dense part in the vortex rotates faster than the denser part. Such a behavior manifests itself as a very peculiar signature when vortices rolled-up at the magnetopause are observed in the Earth's rest, or spacecraft, frame: in the magnetosheath-side part of the vortex, the tenuous magnetospheric plasma has a tailward velocity higher than the magnetosheath plasma. This has indeed been seen in both MHD and hybrid simulations of the KHI (Takagi et al., 2006; Cowee et al., 2010).

Figure 22(a) shows that the scatter plot of the velocity $x$ (or $M$) component $V_x$ versus the density $N$, generated from virtual spacecraft observations of the simulated KH wave or vortex, exhibits a distinct pattern, depending on the phase of the KHI growth. Figure 22(b) shows that the tailward velocity is significantly higher than that of the magnetosheath plasma only in the magnetosheath-side part of the rolled-up vortex. Such predictions were confirmed in the Cluster event of the rolled-up vortices: Figure 23 shows that for part of the low-density ($< 5$ cm$^{-3}$) boundary layer ions, $|V_x|$ is larger than that of the high-density magnetosheath ions ($V_x \sim -250$ km s$^{-1}$) and, importantly, such a lower-density and higher-speed signature was encountered more frequently by the outermost spacecraft (C3) than by the inner ones (C1 and C4) (more green circles than black plusses and blue stars in the lower-left part), consistent with the simulation shown in Fig. 22(b).

The present method has been used to identify rolled-up vortices in the magnetopause region from ion data taken by single spacecraft such as Geotail (Hasegawa et al., 2006b) and Double Star TC-1 (Taylor et al., 2008). It must be kept in mind, however, that its applications should be limited to northward IMF cases, because for southward IMF conditions, when the magnetic shear across the low-latitude magnetopause can be high, lower-density and high-speed tailward flows could result from magnetopause reconnection unrelated to the KHI (Gosling et al., 1986), i.e., regardless of the state of the surface waves.

*4.2.5. Other methods for the surface wave analysis*

When four-spacecraft measurements are available and their inter-spacecraft distance is sufficiently shorter than the scale of the structure in question (in this case, magnetopause vortices), the so-called four-spacecraft timing analysis (Russell et al., 1983; Schwartz, 1998) can be used for the determination of the normal, and the velocity along it, of a discontinuity or surface, such as shocks and current sheets. This method uses as input the information on the time at which each spacecraft crosses the discontinuity/surface and the spacecraft position at the crossing time. The underlying assumptions are that the discontinuity/surface is planar, and the normal direction and the velocity are time-independent over the interval of the crossings. Owen et al. (2004) and Foullon et al. (2008) applied the timing method to Cluster data for an estimation of the normal to undulating magnetopause surfaces (note, however, that in their studies the results from the timing method were not properly used for the estimation of the associated wavelength (Hasegawa, 2009)).

We point out that the timing method can be used to inspect whether magnetopause surface waves are more or less sinusoidal (as depicted in Fig. 24(a)), or are overturned (Fig. 24(b)). Let us discuss the situation of the dusk-flank magnetopause, where it can be assumed that the surface wave propagates anti-sunward along the nominal magnetopause (to the left in Fig. 24). Suppose that the local boundary normal can be estimated by the timing method and its direction is defined positive along the local boundary velocity normal to the surface. Then the normal would always have an anti-sunward (–$M$) component. If the wave is in the linear, viz., quasi-sinusoidal, phase, the normal should have an outward (+$N$) component for inbound crossings, while it should have an inward (–$N$) component for outbound crossings (Fig. 24(a)). On the other hand, if the wave is in the nonlinear, viz., overturned, phase, the normal may have an inward (–$N$) component for some inbound crossings as well as for outbound crossings (Fig. 24(b)). Such abnormal orientations of the boundary normal can be taken as a signature of the wave overturning.

In principle, the orientation and motion of the boundary surface can be estimated by single-spacecraft methods, as described by Sonnerup et al. (2006b), and those results can be used to analyze the state of surface waves in the same way as discussed above. Moreover, information only on the boundary normal is, in fact, sufficient for the present purpose, if it can be assumed that the wave propagates in a specific direction (in this case, anti-sunward). Such an assumption removes the 180° uncertainty in the boundary normal direction that exists for some of the methods, such as the minimum variance analysis of the magnetic field (MVAB) (Sonnerup and Scheible, 1998). This approach has actually been taken by Boardsen et al. (2010) to analyze magnetopause surface waves at Mercury, using only magnetic field data recorded by the MESSENGER spacecraft during its third flyby of the planet. Figure 25 shows the magnetic field data for a 5-min period when the spacecraft traversed the dusk-flank boundary region from the magnetosheath into the magnetosphere. MVAB was applied to the field data taken during the intervals of magnetopause crossing, marked by the arrows in the bottom panel, which occurred quasi-periodically with a time separation of the order of 10 s. Boardsen et al. (2010) found that the magnetopause normal directions had variations consistent with the KH surface waves steepened at their leading edges.

We also remark that the reconstruction of the velocity field based on the GS-like equation for the stream function, reviewed in Subsection 2.2.5, can be used to estimate the amplitude of surface waves or the width of KH vortices, from which the phase of the KHI growth may be inferred. Hasegawa et al. (2007a; 2009b) and Nishino et al. (2011) indeed argued that the widths of the reconstructed vortices of order ~1 $R_E$ are consistent with the roll-up of the KH vortices (e.g., Fig. 7).

*4.3. Excitation of the Kelvin-Helmholtz instability*

According to linear theories of the KHI, the fastest growing wavelength is about ten times the full width of the initial velocity shear layer (e.g., Walker, 1981; Miura and Pritchett, 1982). A puzzling fact, pointed out by Belmont and Chanteur (1989), is that the estimated wavelengths (a few $10^4$ km) of observed magnetopause surface waves (e.g., Kivelson and Chen, 1995) are usually significantly longer than that predicted by linear theories for typical thicknesses (≤1000 km) of the magnetopause current layer (Berchem and Russell, 1982; Phan and Paschmann, 1996). For wavelengths three times longer than the fastest growing wavelength, the growth rate is about three times lower, which is significant from the viewpoint of where on the magnetopause KH vortices can be rolled up and merge. Belmont and Chanteur suggested that the longer wavelengths could result from the coalescence of a few KH vortices with the fastest growing wavelength into one larger vortex. However, Hasegawa et al. (2009b) noted that the time necessary for the coalescence to be completed is too long to account for the longer wavelengths observed at downtail distances of, at most, $x \sim -20\ R_E$. There would have to be other reasons for the longer wavelengths or wave periods (keep in mind that the wave period $T_{KH}$ can be converted to the wavelength $\lambda_{KH}$ via the phase velocity $V_{ph}$ along the nominal magnetopause, $\lambda_{KH} = V_{ph} T_{KH}$). One possibility is that seed perturbations for the KHI are not white noise, but may have such a spectral property that the waves with a period of 1–5 min, as observed, can be excited.

What could be the origin(s) of such seed perturbations? Some compressional fluctuations may exist in the dayside magnetosheath (external origin). Such fluctuations may originate in the upstream solar wind or foreshock region (Fig. 26), where intense wave activities are observed in the ULF frequency range (e.g., Eastwood et al., 2005b). Fluctuations can also be excited in the magnetosheath itself, for example, by the mirror instability (Hasegawa, 1969), in the presence of temperature anisotropy, although magnetosheath mirror-mode structures (e.g., Horbury and Lucek, 2009) may not have the right scale sizes to excite the KH waves as observed. Another possibility is that seed perturbations are generated in the magnetopause itself (internal origin), e.g., by intermittent reconnection, or the SMXR discussed in Section 3.1, which may lead to recurrent FTEs with an average separation time of a few minutes. For northward IMF conditions, when rolled-up KH vortices are often observed (Hasegawa et al., 2006b), the high-latitude magnetopause is the site favorable for reconnection. Since it takes a finite time for Alfvén waves excited at a high-latitude reconnection site to propagate down to the low-latitude magnetopause, where the KHI tends to be excited, and since the reconnected field lines are advected down the tail during the course of wave propagation, it is unlikely that the reconnection-induced fluctuations have some influence on the KHI excitation in the subsolar region. Nonetheless, they may serve as a seed perturbation for the surface wave excitation in the flank boundary regions.

Not inconsistent with the possible dayside origin of the seed perturbations, the wave period and occurrence frequency of large-amplitude KH waves under northward IMF show no clear dawn-dusk asymmetry (Hasegawa et al., 2006b). Figure 27 shows the Geotail paths in GSM coordinates for the years 1995–2003, along which rolled-up KH vortices were identified by the single-spacecraft method described in Subsection 4.2.4. The survey indicates that rolled-up vortices are observed mostly behind the dawn-dusk terminator, with a roughly equal detection probability on the dawn and dusk sides. The results suggest that the magnetopause KH-mode is probably not affected so much by the local condition of the shear layer thickness, but is controlled mostly by external conditions such as magnetosheath fluctuations and the IMF.

It is also noted that the KHI excitation may be aided by the formation of a dense dayside LLBL, which is likely due to dual lobe reconnection for northward IMF conditions (cf. Figs. 3 and 4). As is widely known, the KHI can grow when the following condition (Chandrasekhar, 1961) is fulfilled,

$$\left[(\mathbf{V}_1 - \mathbf{V}_2) \cdot \hat{k}\right]^2 > \frac{\rho_1 + \rho_2}{\mu_0 \rho_1 \rho_2}\left[\left(\mathbf{B}_1 \cdot \hat{k}\right)^2 + \left(\mathbf{B}_2 \cdot \hat{k}\right)^2\right], \tag{15}$$

where the subscripts 1 and 2 are defined to represent the magnetosheath and magnetosphere sides, respectively (note, however, that the above condition, derived for an infinitesimally thin shear layer, can be relaxed when the shear layer has a non-zero thickness (Gratton et al., 2004)). For $\rho_2 \ll \rho_1$, $(\rho_1 + \rho_2)/\rho_1\rho_2 = (1 + \rho_2/\rho_1)/\rho_2 \sim 1/\rho_2$, so it can be said that inequality (15) becomes more easily satisfied with increasing density $\rho_2$ on the magnetosphere side, i.e., the KHI can be excited thanks to the presence of the dense LLBL. Indeed, inequality (15), assessed for the magnetopause crossing by Geotail at ~15 MLT, was satisfied because of the existence of the LLBL, in the Cluster-Geotail conjunctive observations on 20 November, 2001 of the dusk-flank KH waves (Hasegawa et al., 2009b). On the other hand, Sonnerup (1980) suggested that the inner edge of the LLBL is likely to be susceptible to the KHI, and it was indeed shown that inequality (15) is often satisfied at the LLBL inner edge (Ogilvie and Fitzenreiter, 1989). This is because the magnetic fields in the LLBL and on its earthward side are usually nearly parallel to each other, so that the KH mode with a wave vector nearly transverse to the field is little affected by the stabilizing magnetic tension.

Figure 26 schematically summarizes a scenario of how KH waves with wavelengths longer than predicted by theory can be excited under northward IMF conditions. Although, for a northward IMF, evidence exists of the dense dayside LLBL (e.g., Øieroset et al., 2008) and of quasi-periodic fluctuations in the LLBL (Hasegawa et al., 2009b), little is known about the relationship between those fluctuations and magnetosheath waves/structures, and about whether or not those LLBL fluctuations have connections with high-latitude reconnection. These are the issues that should be addressed in future.

## *4.4. Evolution of the Kelvin-Helmholtz instability*

Simulation studies show that as the KHI reaches the nonlinear phase, current sheets of substantial magnetic shears are generated within, or at the edges of, the rolled-up vortex, and/or the resulting or existing current sheets are squeezed by vortex flows and thus become thin (e.g., Otto and Fairfield, 2000; Nykyri and Otto, 2001; Nakamura et al., 2004, 2011; Takagi et al., 2006). 2D simulations also show that further developments could induce turbulent flow within the vortex, characterized by an inverse or forward cascade of energy, i.e., coalescence or breakup of the vortices (e.g., Miura, 1997; Matsumoto and Hoshino, 2004). Reviewed here are some observational signatures of such nonlinear evolution of the KHI.

Figure 28 shows high time resolution data of the electron density and the magnetic field taken by Cluster in the vortices event on 20 November 2001 (cf. Figs. 20 and 21). Despite the fact that under a northward IMF, there was only a weak magnetic shear across the dayside (~15 MLT) magnetopause which was traversed by Geotail upstream of Cluster (Hasegawa et al., 2009b), the trailing, viz., sunward-facing, edges of the KH surface waves seen by Cluster (at $x \sim -4$ $R_E$) had substantial magnetic shears ~60° (Fig. 28(b)). Furthermore, the magnetic field variations observed in and around the vortices are consistent with 3D MHD simulations of the KHI in a magnetotail-like geometry (Hasegawa et al., 2004a; Takagi et al., 2006), where the KH-unstable plasma sheet is sandwiched by the KH-stable northern and southern lobes. In such situations, the KHI grows only along the low-latitude magnetopause between the plasma sheet and the magnetosheath, so that only low-latitude portions of the field lines in the magnetopause region are engulfed in the vortex flows. As a consequence, current sheets with a significant magnetic shear are generated in the boundary regions between the plasma sheet and the lobes: precisely the location where Cluster was situated ($z \sim -3$ $R_E$).

Another key point is that the current sheets are created in regions around the hyperbolic point (cf. Fig. 19) between the vortices, and thus can be compressed by the vortex flows oriented toward that point. Flows toward the current sheet were indeed observed by Cluster. In Fig. 28(d), $V_{i,N}$ tends to be positive in the plasma sheet and negative in the magnetosheath. Consistently, the four-spacecraft timing analysis shows that the current sheet (at ~2035 UT) was rather thin, with a thickness (~240 km) of only a few times the ion inertia length (~100 km). Since such thin current sheets form exactly at the magnetopause where significant density jumps are present (Fig. 28(c)), reconnection induced there can transport solar wind plasma into the magnetosphere via the interconnected field lines. Signatures of such vortex-induced reconnection were indeed identified by Hasegawa et al. (2009b) (Fig. 29).

As proposed by Hasegawa (1976), the KH surface waves could resonantly couple to kinetic Alfvén waves (KAWs), whose field-aligned electric field can accelerate electrons, thereby possibly generating auroral arcs and/or field-aligned fluxes of heated magnetosheath electrons commonly seen in the LLBL (e.g., Phan et al., 1997; Hasegawa et al., 2003). Chaston et al. (2007) reported a KAW turbulence that appeared to have been driven by mode conversion from the KH surface waves, in the same event as studied by Hasegawa et al. (2004a, 2009b). Hasegawa et al. (2009b) also reported a signature suggestive of vortex coalescence beginning around the Cluster location ($x \sim -4$ $R_E$). The wavelet spectrum of the total pressure, shown in Fig. 20(d), has peaks not only at the dominant KH wave period ~200 s but also at ~400 s, twice the dominant period, for the interval around 2015 UT. Since the dayside LLBL seen by Geotail had no clear spectral peak at this longer period, it is possible that Cluster observed an incipient signature of vortex coalescence, viz., an inverse energy cascade.

*4.5. Consequences of the Kelvin-Helmholtz instability*

Although the significance of the KHI in magnetospheric physics has not been fully verified observationally, here we discuss possible consequences of the KHI and some relevant observations. A specific type of reconnection induced by the KH vortex (e.g., Pu et al., 1990; Nakamura et al., 2011), as illustrated in Fig. 29, can occur exactly in the magnetopause current layer and can transport and/or accelerate particles across the magnetopause. This is termed "type I reconnection" by Nakamura et al. (2008), and its signatures have recently been reported by Eriksson et al. (2009) and Hasegawa et al. (2009b). (Note that another type of vortex-induced reconnection (VIR), termed "type II reconnection" by Nakamura et al. (2008), occurs within either the magnetosheath or magnetospheric regime. Therefore, it does not result in the mixing of magnetosheath and magnetospheric particles, although it can inject magnetosheath plasmas into the magnetospheric side of the magnetopause (Nykyri and Otto, 2001). ) Since magnetic shears resulting from the KHI under northward IMF are not very high, and since the energy corresponding to the electron Alfvén speed in the flank magnetopause region is, at most, 100 eV (Hasegawa et al., 2009b), the acceleration associated with VIR may not be sufficient to generate bright aurorae. Moreover, since VIR appears to be initiated only near or behind the terminator, the role of VIR in the plasma transfer is probably unimportant in the subsolar to near-terminator regions (Hasegawa et al., 2009b). However, the ratio of the parallel to perpendicular temperature of cool protons increases with the downtail distance in the dusk-flank plasma sheet (Nishino et al., 2007a), and field-aligned, counter-streaming proton beams and higher parallel than perpendicular ion temperatures, both consistent with reconnection, have been observed around rolled-up KH vortices detected at $x \sim -15$ $R_E$ on the dusk side (Nishino et al., 2007b). These observations may imply that VIR could be fully developed and play a larger role at greater downstream distances.

Figure 30 illustrates a possible scenario of the flank LLBL formation, proposed by Takagi et al. (2006), when the KHI grows three-dimensionally in the tail-flank geometry and VIR occurs in both northern and southern hemispheres. Magnetosheath field lines deformed by the KHI growth may reconnect with the lobe or plasma sheet field lines in the boundary regions between the two lobes and the plasma sheet, thereby creating new closed field lines (bottom panel of Fig. 30). Consequently, a part of the plasmas that were originally in the magnetosheath flux tubes is captured into the newly closed flux tubes. This plasma entry mechanism is akin to that of dual lobe reconnection (cf. Fig. 3), although the location of reconnection is different. Since the portion of the magnetosheath field lines captured has a much shorter length in the VIR scenario, the amount of plasmas transferred per unit reconnected flux is smaller for VIR than for dual lobe reconnection. Nonetheless, VIR may occur repeatedly in association with the possible coalescence/breakup of the vortices during the course of nonlinear KHI evolution, and thus may contribute to the thickening of the flank LLBL with increasing downstream distance.

Another potentially important phenomenon is kinetic Alfvén wave (KAW) turbulence that can cause anomalous diffusion of particles across the magnetic field (Hasegawa and Mima, 1978). An estimate by Chaston et al. (2007) shows that the amplitudes of KAWs observed in the magnetopause region, possibly mode-converted from the KH surface waves, are sufficient for the formation of the flank LLBL (although we suspect that the so-called Taylor hypothesis (e.g., Matthaeus and Goldstein, 1982) should not be used for electromagnetic fluctuations observed in complex structures with substantial velocity fluctuations (Fig. 20(b)), such as KH vortices, and thus their estimate may not be accurate). KAWs can also stochastically heat low-energy ions, primarily in the direction perpendicular to the magnetic field (Johnson and Cheng, 2001). Ion velocity distributions observed in the LLBL, especially on the dayside, show that protons of magnetosheath origin are heated more strongly in the perpendicular, than the field-aligned, direction, and thus are consistent with the KAW heating (e.g., Nishino et al., 2007a; Hasegawa et al., 2009b). It must be kept in mind, however, that the existence of KAWs heating ions or accelerating electrons does not necessarily mean that there is cross-field diffusion via the KAWs. For cross-magnetopause particle transport to occur, the wavenumber vector $\mathbf{k}_\perp$ transverse to the field must have a component tangential to the magnetopause, i.e., the electric and magnetic perturbations of the KAWs must have significant transverse components tangential and normal to the boundary, respectively.

## 5. Global Effects of Magnetopause Processes

While it is well established that magnetopause reconnection for southward IMF plays a central role in controlling magnetospheric structures and dynamics, relatively little is understood concerning which magnetopause process has a global impact in terms of the mass, momentum, or energy transport under northward IMF. Observations show that cool and dense plasmas of solar wind origin, loaded into the plasma sheet under northward IMF, can eventually have access to the inner magnetosphere (Thomsen et al., 2003; Lavraud et al., 2005b), and such preconditioning during northward IMF periods has a large impact on magnetospheric activities such as geomagnetic storms (Lavraud et al., 2006b). In this section, we discuss some observations of the magnetopause and its boundary layers in the context of the possible global effects of magnetopause processes.

### *5.1. Reconnection effects*

#### *5.1.1. Effects on the plasma transport*

Based on THEMIS multipoint observations on 3 June, 2007, Øieroset et al. (2008) showed that a dense LLBL as thick as 0.9 $R_E$ can form on the dayside in 1–2 hours following a northward turning of the IMF. The observed dayside LLBL exhibited no signatures of the KHI and diffusive transport, and appeared to be mostly on closed field lines. They thus concluded that the LLBL was generated through dual lobe reconnection, or reconnection poleward of the cusp in one hemisphere and equatorward of the cusp in the other hemisphere. Hasegawa et al. (2009a) reported similar THEMIS observations on 11 June, 2008 but during an extended northward IMF period, and estimated the thickness of the dayside LLBL (shown in Fig. 4), most likely created through dual lobe reconnection, to be ~0.35 $R_E$. The IMF clock angle, defined as the angle the IMF projected onto the GSM *y-z* plane makes with the *z*-axis, was about –50° in the Øieroset et al. event, while it was ~10° in the Hasegawa et al. event. The observed difference in LLBL thickness is puzzling, because dual lobe reconnection is expected to be more efficient for IMF clock angles closer to zero. The difference may, however, be attributed to a temporal response of the dayside magnetosphere to the northward turning of the IMF. The dayside LLBL may have an overshoot thickness (possibly seen in the Øieroset et al. event) before reaching a quasi-steady state (possibly seen in the Hasegawa et al. event). Another possibility is that the IMF $B_x$ component as well as the dipole tilt angle (Li et al., 2008) plays a role in controlling the rate at which the IMF field lines are reconnected in both hemispheres and, hence, the dayside LLBL thickness (Hasegawa et al., 2009a).

Based on the fact that a stably tailward ion flow was observed in the post-noon LLBL reported by Hasegawa et al. (2009a) (Fig. 4(d)), we estimate the rate of tailward plasma transport in and along the LLBL. For the estimated LLBL width ~0.35 $R_E$, the observed tangential velocity $|V_M|$ ~100 km s$^{-1}$ (at around 18:40 UT in Fig. 4(d)), number density $N$ ~10 cm$^{-3}$ (Fig. 4(a)), and the assumed flux tube length of order ~10 $R_E$, the number of transported ions (and equivalently electrons) is ~2.8 × 10$^{26}$ s$^{-1}$, given that the LLBL as observed exists on both dawn and dusk sides. We note that this rate is sufficient to fill a volume 50 $R_E$ × 50 $R_E$ × 5 $R_E$ of the plasma sheet with density $N$ ~1.0 cm$^{-3}$ in just ~3.2 hours; if the dayside LLBL plasma is delivered into the nightside plasma sheet, thereby forming the CDPS as observed, dual lobe reconnection can be the dominant plasma entry mechanism for northward IMF. Imber et al. (2006) also reached a similar conclusion, by estimating the solar wind mass loading rate based on ground-based (SuperDARN radar) observations of sunward ionospheric convection across the dayside open/closed field line boundary, resulting from dual lobe reconnection.

The question then is how the dayside LLBL plasmas, most of which are probably loaded by reconnection, can be carried into the nightside plasma sheet where the cool and dense plasmas are often observed during extended northward IMF periods (Terasawa et al., 1997; Wing and Newell, 2002). The paths of reconnection-driven convection under a northward IMF may not necessarily extend to the midnight portion of the near-tail plasma sheet. We infer that viscous drag and diffusion, possibly enhanced by the magnetopause/LLBL KHI, could play a role in the transport of the dayside-loaded plasmas, first into the flank LLBL and then toward the plasma sheet center. We emphasize that in a steady-state picture of the magnetosphere driven by dual lobe reconnection under a due northward IMF (see, e.g., Plate 3 in Song et al., 1999), the captured LLBL plasmas flow exclusively tailward in the equatorial region, provided that there is no plasma diffusion across the topological boundary in the streamline pattern. Consistently, a global MHD simulation conducted for a real event during a northward IMF period shows that once arriving at the low-latitude region, a parcel of the captured magnetosheath plasma continues to travel tailward, though slowly (Li et al., 2005). On the contrary, spacecraft often encounter an inner LLBL or plasma sheet regions where dense magnetosheath-like ions stream sunward or subsolar-ward under a northward IMF (e.g., Fujimoto et al., 1998; Hasegawa et al., 2003). Moreover, Wang et al. (2007) showed that plasma flows in the nightside plasma sheet are on average earthward and flankward, although the data used in their analysis were not limited to dense plasma regions. These observations seem to demonstrate that plasma transport processes other than dual lobe reconnection (Song and Russell, 1992) and reconnection poleward of the cusp in one hemisphere and equatorward of the cusp in the other hemisphere (Øieroset et al., 2008) are operating under northward IMF conditions (see also observations reported by Taylor and Lavraud (2008) and Hasegawa et al. (2009b), which are suggestive of processes other than dayside reconnection). Such processes could be vortex-induced reconnection and cross-field diffusion associated with, for example, KAWs, as discussed in Section 4.5. In particular, the VIR mechanism, as illustrated in Fig. 30, could be able to drive a sunward flow of dense plasmas. We note, however, that the dense sunward flows themselves may be generated, without violating the frozen-in condition within the magnetosphere, if the low-latitude magnetosphere is in a highly non-steady or turbulent state (e.g., White et al., 2001; Borovsky and Funsten, 2003).

The length of the reconnection X-line on the poleward-of-the-cusp magnetopause can be estimated for the event studied by Hasegawa et al. (2009a), assuming that reconnection is fast with the normalized reconnection rate of ~0.1, so that the inflow speed $V_N$ ~35 km s$^{-1}$ for the observed Alfvén speed $V_A$ ~350 km s$^{-1}$: from the magnetic flux conservation, the X-line length $L = 2D|V_M|/V_N$ = 2×0.35 R$_E$×100 km s$^{-1}$/35 km s$^{-1}$ = 2.0 R$_E$, where $D$ is the LLBL thickness and the "2" takes into account both dawn and dusk sides. The result is in approximate agreement with the estimate (at least 3.6 R$_E$) based on proton aurora observations for the event shown in Fig. 17 (Phan et al., 2003) and with the extent ~2.6 R$_E$ of the merging line estimated by Imber et al. (2006). It can thus be concluded that the length of the high-latitude X-line is substantially shorter than that of the low-latitude X-line (possibly of order 10 R$_E$) (Phan et al., 2006, and references therein). This is basically consistent with the observed difference in the cross-polar cap potential between the southward and northward IMF conditions (e.g., Ruohoniemi et al., 2005; Haaland et al., 2007).

*5.1.2. Effects on the magnetospheric activity*

One intriguing possible consequence of sequential multiple X-line reconnection (SMXR), as discussed in Section 3.1, is that it could regulate the rate of energy transfer from the solar wind into the magnetosphere (Hasegawa et al., 2010a). If there is only a single X-line in the dayside low-latitude magnetopause (Fig. 14(a)), reconnection there occurs exclusively between the IMF and closed terrestrial field lines. The consequence is change in magnetic topology from the closed to open field lines, namely, erosion of the dayside field lines, and subsequent transport and storing of magnetic flux into the geomagnetic tail. This accumulation of magnetic energy in the tail forms the basis for the majority of active phenomena in the magnetosphere, including substorms. Let us now discuss the situation when there exists more than one X-line (Fig. 14(b)). Reconnection at some X-line (or some segment of an X-line) during a certain interval may occur between the IMF and closed field lines (as in the lower right panel of Fig. 14(b)). However, reconnection at other X-lines (or other segments of an X-line) or during other intervals may occur between the IMF and already open field lines, which creates another set of IMF-type and open field lines, or between two open field lines, which creates one closed and one IMF-type field lines (or another set of two open field lines if they are both anchored to the same hemisphere), or even on a single open field line (as in the lower middle panel of Fig. 14(b)), which creates one island-type field line and one open field line. In other words, reconnection in the presence of multiple X-lines does not necessarily increase, or occasionally may even reduce, the amount of open magnetic flux. In such circumstances, a smaller number of the IMF field lines impinging on the dayside magnetopause would be reconnected, i.e., a larger number of the IMF field lines would be deflected flankward and be convected anti-sunward, without experiencing reconnection.

Thus, it can be concluded that there are three controlling factors for the reconnection-induced solar wind energy transfer: (i) the electric field or local rate of reconnection, and (ii) the length and (iii) number of X-lines. Suppose now that a single subsolar X-line tends to form under smaller dipole tilt conditions and more X-lines form under a larger dipole tilt, as in Raeder's simulations, and that the global reconnection rate, i.e., the total amount of magnetic flux reconnected per unit time, at the most subsolar-ward X-line does not depend on the dipole tilt. Then, the above discussion suggests that more open field lines are created, i.e., the transfer of solar wind energy into the tail is more efficient, for smaller dipole tilt conditions.

It is widely known that geomagnetic activity indices such as *Kp* and *Dst*, the proxies of magnetospheric activities, exhibit semiannual variations and that magnetospheric phenomena, such as geomagnetic storms and substorms, tend to be more active near equinoxes (e.g., Russell and McPherron, 1973; Mursula and Zieger, 1996). A substantial part of such seasonal or dipole tilt dependences has been attributed to the so-called Russell-McPherron effect (Russell and McPherron, 1973; McPherron et al., 2009): when the dipole tilt is smaller in GSM coordinates, i.e., when the dipole axis is tilted more toward the plus or minus *y*-direction in GSE coordinates, the GSM $B_z$ component of the IMF becomes more frequently and/or more strongly negative near equinoxes, such that more solar wind energy is transferred to the magnetosphere through dayside reconnection. However, an analysis by Hakkinen et al. (2003) suggests that not all daily and seasonal variations can be explained by the Russell-McPherron effect and other effects proposed to date. We suggest that the dipole tilt-dependent solar wind energy transfer, possibly resulting from the SMXR and, as discussed in the previous paragraphs, could provide a new explanation for the semiannual variation of the geomagnetic activities, although its relative contribution may be secondary. Future studies should elucidate the role of the SMXR, e.g., by investigating if and how the size of, and the total magnetic flux within, the polar cap depends on the dipole tilt angle under similar IMF conditions, and by revealing the detailed properties of FTEs.

*5.2. Kelvin-Helmholtz instability effects*

Figure 31 shows the transition of proton energy distributions from the magnetosheath to the magnetosphere in and around the KH vortices observed nearly simultaneously on the dawn and dusk flanks (Nishino et al., 2011), which may give some clues as to kinetic processes operating in the vortices. Observed on the dusk side (panel b) are two distinct ion populations, i.e., cool magnetosheath-like and hot magnetospheric populations. Energies of the cool component at its energy flux peaks remain lower than 1 keV throughout the transition region. On the dawn side (panel a), however, the energy distributions have only single peaks and, interestingly, the peak energy increases up to a few keV as one moves from the magnetosheath into the magnetosphere. This dawn-dusk asymmetry implies that the process of ion heating is more efficient on the dawn side than on the dusk side. Such asymmetries (see also Hasegawa et al., 2003, 2004c; Wing et al., 2005) may be unrelated to the KHI, and may simply be associated with the asymmetric ULF activities in the foreshock or magnetosheath region (e.g., Engebretson et al., 1991), which could asymmetrically excite KAWs in the LLBL through the process discussed, e.g., by Johnson and Cheng (1997). Nonetheless, by revealing the observed characteristics of particle and wave energy spectra (e.g., Cornilleau-Wehrlin et al., 2008) and of the ion to electron temperature ratio (e.g., Lavraud et al., 2009), we would be able to give constraints on the possible mechanisms of particle transport in coordinate and/or velocity space.

ULF waves in the Pc5 (2-7 mHz) band are believed to be the key to the acceleration of outer radiation belt electrons (e.g., Elkington, 2006), and the KHI is often invoked as one of the dominant mechanisms for Pc5 wave generation in the magnetosphere, because the dayside Pc5 wave intensity has a clear positive correlation with the solar wind speed (e.g., Engebretson et al., 1998). Rae et al. (2005) reported an event under a high solar wind speed condition, in which the Pc5 waves appeared to have been excited at the magnetopause by the KHI. However, the KH wave frequencies are often in the Pc4 (7-22 mHz) range (Hasegawa et al., 2006b), and a multi-satellite observation reported by Takahashi et al. (1991) is supportive of the KHI not being the driver of the Pc5 pulsations. Moreover, lower frequency (namely, longer wavelength) KH modes, corresponding to the Pc5 band, have a lower growth rate, so that their amplitudes are not likely to be large enough to generate intense Pc5 waves in the dayside or near-Earth tail region, unless they are strongly driven externally. In addition, rolled-up KH vortices tend to be observed during northward IMF periods (Hasegawa et al., 2006b), and theoretical arguments made in Section 4.3 and by Miura (1995) also support such an IMF control of the KHI. The KHI-induced ULF waves, even if existing, probably will not lead to the acceleration of radiation belt electrons, which seems to occur predominantly under southward IMF (e.g., Baker et al., 1998).

We surmise that the strong correlation between the magnetospheric Pc5 wave activity and the solar wind speed can be due to a possible increase in the Pc5 wave intensity in the foreshock or magnetosheath region with increasing solar wind speed. This is because, as the solar wind becomes faster, the upstream Mach number generally becomes higher, so that the foreshock waves/fluctuations (e.g., Eastwood et al., 2005b) can be more intense and/or ions could be heated more strongly in the direction transverse to the magnetic field when crossing the bow shock, thereby driving anisotropy-driven instabilities (e.g., Schwartz et al., 1996; Shoji et al., 2009). Such instabilities can, in turn, excite ULF fluctuations in the magnetosheath.   Note also that the fast solar wind emanating from coronal holes often contains large amplitude Alfvén waves (e.g., Belcher and Davis, 1971), which, after interacting with and crossing the bow shock, can evolve into magnetosheath structures with dynamic pressure variations (Lin et al., 1996a, 1996b).   Such foreshock or magnetosheath fluctuations might have triggered the surface waves and Pc5 pulsations as reported by Rae et al. (2005).   To conclude, there seems to be no strong evidence for magnetospheric Pc5 waves being excited directly by the KHI.

## 6.   Summary and Open Questions

### *6.1. Summary*

We have reviewed our recent, and other relevant, studies on the structure and dynamics of the magnetopause and its boundary layers from both local and global points of view.   The main points of this monograph can be summarized as follows.

1. Over the past 15 years, various types of techniques have been developed for visualization of two-dimensional (2D) or 3D structures of the plasma and fields from *in situ* data taken by one or more spacecraft when the structures move past the spacecraft.   These reconstruction methods are based on some governing equation(s), such as the Grad-Shafranov equation, or its variants, and the MHD or Hall-MHD system of equations with time dependence neglected.   Applications show that either of the methods can be used to analyze current sheets with, or without, active reconnection sites embedded, magnetic flux ropes, flow vortices associated with the Kelvin-Helmholtz instability (KHI), etc., and to extract information on the size, shape, and orientation of, and key processes governing, those structures.   In principle, it is also possible to recover time evolution of the structures based on similar concepts as used for the reconstruction of time-independent structures.   In future, it could become possible to reconstruct the structures in full 4D space-time using data from closely separated multiple spacecraft as the initial conditions.

2. Both remote and *in situ* observations show that magnetopause reconnection can be continuously active, at least globally.   Such evidence exists for both southward and northward interplanetary magnetic field (IMF) cases. However, there is no convincing evidence that magnetopause reconnection can be quasi-steady. Questions remain about whether magnetic reconnection is inherently non-steady, or whether the non-steadiness of magnetopause reconnection is an effect of the boundary conditions, such as the flow and spatiotemporal variability in the magnetosheath and the ionosphere to which reconnected field lines may be anchored.

3. Evidence exists that more than one X-line can form in the same (low-latitude or high-latitude) portion of the magnetopause. Since multiple X-line reconnection at the magnetopause inevitably progresses in a time-dependent manner and involves complicated changes of magnetic topology, it is inferred that such reconnection may regulate or suppress the energy transfer from the solar wind into the magnetosphere. An important question from the magnetospheric dynamics point of view is what conditions are preferred by multiple X-line reconnection.

4. A number of single- and multi-spacecraft methods have been developed for the identification and analysis of flow vortices generated by the magnetopause KHI. For northward IMF, evidence has been reported of the nonlinear evolution of the KHI, such as the formation of rolled-up vortices, 3D deformation of the field lines which gives birth to current sheets at the edges of the vortices, thinning of the current sheets down to the scale of ion inertia length, and reconnection in the thinned current sheets which can generate meso-scale flux ropes. It is yet unclear if the KHI can develop to a strongly nonlinear stage for southward IMF (see, however, Hwang et al. (2011) for a probable case under a southward IMF), although there are explanations for northward, rather than southward, IMF conditions being favorable for the KHI growth.

5. While dual lobe reconnection may well be the dominant mechanism for solar wind plasma "entry" into the magnetosphere under northward IMF, there are also signatures suggestive of mechanisms other than reconnection, such as the KHI and kinetic Alfvén waves (KAWs), contributing to the plasma transport in or across the flank magnetopause/LLBL. However, the question of whether they play a significant role in the transport on a global scale remains unanswered.

*6.2. Other remaining issues*

There are a number of fundamental questions that remain unanswered or have not even been addressed:

(1) What controls the thickness of the magnetopause current layer? It may be determined by competition or coupling among diffusion, convection or erosion of field lines due to the viscous interaction, reconnection, etc. and compression of the current sheet by the magnetosheath or KHI-driven vortex flow (see, e.g., Paschmann et al., 2005; Nakamura et al., 2010a). This issue is also related to the question of whether intermediate shocks can form in the magnetopause (e.g., Hau and Sonnerup, 1992; Karimabadi et al., 1995), why magnetosheath particles appear to be heated when crossing the RD-type magnetopause, and how, and what fraction of, particles are reflected at the magnetopause. However, understanding this at a fundamental level likely requires unveiling the microphysics behind the viscosity and diffusivity and thus may require future high-resolution measurements. It is also worth noting that the magnetopause at Mercury seems to be rather thin (e.g., Boardsen et al., 2010), despite the possible presence of heavy ions such as $Na^+$ (Zurbuchen et al., 2008) and the potential importance of kinetic effects on the magnetopause structure (Cai et al., 1990; Nakamura et al., 2010a). Therefore, comparative studies of planetary and moon magnetopauses are crucial for unraveling the factors governing the magnetopause thickness.

(2) What is the typical thickness of velocity shear layers at the magnetopause or in the LLBL, and what controls their thickness? While the thickness of the magnetopause, as well as the magnetotail current layer, has been investigated over the years, to the best of our knowledge, there is no observational study on the width and fine-scale structures of the velocity shear layer. The shear layer structure is intimately related to the question of at what wavelength the fastest growing KH mode is excited and at what rate it grows (e.g., Miura and Pritchett, 1982; Nakamura et al., 2010a). Answering this question will help elucidate the excitation mechanism of magnetopause surface waves in general and the details of the viscous interaction involved.

(3) What condition controls the reconnection rate (e.g., Bosqued et al., 2005; Birn et al., 2010) and the length, orientation, and motion of reconnection X-line? This question also concerns the non-steadiness of magnetopause reconnection (Summary point 2).  The X-line orientation estimated for a magnetopause reconnection event (Teh and Sonnerup, 2008) is consistent with a theoretical prediction (Swisdak and Drake, 2007) which is determined only by the local conditions of the current sheet.  However, global simulations suggest that the X-line behavior cannot be described as functions of the local conditions alone, but are affected by global geometries of the dayside magnetosphere and IMF (e.g., Raeder, 2006; Dorelli et al., 2007).  Also, the azimuthal extent of FTEs, which would be equivalent to that of the associated X-line(s), is in some cases rather large ($\geq 2$ $R_E$) (Fear et al., 2008), but in other cases can be small (Fear et al., 2010).  We thus need to clarify the relative importance of local and global conditions, by investigating how the X-line properties depend on the dipole tilt angle or season (cf. Subsection 5.1.2), the nature of magnetosheath turbulence (Coleman and Freeman, 2005), etc.  There is also the interesting question of how plasma $\beta$ (ratio of the plasma to magnetic pressure), magnetic shear, sheared flows, etc. affect the nature of magnetopause reconnection (e.g., Paschmann et al., 1986; Roytershteyn and Daughton, 2008).  While solar wind observations (Phan et al., 2010) confirmed a theoretically expected dependence of the occurrence of reconnection on a combination of $\beta$ and magnetic shear (Swisdak et al., 2010), it is unclear if such a dependence exists for magnetopause reconnection.

Some of the above issues can be addressed by currently available data from, for example, Cluster and THEMIS, and others will probably need future missions such as MMS and SCOPE.  The following are the questions regarding the global effects of magnetopause processes, which, of course, cannot be entirely independent of the above fundamental issues:

(4) Can solar wind plasmas captured into closed flux tubes on the dayside under northward IMF be transported into the midnight portion of the plasma sheet and, if so, how?  What are the roles of flank magnetopause and LLBL processes (such as the KHI and kinetic waves and associated turbulence) in the transport of plasma from the dayside to tail-flank LLBL, across the flank magnetopause, and/or across the tail LLBL?  It might be that only an MHD turbulence and associated eddy diffusion (with no cross-field diffusion) can basically account for the observed formation of the CDPS, but, if this is the case, what process is responsible for the observed dawn-dusk asymmetry in the LLBL/CDPS ion velocity distributions (Hasegawa et al., 2004b; Wing et al., 2005)?  Specific questions that should be addressed are: to what extent are transient structures in the solar wind or magnetosheath important (e.g., Tian et al., 2010)? What are the effects of the magnetosphere-ionosphere coupling (e.g., Sonnerup and Siebert, 2003)? For example, to what extent is the growth of the KHI or MHD turbulence suppressed by such coupling (e.g., Yamamoto, 2008; Borovsky and Funsten, 2003)? How, where, and how strongly are KAWs and other kinetic waves excited (see, however, Yao et al., 2011)? A number of processes exist for the KAW excitation, such as through reconnection (Chaston et al., 2005, 2008) and mode conversion from KH surface waves (Hasegawa, 1976; Chaston et al., 2007) or from compressional ultra-low frequency (ULF) waves in the magnetosheath (Johnson and Cheng, 1997), but their relative contributions and whether the KAWs have appropriate wavenumber vectors (namely, electromagnetic fluctuations directed in right directions) for causing cross-field transport are poorly understood.

(5) How do magnetopause processes affect the generation of magnetospheric ULF waves? There seems to be a myth that a considerable fraction of the Pc5 ULF waves is excited by the magnetopause KHI, but its evidence is weak (Section 5.2). There is probably no doubt that magnetopause surface waves have some relation to the dayside ULF waves. However, a full understanding of how they are excited (by flow shear at the magnetopause, or pressure/density variations in the solar wind or magnetosheath (e.g., Kepko and Spence, 2003; Claudepierre et al., 2010), or something else) is still lacking.

Last but not least, the following issues could be addressed with observations of the distant tail magnetopause region or of the boundary region between the lobe and magnetosheath, for example, by dual ARTEMIS spacecraft (Sibeck et al., 2011):

(6) What are the properties of the KH waves at greater downstream distances? Do they evolve continuously to form larger vortices (e.g., Miura, 1999), shocklets (Miura, 1992; Lai and Lyu, 2006), and/or an eddy turbulence (Matsumoto and Hoshino, 2004)? What is the nature in the distant tail region of outflow jets anti-sunward of tailward-of-the-cusp reconnection? Does turbulence develop there as the jets travel anti-sunward, and are particles energized there? It is not easy to address the latter questions with observations of the dayside or low-latitude magnetopause. This is because the field lines reconnected there are connected to the ionosphere, so that the turbulence behavior may be modified by coupling with, or by waves reflected off, the ionosphere, and because a considerable amount of high-energy particles already exists on the magnetospheric side of the low-latitude magnetopause, making it difficult to discriminate between those energized through reconnection or turbulence and those preexisting in the ring current region. On the other hand, being magnetically detached from the ionosphere, the jets anti-sunward of a tailward-of-the-cusp reconnection site would not involve Alfvén waves reflected off the ionosphere. Also, the lobe region usually lacks energetic particles. We note that reconnection in the distant magnetopause and that in the solar wind may be similar, in that both occur with a kind of open boundary condition. Therefore, studies of reconnection exhausts in the distant magnetopause region may help to solve the mystery of the lack of non-thermal particles and electron heating signatures in reconnection exhausts in the solar wind (Gosling, 2011). Are the observed solar wind exhausts merely a fossil signature of reconnection that occurred closer to the sun, or does this lack have any association with the boundary conditions?

**Appendix A.   LMN Boundary Coordinate System**

The LMN boundary coordinate system (Russell and Elphic, 1978), when based on some magnetopause model, is defined as follows: $N$ is along the model magnetopause normal and is directed radially outward, $M$ is along the cross product of $N$ and the GSM $z$-axis and is directed generally dawnward on the dayside, and $L$ completes the right-hand orthogonal system and is directed generally northward (see also Figure 7 in Russell (1990)). For widely-used models of the terrestrial magnetopause location, see Fairfield (1971), Roelof and Sibeck (1993), and Shue et al. (1998).

For analysis of a particular magnetopause crossing, one may also use the LMN coordinate system defined based on the results from minimum variance analysis of the magnetic field (MVAB) (Sonnerup and Scheible, 1998): the $L$, $M$, and $N$ axes are then defined to be parallel to the maximum, intermediate, and minimum variance directions. This local coordinate system can be used to study current sheets in any regions, such as the solar wind and magnetotail. A caution is that MVAB often does not provide a good estimate of the current sheet normal direction, while the $L$-axis can be rather well determined. In such cases, one may assume, for a TD-like current sheet, that the mean field component along the normal $\langle B_N \rangle = 0$ (Sonnerup and Scheible, 1998) or, if multi-point measurements are available, four-spacecraft timing analysis (Russell et al., 1983; Schwartz, 1998) may be used to estimate the normal ($N$-axis). In the latter case, the $M$-axis can be taken to be along the cross-product of $N$ (from the timing analysis) and *L* (from MVAB).

**Appendix B.   DeHoffmann-Teller frame**

In the deHoffmann-Teller (HT) frame, the electric field should in principle vanish and magnetic field lines look stationary, so that only field-aligned flows may exist in a plasma that satisfies the frozen-in condition.  The HT frame can exist for a steady, static plasma structure in any spatial dimension.  For those structures which are time-dependent or are steady but in a dynamical equilibrium, different segments of the structure may move at different velocities, so that a common HT frame cannot be found. The velocity of the HT frame, called the HT velocity, can be determined by a least-squares technique, based on single-spacecraft measurements of the plasma velocity (or electric field) and magnetic field (Khrabrov and Sonnerup, 1998), and represents the motion of the structure in question.  Khrabrov and Sonnerup (1998) also describe an extended version of the method which takes into account a non-zero, constant acceleration of the HT frame.

**Acknowledgements**
A large number of colleagues have contributed to the research reviewed here, but I would particularly like to thank Bengt Sonnerup for close collaborations and inspiring discussions.  I acknowledge fruitful discussions with T. Izutsu, J. R. Johnson, B. Lavraud, J. P. McFadden, T. Nagai, T. K. M. Nakamura, J. Raeder, J. A. Slavin, and A. Vaivads, and also thank T. Izutsu and T. K. M. Nakamura for comments on a draft version of the manuscript.   I am indebted to the international teams that have made the Cluster, Geotail, and THEMIS missions successful.  Part of the work at JAXA was supported by Grant-in-Aid for Scientific Research KAKENHI 21740358.

Two-dimensional particle simulations, J. Geophys. Res., 116, A03227, doi:10.1029/2010JA016046, 2011.

Nishino, M. N., M. Fujimoto, T. Terasawa, G. Ueno, K. Maezawa, T. Mukai, and Y. Saito, Geotail observations of temperature anisotropy of the two-component protons in the dusk plasma sheet, Ann. Geophys., 25, 769-777, 2007a.

Nishino, M. N., M. Fujimoto, G. Ueno, T. Mukai, and Y. Saito, Origin of temperature anisotropies in the cold plasma sheet: Geotail observations around the Kelvin-Helmholtz vortices, Ann. Geophys., 25, 2069-2086, 2007b.

Nishino, M. N., H. Hasegawa, M. Fujimoto, et al., A case study of Kelvin-Helmholtz vortices on both flanks of the Earth's magnetotail, Planet. Space Sci., 59, 502-509, doi:10.1016/j.pss.2010.03.011, 2011.

Nykyri, K., and A. Otto, Plasma transport at the magnetospheric boundary due to reconnection in Kelvin-Helmholtz vortices, Geophys. Res. Lett., 28, 3565-3568, 2001.

Ogilvie, K. W., R. J. Fitzenreiter, and J. D. Scudder, Observations of electron beams in the low-latitude boundary layer, J. Geophys. Res., 89, 10,723-10,732, 1984.

Ogilvie, K. W., and R. J. Fitzenreiter, The Kelvin-Helmholtz instability at the magnetopause and inner boundary layer surface, J. Geophys. Res., 94, 15,113-15,123, 1989.

Øieroset, M., T. D. Phan, V. Angelopoulos, J. P. Eastwood, J. McFadden, D. Larson, C. W. Carlson, K.-H. Glassmeier, M. Fujimoto, and J. Raeder, THEMIS multi-spacecraft observations of magnetosheath plasma penetration deep into the dayside low-latitude magnetosphere for northward and strong $B_y$ IMF, Geophys. Res. Lett., 35, L17S11, doi:10.1029/2008GL033661, 2008.

Øieroset, M., T. D. Phan, J. P. Eastwood, M. Fujimoto, W. Daughton, M. A. Shay, V. Angelopoulos, F. S. Mozer, J. P. McFadden, D. E. Larson, and K.-H. Glassmeier, Direct evidence for a three-dimensional magnetic flux rope flanked by two active magnetic reconnection X lines at Earth's magnetopause, Phys. Rev. Lett., 107, 165007, doi:10.1103/PhysRevLett.107.165007, 2011.

Omidi, N., and D. G. Sibeck, Flux transfer events in the cusp, Geophys. Res. Lett., 34, L04106, doi:10.1029/2006GL028698, 2007.

Onsager, T. G., J. D. Scudder, M. Lockwood, and C. T. Russell, Reconnection at the high-latitude magnetopause during northward interplanetary magnetic field conditions, J. Geophys. Res., 106, 25,467-25,488, 2001.

Otto, A., and D. H. Fairfield, Kelvin-Helmholtz instability at the magnetotail boundary: MHD simulation and comparison with Geotail observations, J. Geophys. Res., 105,
79

**Tables**

**Table 1.** Methods for the reconstruction of two-dimensional (2D) or three-dimensional (3D) plasma/magnetic field structures in space.

| Structures recoverable | Assumption(s)[a] | Input data[b] |
|---|---|---|
| 2D magneto-hydrostatic structures | Magneto-hydrostatic | **B**, plasma moments (**v**, $p$)[c] |
| 2D structures with field-aligned flow | Steady, field-aligned, isentropic, MHD flow | **B**, plasma moments ($\rho$, **v**, $p$) |
| 2D structures with flow transverse to the magnetic field | Steady, isentropic, MHD flow; the magnetic field is along the invariant axis ($z$). | **B**, plasma moments ($\rho$, **v**, $p$) |
| Ideal or resistive 2D MHD structures | Steady, MHD, isentropic (in the ideal case) | **B**, plasma moments ($\rho$, **v**, $p$), **E** in the resistive case |
| Ideal or resistive 2D Hall-MHD structures | Steady, Hall-MHD, isentropic (in the ideal case) | **B**, **E**, ion moments ($\rho$, **v**, $p$) |
| Slowly evolving 2D magneto-hydrostatic structures | Magneto-hydrostatic, incompressible, frozen-in-flux | **B**, plasma moments (**v**, $p$) |
| 3D magneto-hydrostatic structures | Magneto-hydrostatic | **B**, plasma moments (**v**, $p$)[c] |

[a]Besides two-dimensionality or three-dimensionality of the structures.
[b]**B**: magnetic field; **E**: electric field; $\rho$: mass density; **v**: velocity; $p$: plasma pressure.
[c]Pressure data is not necessary under the force-free assumption (Teh et al., 2010a).

| Field-line or streamline invariants[d] | Applicable structures[e] | References |
|---|---|---|
| $B_z(A)$, $p(A)$, $P_t(A)$ | Current sheets of TD-type; fossil magnetic flux ropes | Hau & Sonnerup, 1999; Hu & Sonnerup, 2002, 2003; Hasegawa et al., 2005 |
| $S(A)$, $H(A)$, $G(A)$, $C_z(A)$ | Current sheets of RD-type; flux ropes with field-aligned flow | Sonnerup et al., 2006a; Teh et al., 2007 |
| $S(\psi)$, $\tilde{H}(\psi)$, $F(\psi)$ | Flow vortices, e.g., of Kelvin-Helmholtz type | Sonnerup et al., 2006a; Hasegawa et al., 2007a, 2009b; Eriksson et al., 2009 |
| $S(\psi)$ in the ideal case | Current sheets of TD- and RD-type; magnetic flux ropes | Sonnerup & Teh, 2008; Teh & Sonnerup, 2008; Eriksson et al., 2009 |
| $S_i(\psi)$ & $S_e(\psi)$ in the ideal case | Current sheets of ion scale; ion diffusion regions of reconnection | Sonnerup & Teh, 2009; Teh et al.. 2011b |
| $B_z(A)$, $p(A)$, $P_t(A)$, $\rho(A)$ | Current sheets of TD-type; slowly evolving flux ropes | Sonnerup & Hasegawa, 2010; Hasegawa et al., 2010b |
| $p = const.$ along field lines | Current sheets of TD-type; fossil magnetic flux ropes | Sonnerup & Hasegawa, 2011 |

[d]$A$: partial magnetic vector potential $((B_x, B_y) = (\partial A/\partial y, -\partial A/\partial x))$; $\psi$: compressible stream function $((\rho v_x, \rho v_y) = (\partial \psi/\partial y, -\partial \psi/\partial x))$; $B_z$: magnetic field component along the invariant ($z$) axis; $P_t = p + B_z^2/(2\mu_0)$ : transverse pressure; $S = c_v \ln[T/\rho^{(\gamma-1)}]$ ($\gamma = c_p/c_v$: ratio of specific heats): entropy; $H = c_p T + v^2/2$ ($v^2 = v_x^2 + v_y^2 + v_z^2$): total enthalpy; $G = \mu_0 \rho v/B = M_A\sqrt{\mu_0 \rho}$: mass flux invariant; $C_z = (1 - M_A^2)B_z$ : axial invariant; $\tilde{H} = c_p T + (v_x^2 + v_y^2)/2 + B^2/(\mu_0 \rho)$ : generalized total enthalpy; $F = B/\rho$: frozen-flux function; $i$: ion; $e$: electron.

[e]TD: tangential discontinuity; RD: rotational discontinuity.

**Table 2a.** Temporal nature of magnetopause reconnection (local view).

| Local view (in the rest frame of a specific X-line) | | |
|---|---|---|
| Continuous ($E_{RX} \neq 0$)[a] | | Transient ($E_{RX} \geq 0$) |
| Steady ($dE_{RX}/dt = 0$) | Non-steady ($dE_{RX}/dt \neq 0$) | |
| Possible only when the X-line/separator is stationary in the planet/moon rest frame as well. No FTEs form. | May create FTEs through temporal modulation of the reconnection rate (Phan et al., 2004). | May create FTEs through intermittent bursts of reconnection (Scholer, 1988; Southwood et al., 1988). |

[a]$E_{RX}$: reconnection rate or reconnection electric field.

**Table 2b.** Temporal nature of magnetopause reconnection (global view).

| Global or non-local view (in the rest frame of a planet or a moon) | | |
|---|---|---|
| Continuous ($\Phi_{RX} \neq 0$)[b] | | Transient ($\Phi_{RX} \geq 0$) |
| Steady ($d\Phi_{RX}/dt = 0$) | Non-steady ($d\Phi_{RX}/dt \neq 0$) | |
| Possible only when the X-line/separator is stationary. No FTEs form. | Each X-line/separator can be either continuously or transiently active. The reconnection rate remains non-zero somewhere on the magnetopause (Hasegawa et al., 2008). | There exists a moment when the reconnection rate is zero everywhere. |

[b]$\Phi_{RX}$: total magnetic flux reconnected per unit time at the magnetopause.

**Table 3.** Possible roles of the Kelvin-Helmholtz instability at the magnetopause or in the low-latitude boundary layer (LLBL).

| Role | Possible consequence(s) | Proponent(s) | Evidence | Predominance |
|---|---|---|---|---|
| Mass transport | Formation of flank LLBL & CDPS[a] for northward IMF | e.g., Fujimoto & Terasawa, 1994 | Some (Chaston et al., 2007; Hasegawa et al., 2009b) | Possibly not (possibly dominant double lobe reconnection) |
| Momentum transport | Driving of magnetospheric convection | e.g., Miura, 1984 | Some (Sundberg et al., 2009) | Unknown, but works only for northward IMF |
| Energy transport | Excitation of ULF pulsations | e.g., Southwood, 1968 | Not clear (Rae et al., 2005) | Unknown |
| | Generation of spatially periodic dayside aurora | Lui et al., 1989 | Not clear | Unknown |

[a]CDPS: cold and dense plasma sheet where proton temperature is less than 1 keV and the density exceeds 1 cm$^{-3}$ (e.g., Terasawa et al., 1997).

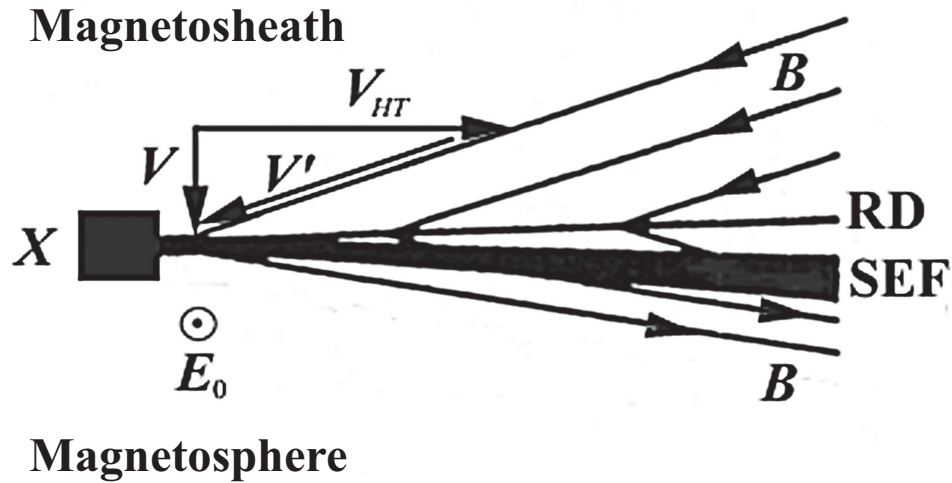

**Figure 1.** A magnetohydrodynamic (MHD) view of steady, two-dimensional (2D) magnetic reconnection at the magnetopause. The sketch shows only the right half of the reconnection layer (in the reconnection X-line rest frame) that contains an upstream rotational discontinuity (RD) and a slow-mode expansion fan (SEF) (Levy et al., 1964). In this frame, the reconnection electric field $E_0$ is non-zero and constant in space and time and the inflow velocity is $V$. In the deHoffmann-Teller (HT) frame (see Appendix B), which moves to the right with speed $V_{HT}$, the plasma flow $V'$ is along the magnetic field and thus the out-of-plane component of the convection electric field vanishes in regions around the RD well away from the SEF and the X-line (from Sonnerup and Teh, 2008).

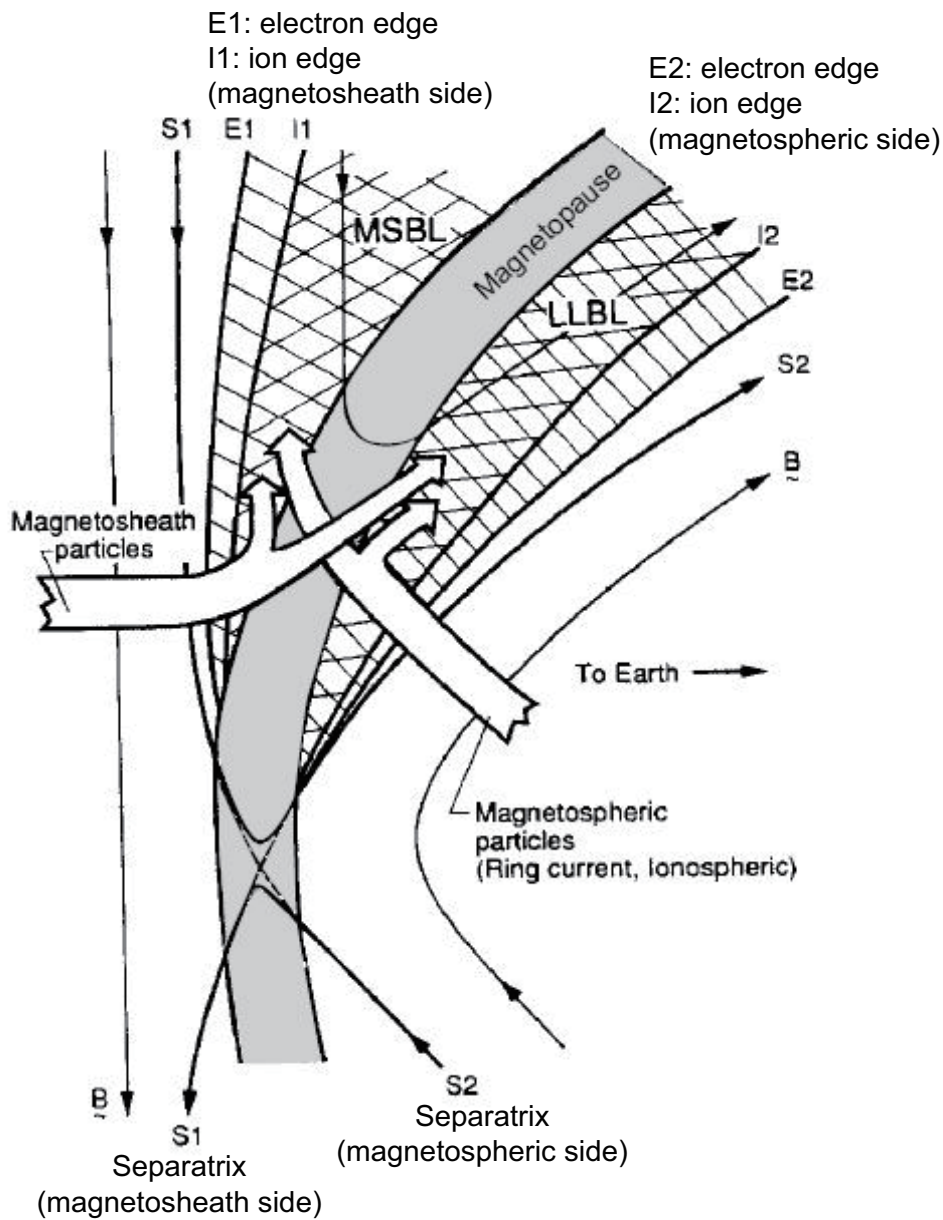

**Figure 2.** Schematic drawing of the dayside, low-latitude magnetopause region during steady reconnection under southward interplanetary magnetic field (IMF) conditions (view from dusk). The magnetopause is identified as a current layer of RD-type. Magnetosheath particles either reflect off the magnetopause and enter the magnetosheath boundary layer (MSBL), or cross the magnetopause and enter the low-latitude boundary layer (LLBL). Conversely, magnetospheric particles either reflect off the magnetopause and enter the LLBL, or cross the magnetopause and enter the MSBL (from Gosling et al., 1990).

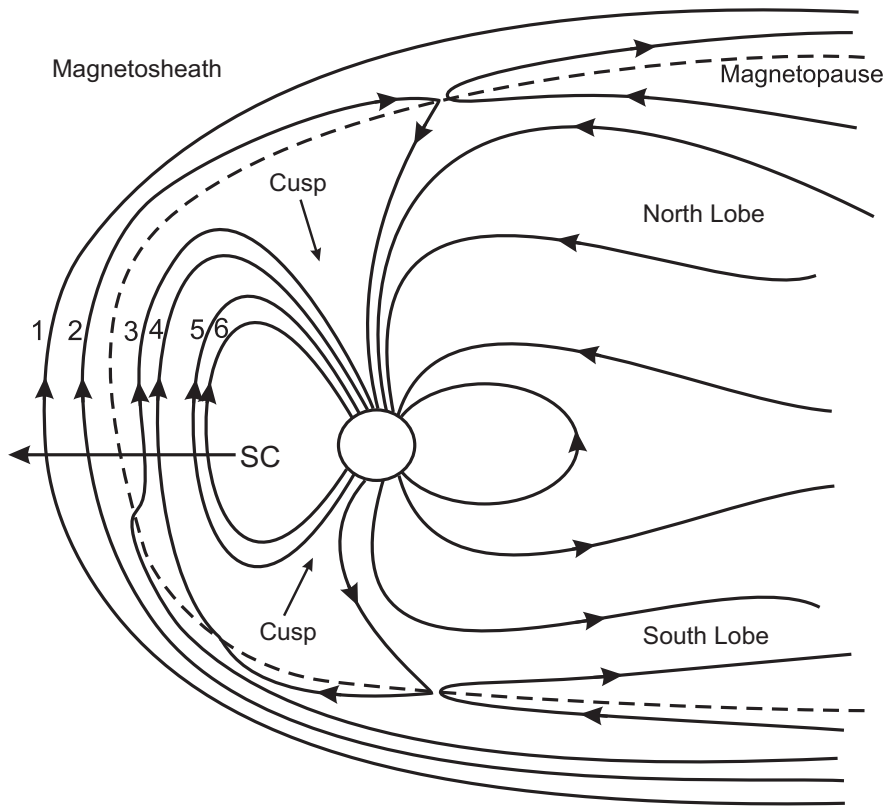

**Figure 3.** Schematic drawing of Earth's magnetosphere for predominantly northward IMF when reconnection is occurring steadily at the magnetopause poleward of the cusp in both hemispheres (view from dusk). In this sketch (adapted from Bogdanova et al., 2008) illustrating the formation of the dayside LLBL through dual lobe reconnection, an IMF field line reconnects first in the northern hemisphere and subsequently in the south. Line 1 shows a field line of the plasma depletion layer (PDL), which is draped around the dayside magnetosphere and has not experienced reconnection. Line 2 is a field line formed through reconnection between the PDL and lobe field lines poleward of the northern cusp. This field line is open and is on the magnetosheath side of the magnetopause at low latitudes. Line 3 evinces an open field line which is formed through reconnection poleward of the northern cusp and is now earthward of the subsolar magnetopause; an MHD (Alfvén or slow-mode) wave launched from the reconnection site and propagating along this field line now reaches south of the equator. Line 4 is a field line newly closed by reconnection poleward of the southern cusp. Line 5 is a field line which fully sinks in the magnetosphere and contains plasmas of magnetosheath origin. Line 6 is a field line which does not experience reconnection but may contain a mixture of magnetosheath and magnetospheric plasmas as a result of cross-field diffusion. The line segment with arrow shows the approximate path of the THEMIS-E spacecraft (SC) during the observations shown in Figure 4.

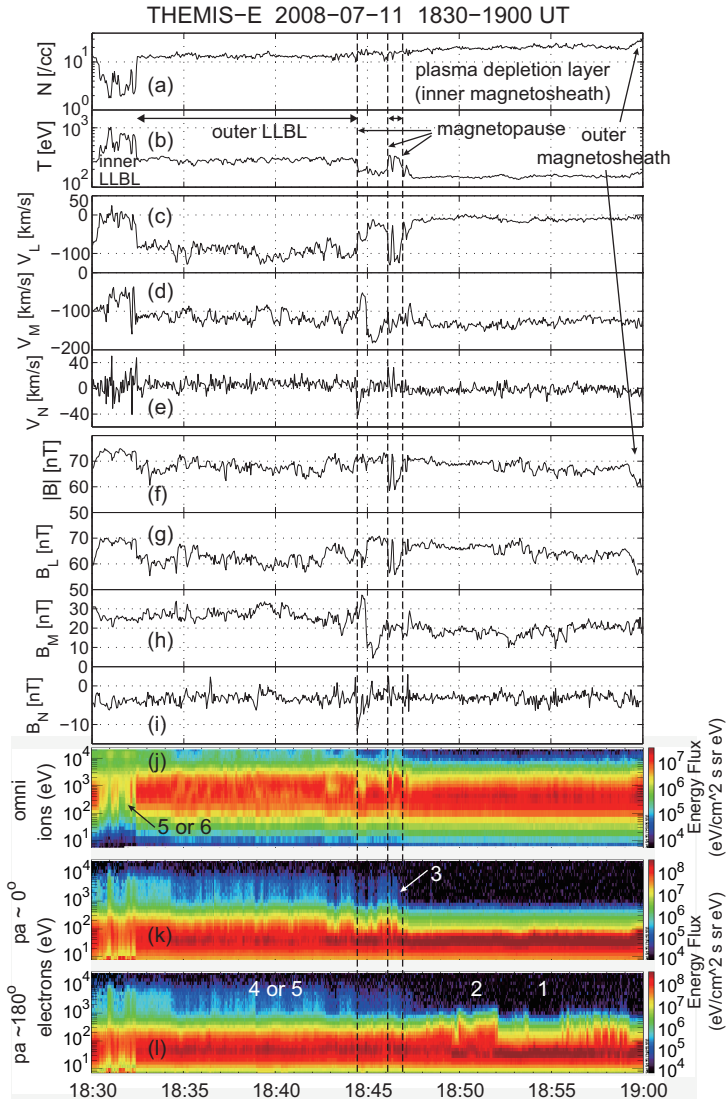

**Figure 4.** Plasma and magnetic field observations by the THEMIS-E spacecraft in the low-latitude, post-noon magnetopause region during strongly northward IMF when the magnetopause is hardly identifiable from magnetic field data alone. The THEMIS-E position at 1845 UT was (8.2, 5.0, –3.5) $R_E$ in the geocentric solar magnetospheric (GSM) coordinate system. (a) Ion (mostly proton) density, (b) ion temperature, (c-e) three *LMN* components of ion velocity (see Appendix A for definition of LMN coordinates), (f-i) magnitude and *LMN* components of the magnetic field, (j) energy-versus-time (E-t) spectrogram of omni-directional ions, and (k-l) E-t spectrograms of magnetic field-aligned and anti-field-aligned electrons. The numbers in the bottom three panels represent the regions threaded by the numbered field lines in Figure 3. The magnetopause under predominantly northward IMF is discernible most clearly by an increase in ion temperature from the PDL to the LLBL (adapted from Hasegawa et al., 2009a).

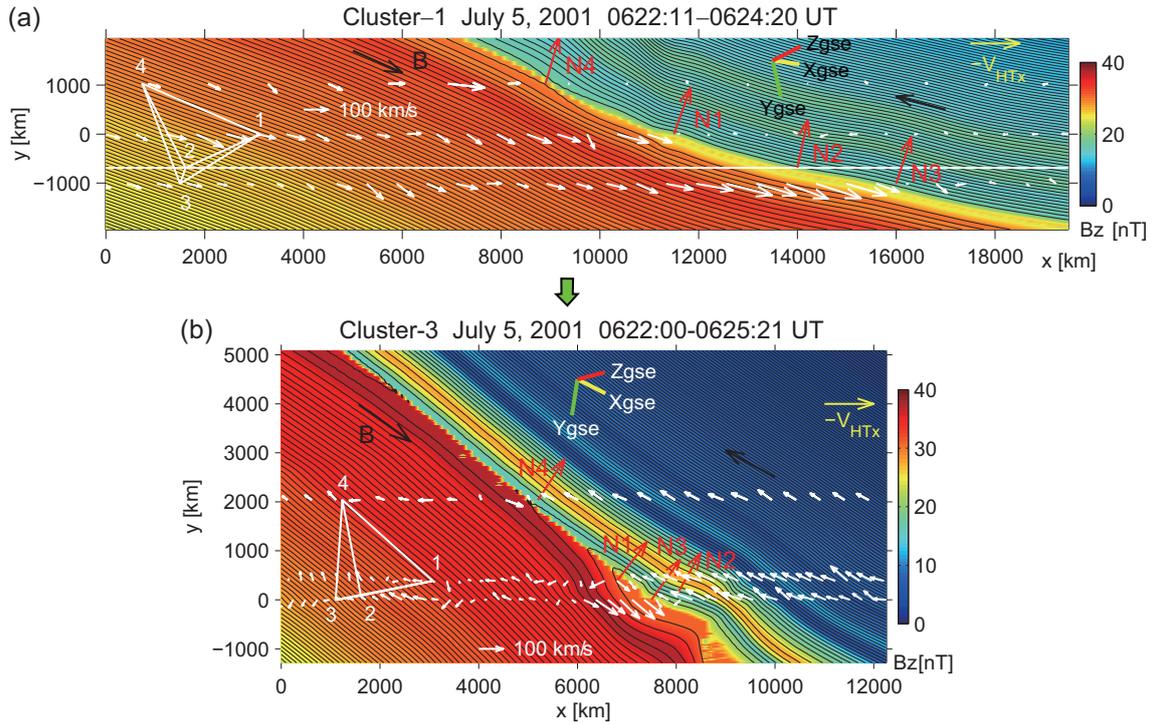

**Figure 5.** 2D magnetic field structures of a dawn-flank magnetopause visualized by magneto-hydrostatic Grad-Shafranov (GS) reconstruction (subsection 2.2.3), with the magnetosheath on the upper-right side and the magnetotail on the lower-left side. Black lines are the in-plane, namely transverse, magnetic field lines, whose direction is indicated by the black arrows. Colors show the out-of-plane, namely axial, field component $B_z$. White arrows show the *x-y* projection of ion velocity vectors, measured along the satellite paths and transformed into the HT frame. Red arrows are the projections of the magnetopause normal vectors, estimated from minimum magnetic variance analysis constrained to give the mean magnetic field component along the normal $\langle B_N \rangle = 0$ (Sonnerup and Scheible, 1998). In the maps, the four Cluster spacecraft (shown as a tetrahedron with numbers at its vertices) moved to the right; Cluster 3 (C3) crossed the magnetopause about 30 seconds later than did C1. The upper map is recovered from C1 data while the lower map is from C3 data; the magnetopause structure evolved significantly over the 30 second interval as a result of ongoing reconnection. C3 was at ~(−6.8, −15.0, 6.2) $R_E$ in the geocentric solar ecliptic (GSE) coordinate system (adopted from Hasegawa et al., 2004c). Yellow, green, and red bars are the projections onto the plane of the GSE *x*, *y*, and *z* axes, respectively.

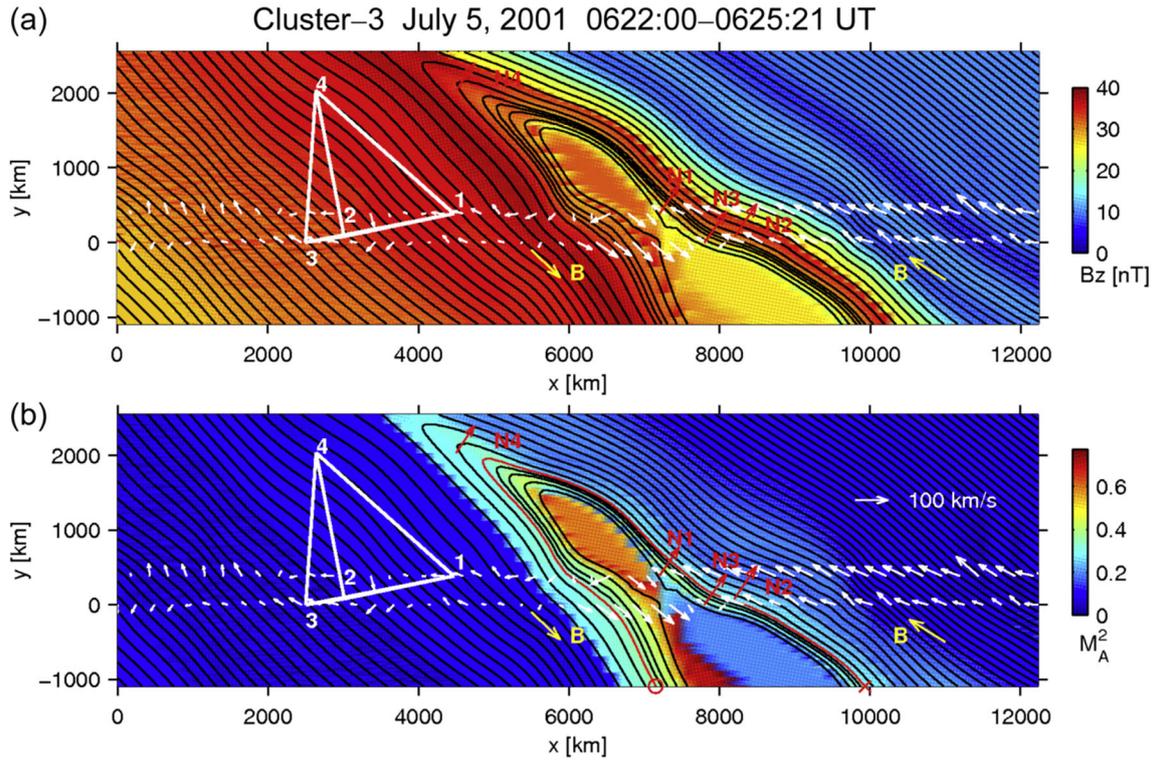

**Figure 6.** 2D field maps recovered from C3 data for the same event (on 5 July 2001) as in Figure 5, by use of GS-like reconstruction for MHD structures with field-aligned plasma flow (subsection 2.2.4). The coordinate system and format are the same as in the lower map in Figure 5, except that colors now show the axial field component in the top panel (a) and squared Alfvén Mach number $M_A^2$ in the bottom (b). The axial field $B_z$ is no longer a field line invariant as in the magneto-hydrostatic GS equilibrium and $M_A^2$ varies along the transverse field lines in this RD-type magnetopause where dynamical effects of the field-aligned flows cannot be neglected (from Teh et al., 2007).

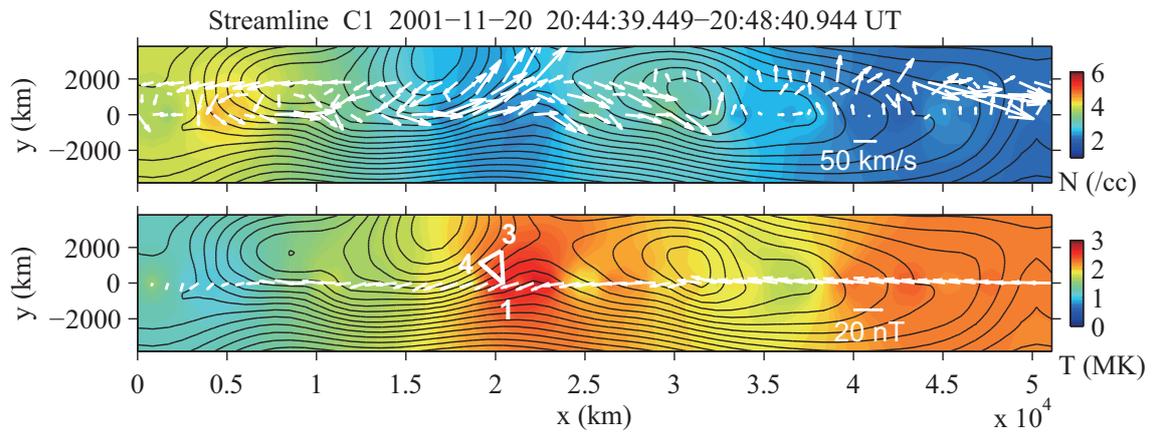

**Figure 7.** 2D structures of an LLBL observed behind the dusk-flank terminator, recovered from GS-like reconstruction of the velocity field for structures with flow transverse to the magnetic field (subsection 2.2.5). Black lines represent the reconstructed streamlines. Colors show the density in the top panel and ion temperature in the bottom. White arrows show the ion velocity vectors, measured by C1, C3, and C4 and transformed into the frame co-moving with the structure (top panel), and the transverse field vectors measured by C1 (located at (–3.5, 18.5, –2.7) $R_E$ in GSM coordinates) (bottom panel). The subsolar point is to the right and the magnetosheath is on the upper side (view from north). A train of Kelvin-Helmholtz (KH) vortices was embedded in this LLBL (from Hasegawa et al., 2009b).

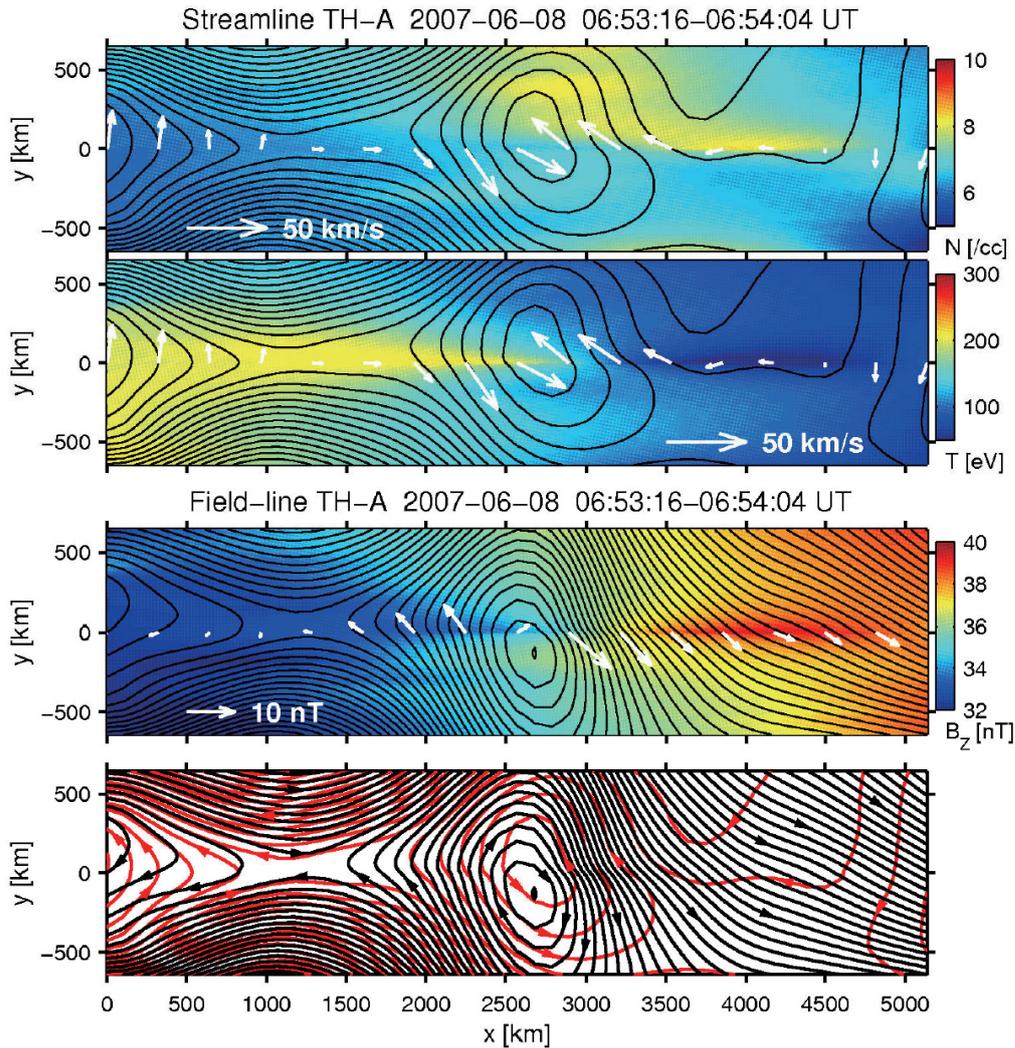

**Figure 8.** 2D maps of a magnetopause recovered by use of ideal MHD reconstruction (subsection 2.2.6) that allows for simultaneous creation of the velocity and magnetic field maps. Black lines show the reconstructed streamlines in the top two panels and the transverse field lines in the third panel. The bottom panel is an overlay of the streamlines (red) and field lines (black). In the first and second panels, colors show the recovered density and ion temperature, respectively, and white arrows are the measured velocity vectors transformed into a co-moving frame. In the third panel, colors show the axial field and the arrows are the measured field vectors. The subsolar region is to the bottom right and the magnetosheath is on the top side (view from north). The meso-scale magnetic flux rope in the map was roughly collocated with a vortex of similar size and was observed on the trailing (sunward-facing) edge of a larger-scale surface wave propagating along the post-noon magnetopause (THEMIS-A was at (8.1, 10.5, –3.5) $R_E$ in GSM coordinates) (from Eriksson et al., 2009).

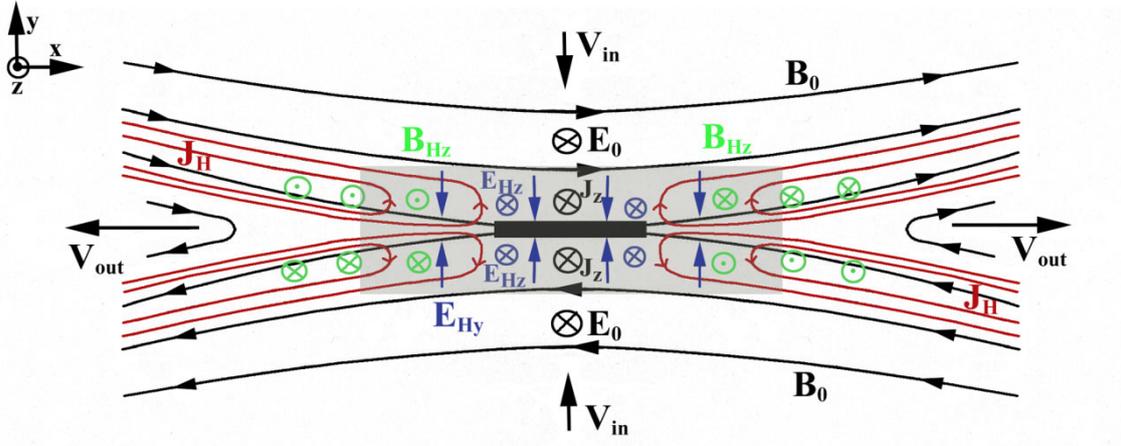

**Figure 9.** Schematic drawing of the ion diffusion region (gray box) surrounding a reconnection site that can be recovered by Hall-MHD reconstruction (subsection 2.2.7). Hall electric field components shown are the transverse field $E_{Hy} = j_z B_x/ne$ (blue arrows) and the axial field $E_{Hz} = -j_{Hy} B_x/ne$ (blue crosses); the field component $E_{Hx}$ is not shown. Hall driven current loops (red lines) induce a quadrupolar axial field (green symbols). Black rectangle (not to scale) at center represents the electron diffusion region (from Sonnerup and Teh, 2009).

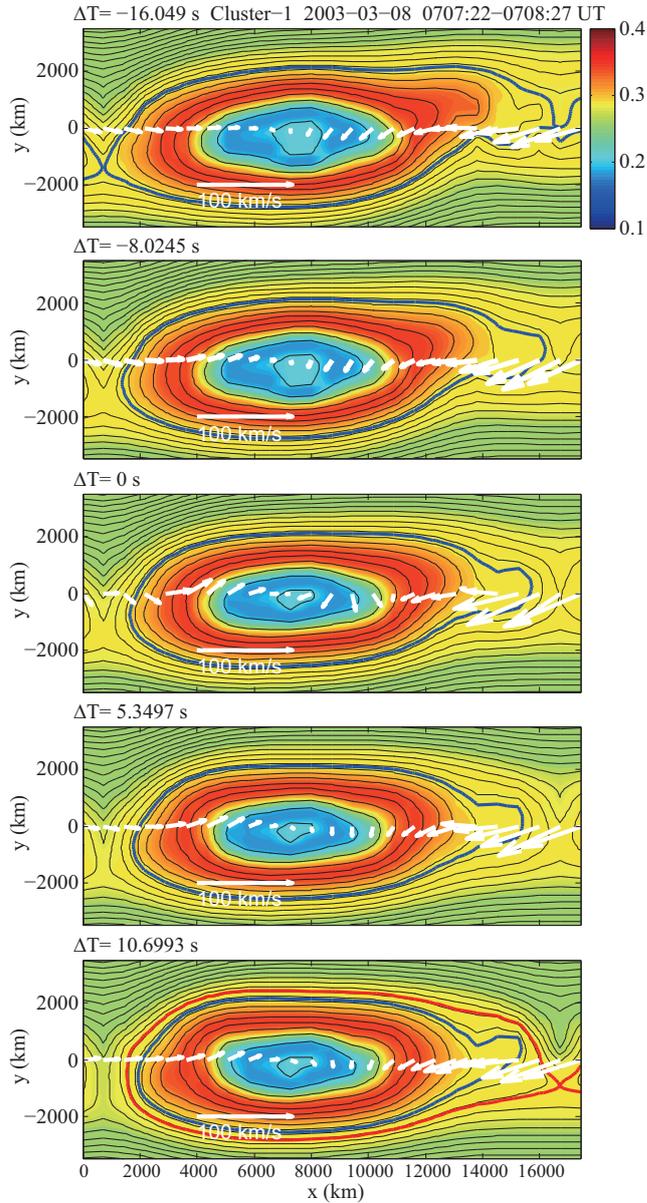

**Figure 10.** Evolution of a flux transfer event (FTE) observed at the dayside high-latitude magnetopause, recovered from C1 data (taken at (7.2, 2.8, 7.7) $R_E$ in GSE coordinates) by use of the reconstruction method for slowly evolving GS equilibria (subsection 2.2.8). Colors show the plasma pressure in nPa, and white arrows show the transverse velocity vectors (actually measured for $\Delta T = 0$ and recovered for other times). The magnetosheath, where $B_x$ in reconstruction coordinates is positive, and the magnetosphere, where $B_x$ is negative, are at the top and bottom, respectively, and the subsolar region is to the right (view from dawn). The blue loops show the field lines that are on the separatrix at $\Delta T = -16.049$ s, and the red loop shows the one connected to the subsolar-ward X point at $\Delta T = +10.6993$ s (from Hasegawa et al., 2010b).

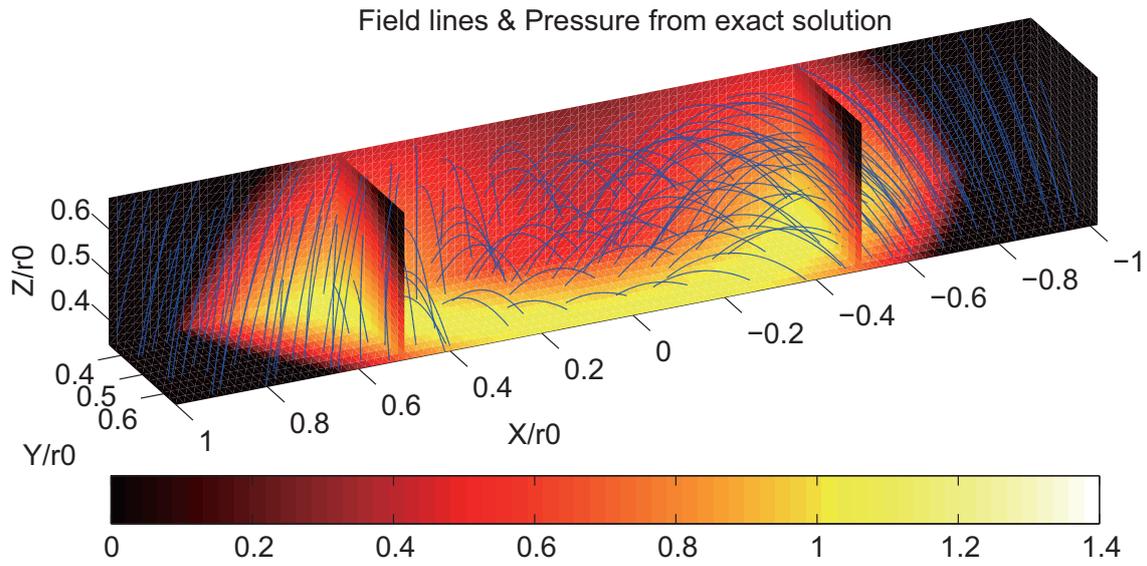

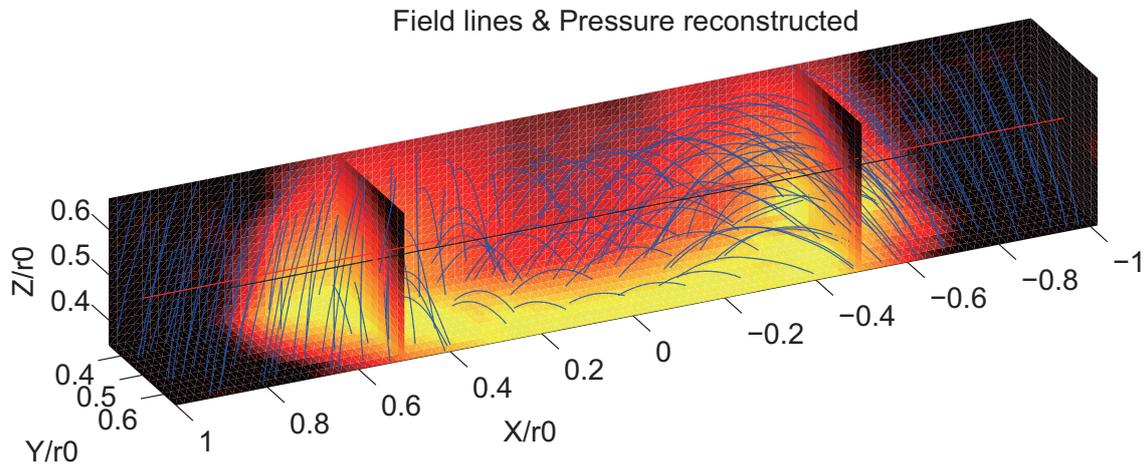

**Figure 11.** Three-dimensional (3D) view of exact (top) and reconstructed (bottom) field lines (blue curves) in a benchmark test of the method for reconstruction of 3D magneto-hydrostatic structures (subsection 2.2.9). Plasma pressure is represented by color, and the black solid and red dashed lines denote the paths of two virtual spacecraft that made measurements and provided the spatial initial values for the 3D reconstruction (from Sonnerup and Hasegawa, 2011).

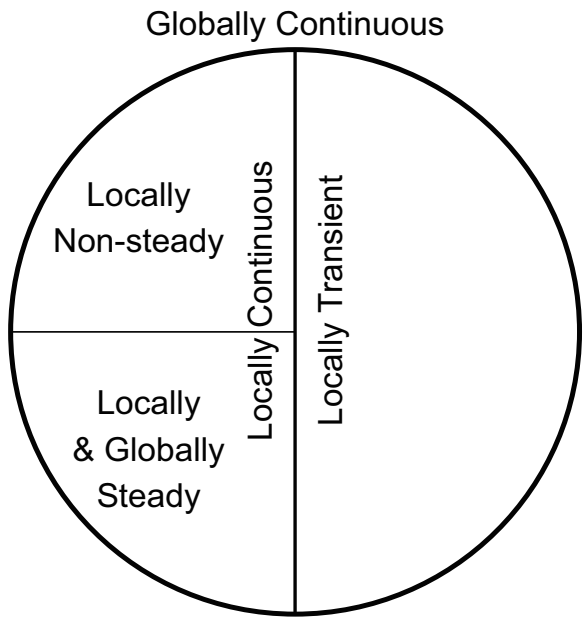

**Figure 12.** Types of magnetopause reconnection from a local viewpoint, which may be categorized as globally continuous reconnection (see also Table 2). Note that an ensemble of locally transient reconnection may be viewed as globally continuous reconnection, if the reconnection rate remains non-zero somewhere within a spatial domain under consideration.

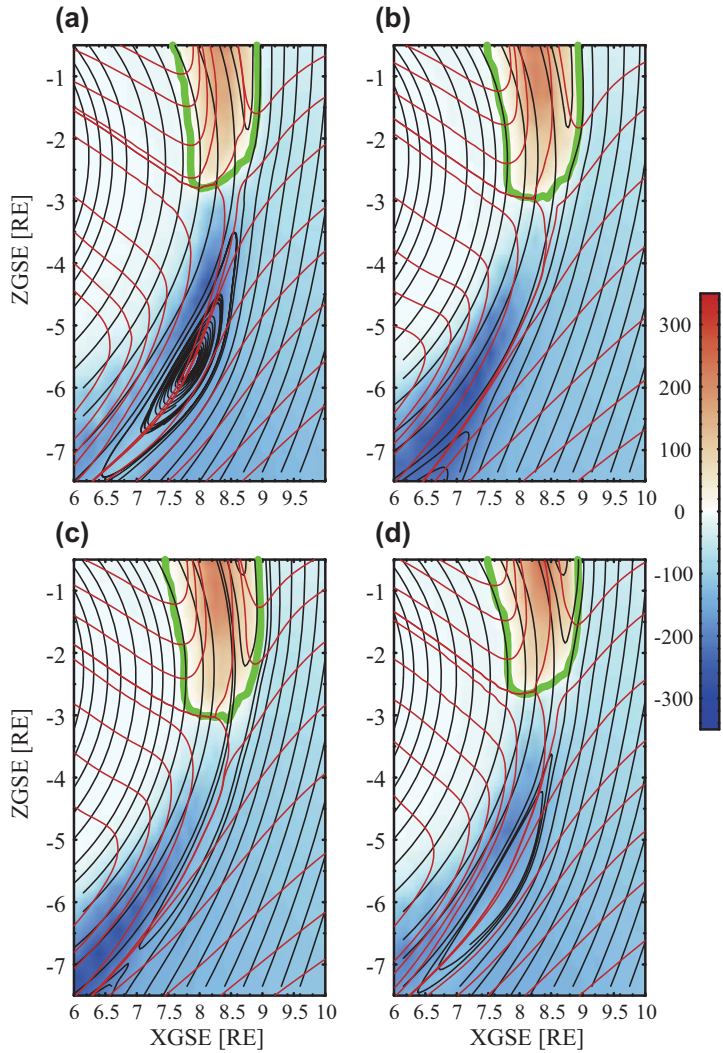

**Figure 13.** Four sequential snapshots (in the noon-midnight meridian plane viewed from dawn) of the dayside magnetopause region in a global MHD simulation of the solar wind-magnetosphere interaction under the condition when the geomagnetic dipole axis is tilted sunward in the northern hemisphere and the IMF is nearly southward. Colors show the $z$ component of the plasma velocity $v_z$ in km s$^{-1}$. The thick green line marks the contour of $v_z = 0$. Black lines are magnetic field lines; these lines are parallel to the $x$-$z$ projection of the field vector in the plane $y=0$. Red lines are streamlines; these lines are aligned with the $x$-$z$ projection of the instantaneous velocity vector in the plane $y=0$. Reconnection lines form slightly southward of the flow stagnation point in the subsolar region (a) and thus move southward along the magnetopause (b, c). Subsequently, a new X-line forms near the location of the old X-line formation (d), the result being the creation of a magnetic flux rope between the two X lines. This flux rope is swept into the winter hemisphere (southward here), and can be identified as one type of FTE (from Raeder, 2006).

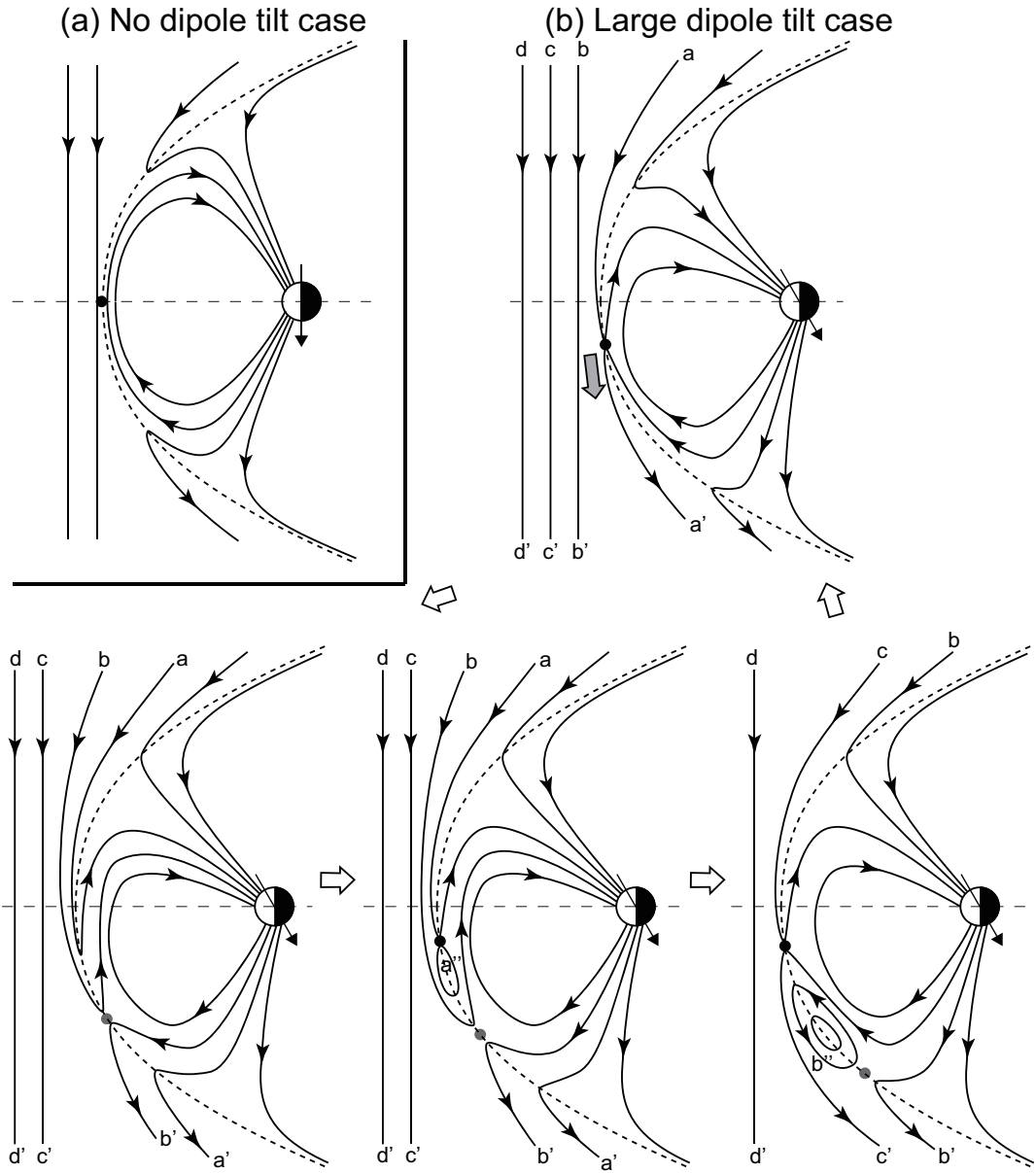

**Figure 14.** Schematic drawing (view from dusk) illustrating possible consequences of what Raeder (2006, 2009) calls sequential multiple X-line reconnection (SMXR) under large geomagnetic dipole tilt, as shown in Figure 13. In the case of no or small dipole tilt (a), subsolar reconnection can occur continuously, leading to efficient erosion of dayside geomagnetic field lines and subsequent transport of open magnetic flux into the magnetotail. In the case of large dipole tilt (b), reconnection at a newly formed subsolar X-line tends to occur, at least in its initial phase, on already open field lines or between the IMF and open field lines; no open field lines are newly created. A possible consequence is less efficient transport and storage of open magnetic flux into the tail for larger dipole tilt (from Hasegawa et al., 2010a).

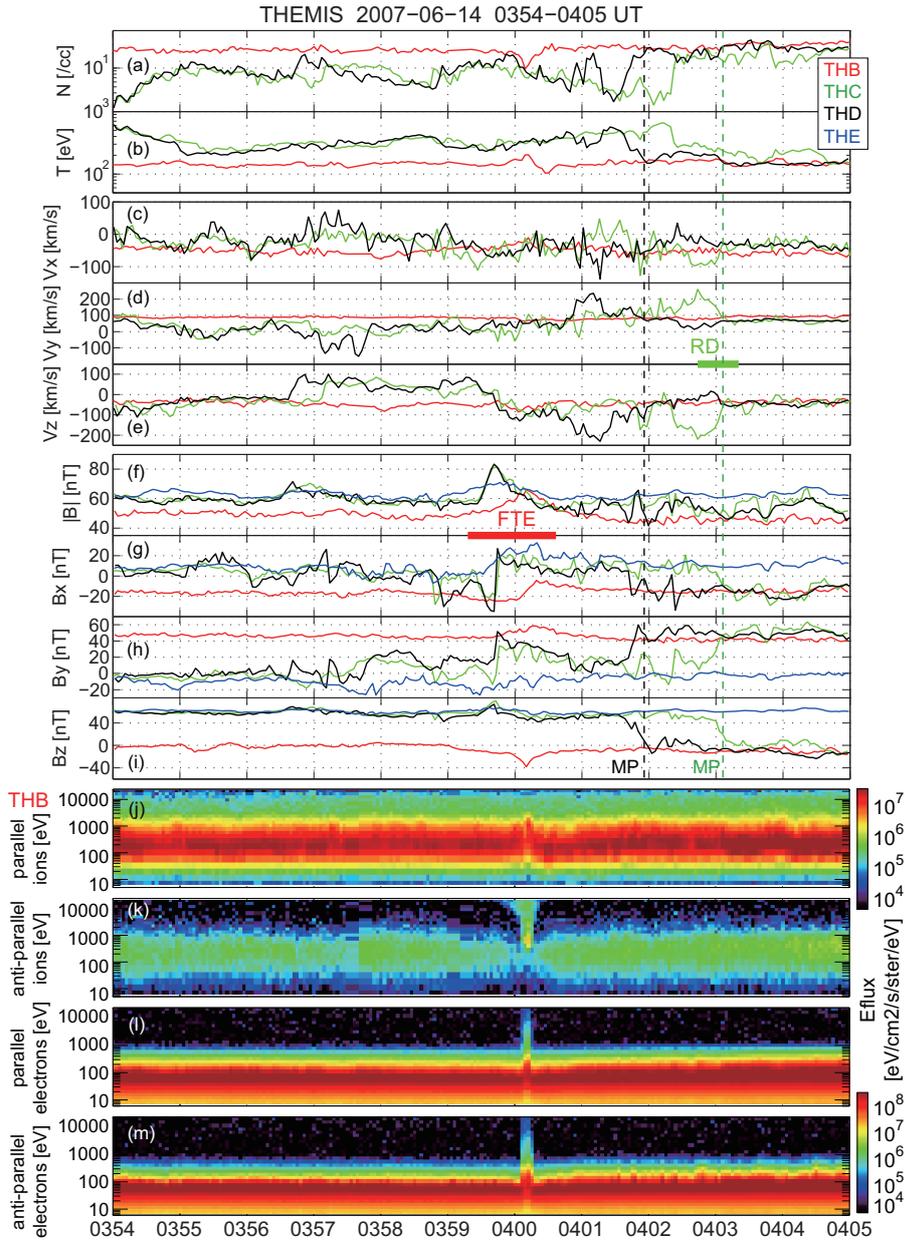

**Figure 15.** Observations near solstice of an FTE generated by multiple X-line reconnection at the dayside, low-latitude magnetopause. (a–i) Plasma (ion) and magnetic field data in GSM coordinates for 14 June 2007, 0354–0405 UT, recorded by four (THB, THC, THD, and THE) of the THEMIS satellites which encountered the FTE at ~0400 UT. The green bar marks the THC interval of the Walén test (section 2.1), from which the magnetopause is found to be of RD-type. Green and black vertical dashed lines mark approximate times of magnetopause crossing by THC and THD, respectively. The bottom four panels show E-t spectrograms of (j) field-aligned and (k) anti-field-aligned streaming ions and (l, m) those of electrons, observed by THB on the magnetosheath side of the magnetopause (from Hasegawa et al., 2010a).

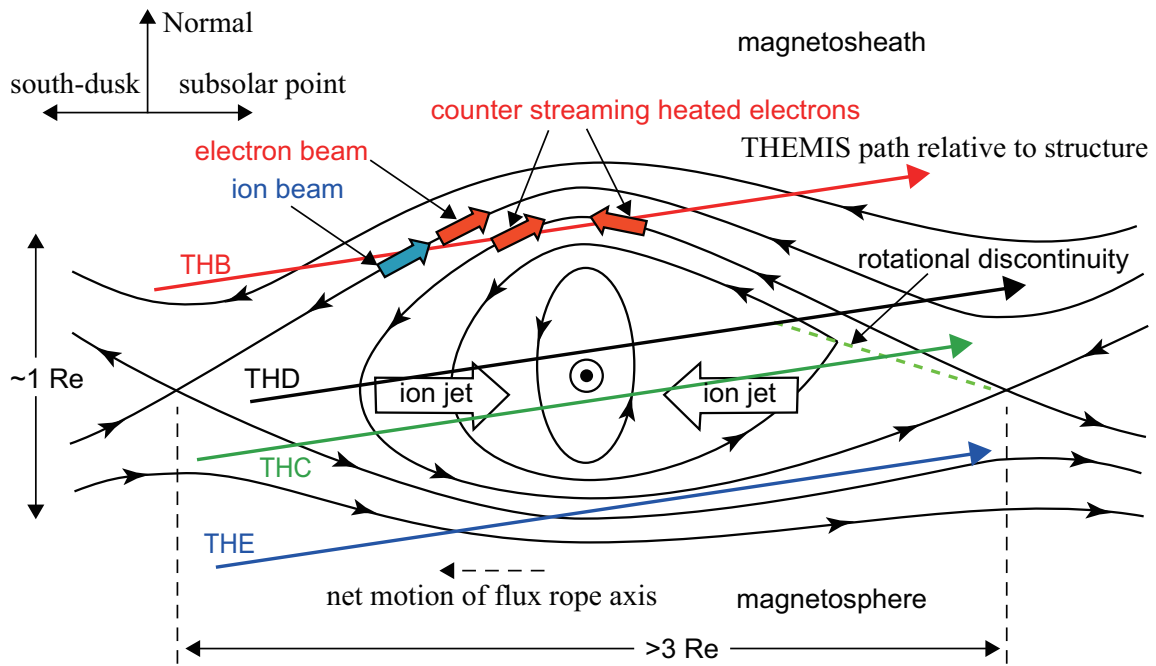

**Figure 16.** Schematic drawing (view from north-dusk) summarizing the FTE on 14 June 2007 shown in Figure 15. The observed features (multiple X-lines with the subsolar one formed later than the south-dusk one, and the motion toward the winter hemisphere of the FTE flux rope) agree well with those expected from Raeder's SMXR model (Figures 13 and 14) (from Hasegawa et al., 2010a).

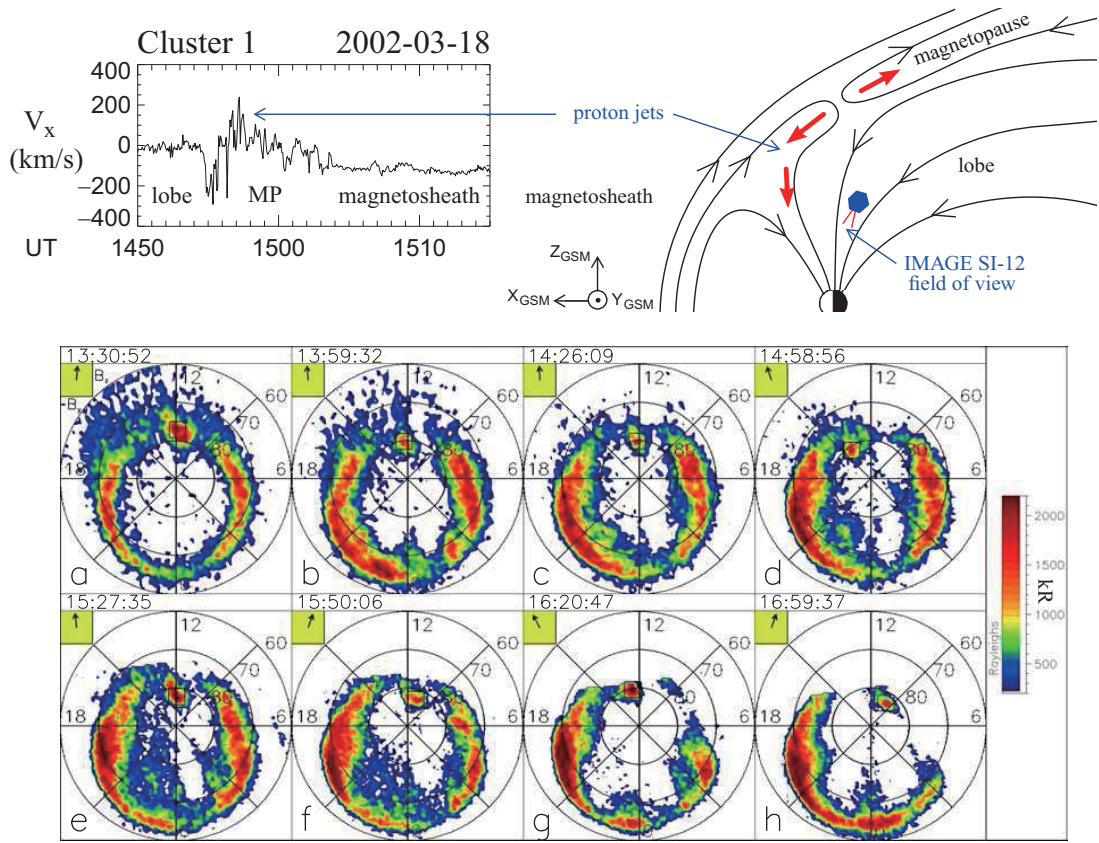

**Figure 17.** Observations of globally continuous reconnection at the magnetopause poleward of the northern cusp during an extended northward IMF period. The simultaneous observations of reconnection jets at the high-latitude magnetopause by Cluster (top left panel) and a proton auroral spot on the field lines linked to the magnetopause by the IMAGE satellite (red spot in the noon area in the bottom panels) demonstrate that the proton spot is a remote signature of reconnection. While the jets were observed for only 5 minutes when Cluster traversed the magnetopause, IMAGE detected the proton spot uninterruptedly for ~4 hours, implying continuous reconnection (from Phan et al., 2005).

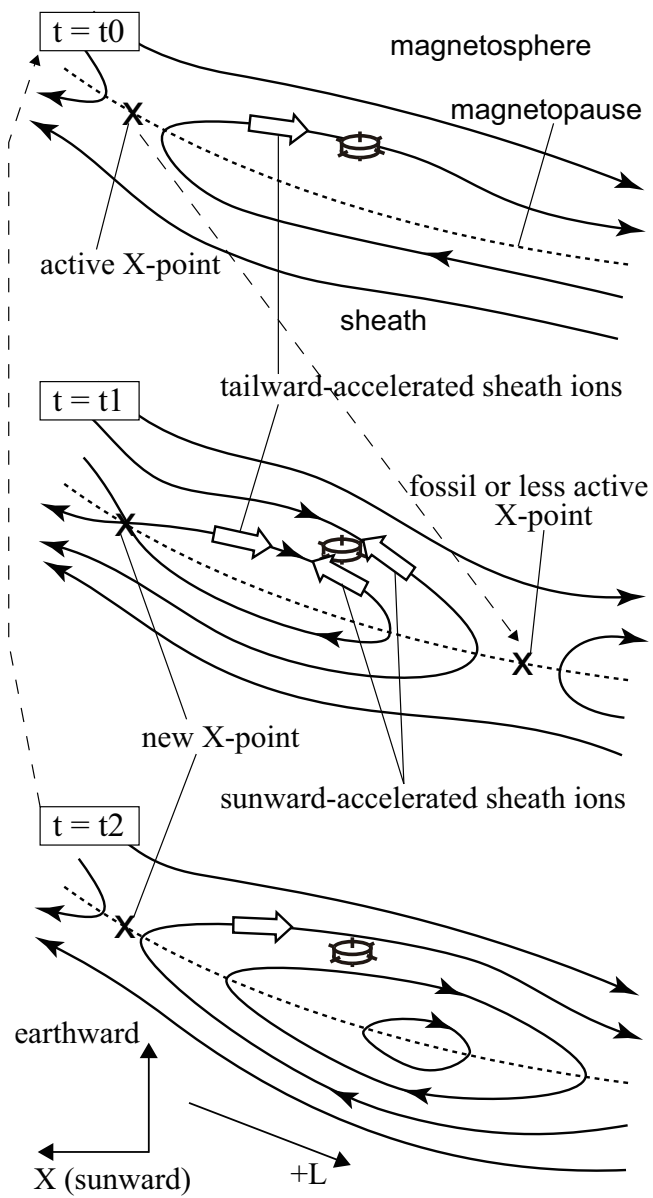

**Figure 18.** Schematic drawing of magnetic field lines around the high-latitude magnetopause in the southern hemisphere (view from north-dusk), which illustrates tailward retreat and reformation of X-line(s) during quasi-continuous reconnection tailward of the southern cusp under northward and dawnward IMF. Cluster observations on 19-20 November 2006 are consistent with this scenario; individual X-lines may have been only transiently active, but reconnection was almost persistently occurring somewhere within this south-dusk sector of the tailward-of-the-cusp magnetopause (from Hasegawa et al., 2008).

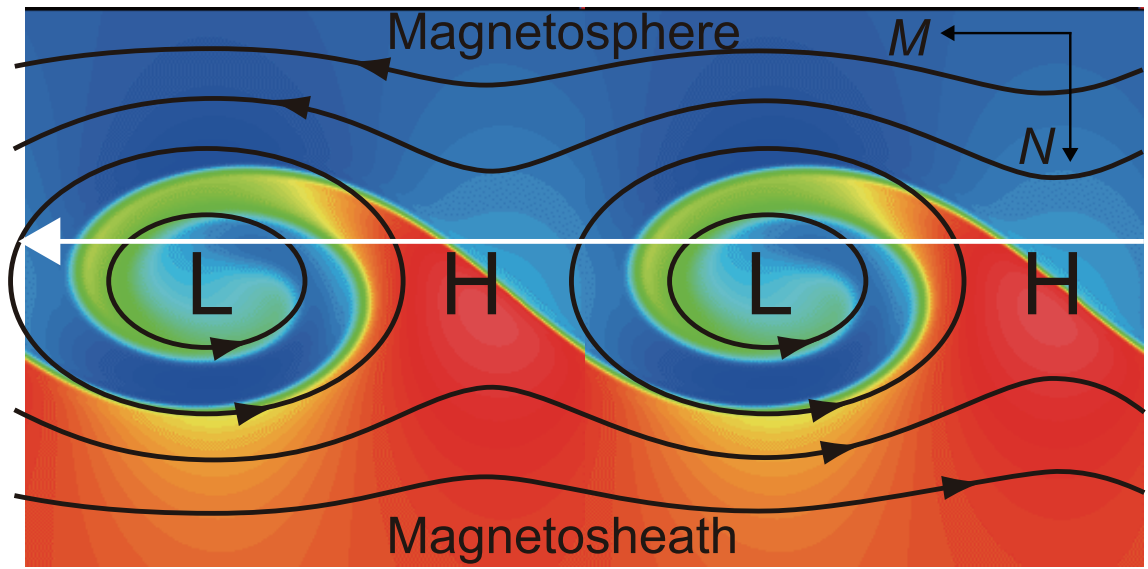

**Figure 19.** Schematic drawing of rolled-up KH vortices at the dusk-flank magnetopause, showing the relationship between the streamline pattern (black lines) and total (magnetic plus plasma) pressure and density (red, dense; blue, tenuous) distributions, when viewed from north and in the vortex rest frame. The subsolar region is to the left. The total pressure is minimized at the center (L) of the vortices, while it is maximized at the hyperbolic point (H), which is a flow stagnation point in the vortex rest frame and around which the streamlines form hyperbola (e.g., Miura, 1997). When a spacecraft, nearly standing still in the Earth's rest frame, observes a train of rolled-up vortices traveling tailward along the magnetopause (white arrow), it is expected that each of clear magnetosphere to magnetosheath transitions, characterized by a large and rapid density jump, coincides approximately with a maximum in the total pressure (see confirmation in Figure 20).

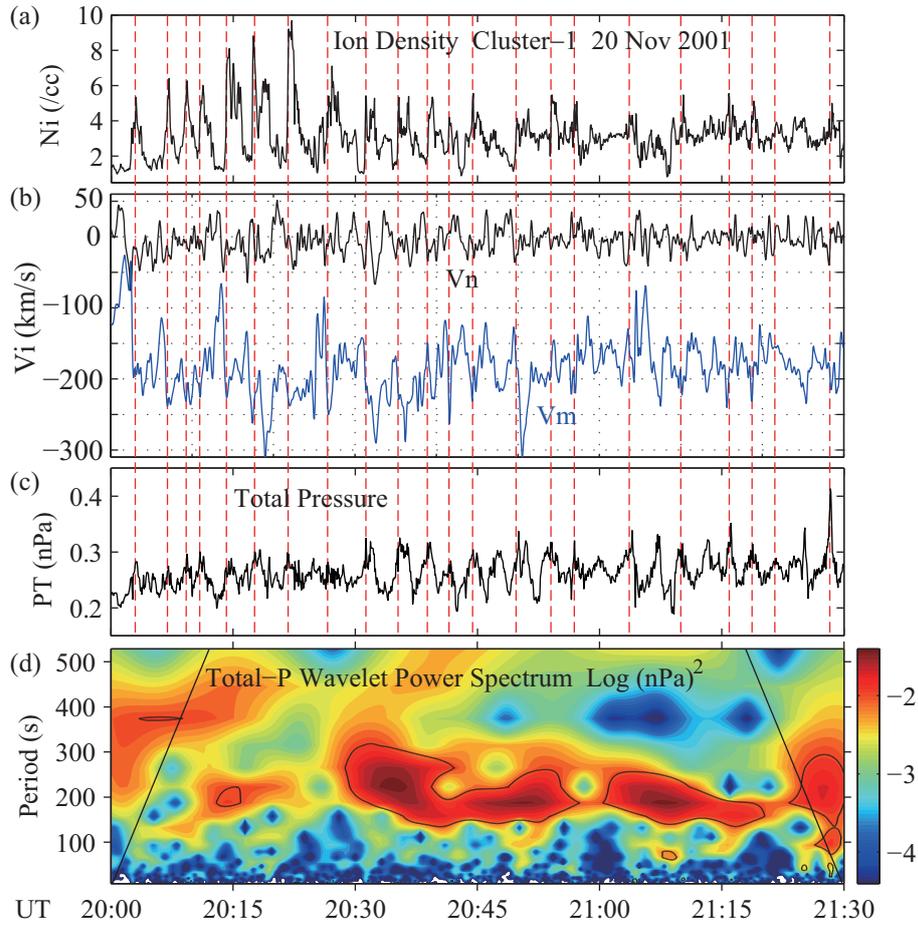

**Figure 20.** Quasi-periodic fluctuations of the bulk plasma parameters during Cluster 1 (C1) observations of rolled-up KH vortices at the dusk-flank magnetopause. The event is the same as shown in Figure 7. (a) Ion density, (b) *M* and *N* components of the smoothed velocity (in the Earth's rest frame), (c) total (magnetic plus ion) pressure, and (d) wavelet spectra of the total pressure. The spectra above the oblique lines may be dubious. Note that density jumps from the magnetospheric to magnetosheath values nearly coincide with total pressure maxima (red vertical dashed lines), as expected in the developed KH vortices (cf. Figure 19) (adopted from Hasegawa et al., 2009b).

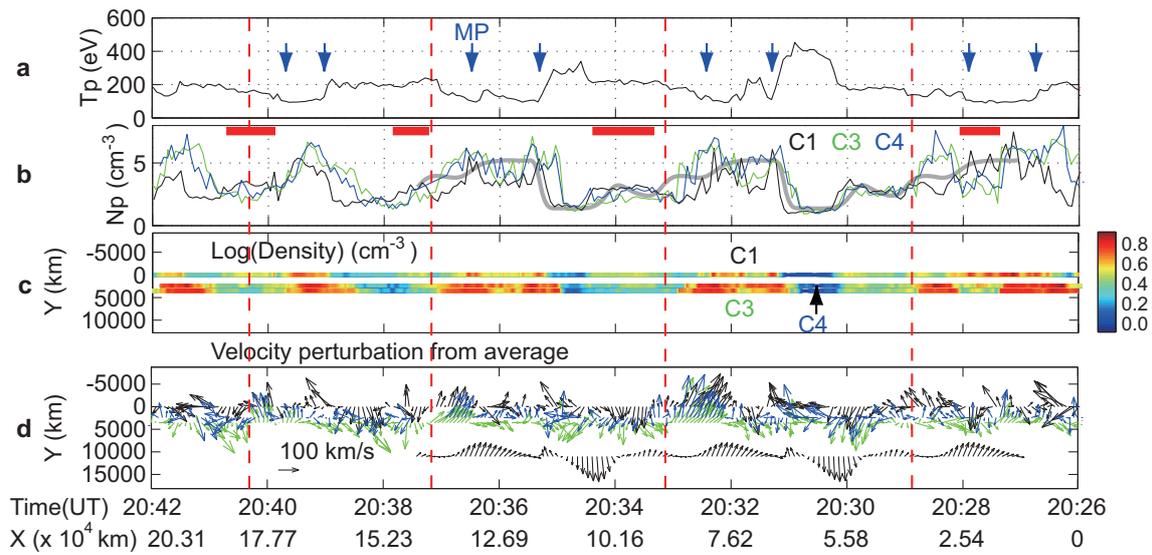

**Figure 21.** Evidence for overturning of large-scale magnetopause surface waves (roll-up of KH vortices) observed by Cluster on 20 November 2001. The event is the same as shown in Figures 7 and 20. The four satellites were separated by ~2000 km from each other. Time progresses to the left and is translated into the position of the spacecraft. (a) Ion temperature from C1, with approximate locations of the magnetopause marked by the blue arrows. (b) Plasma density variations (at C1, C3, and C4) showing their similarity to that predicted by a 3D MHD simulation of the KH instability (KHI) (thick gray curve). (c) Plasma density color-coded and projected along the spacecraft paths (view from north). The *y* axis is orthogonal to both the *x* axis (roughly sunward) and the average magnetic field direction (roughly northward), and points toward the magnetosheath along the nominal magnetopause normal (roughly duskward). (d) The *x-y* projection of the velocity deviations from its average values, along with the behavior predicted by the simulation (black arrows in the lower part). The red vertical dashed lines mark the approximate centers of the vortices, and the red bars mark the KH-wave phases at which the density increases toward the magnetosphere (adopted from Hasegawa et al., 2004a).

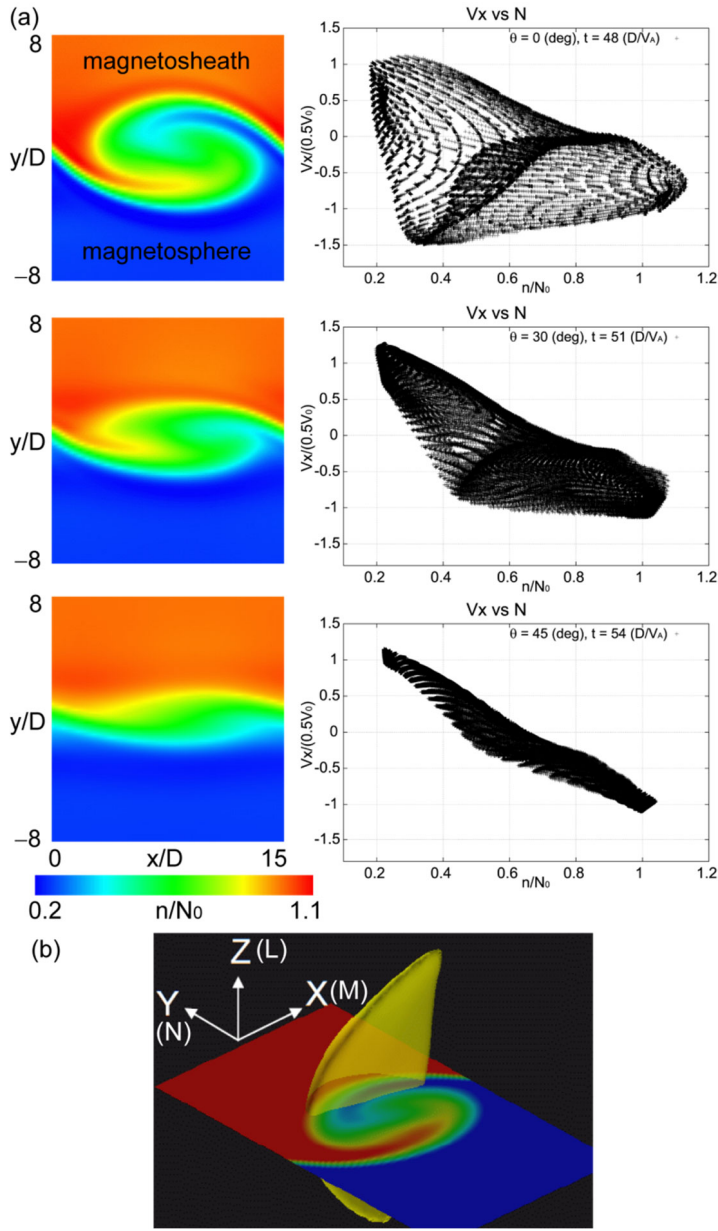

**Figure 22.** (a) Scatter plots of the $x$ (sunward) component of the velocity $V_x$ versus plasma density $n/N_0$, at certain stages of 3D MHD simulations of the KHI (Takagi et al., 2006). The left panels show the normalized density in the $x$-$y$ plane in color. The initial value of $V_x$ is $-1$ and $+1$ on the magnetosheath side ($y > 0$) and on the magnetospheric side ($y < 0$), respectively. Note that within the rolled-up vortex (top), a significant fraction of low-density plasmas ($n/N_0 < 0.5$) has an anti-sunward speed higher than that of the magnetosheath plasma characterized by $V_x = -1$ and $n/N_0 = 1$. (b) The yellow isosurface highlights the region in 3D where $V_x < -1.2$ for the case of rolled-up vortex (from Hasegawa et al., 2006b).

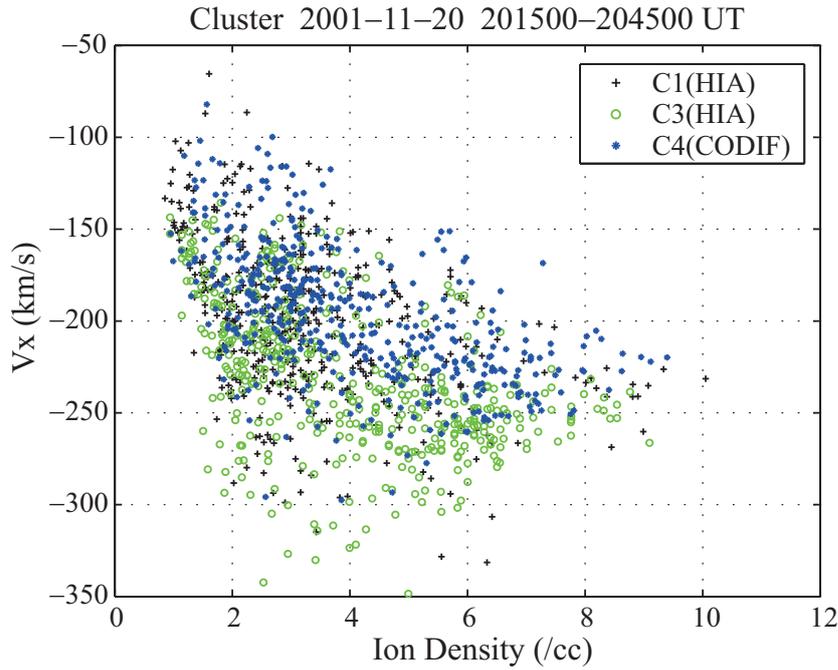

**Figure 23.** Scatter plot of the velocity component, $V_x$, tangential to the nominal magnetopause versus ion density seen in the rolled-up vortices detected by Cluster. The event is the same as shown in Figures 7, 20, and 21. Here the $-x$ direction is defined to be along the ion velocity (in GSM) averaged over the interval under investigation (2015–2045 UT), and is roughly along the $-M$ direction (anti-sunward) in LMN coordinates. The observation confirms the prediction by simulation (top panel in Figure 22) that in a rolled-up vortex, a fraction of the low-density magnetospheric plasmas flows faster than the magnetosheath plasma (from Hasegawa et al., 2006b).

(a) Linear Phase

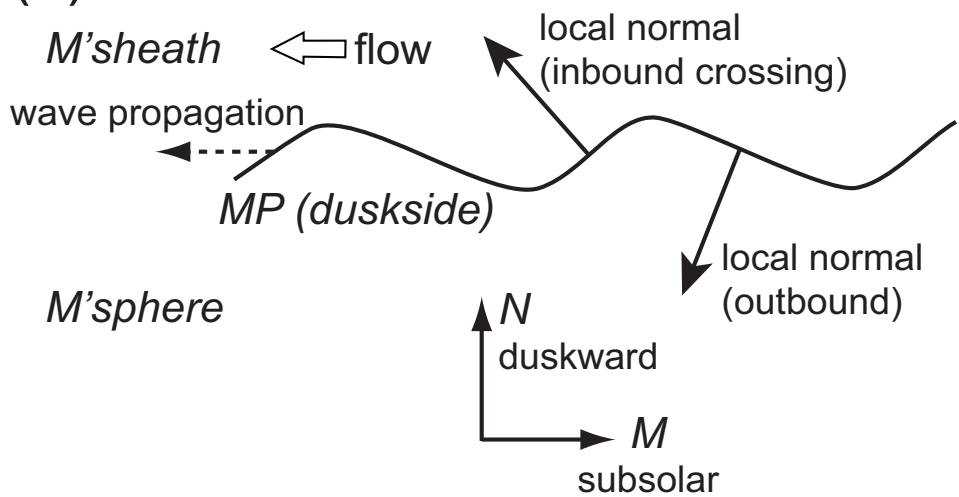

(b) Nonlinear Phase

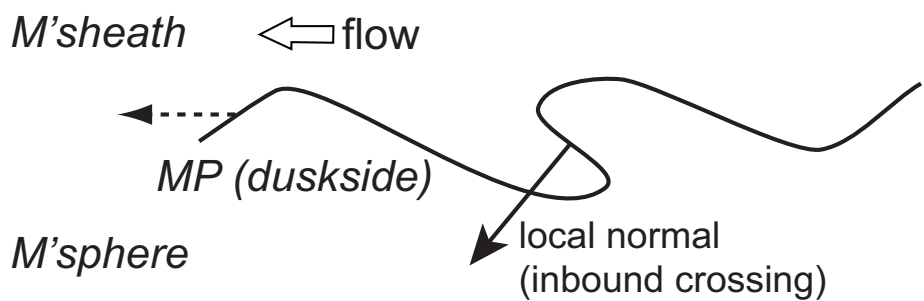

**Figure 24.** Schematic drawing illustrating the orientation and motion of the local wave surfaces in the linear (a) and nonlinear (b) stages of the KHI excited along the duskside magnetopause (view from north). In the linear stage the local normal (solid arrows) systematically points toward the magnetosheath and magnetosphere for inbound (magnetosheath-to-magnetosphere) and outbound (magnetosphere-to-magnetosheath) crossings, respectively, while in the nonlinear stage the normal can point toward the magnetosphere even for inbound crossings. Here, the normal is defined positive in the direction of local wave surface motion.

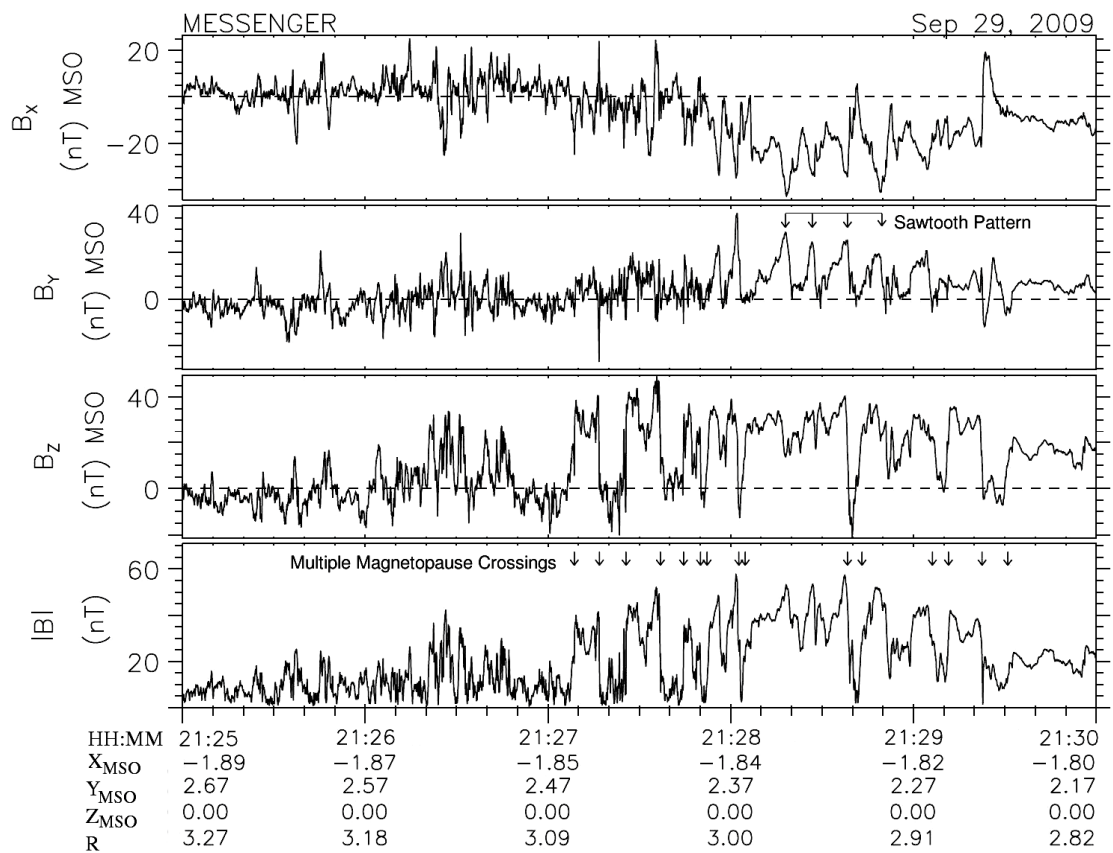

**Figure 25.** Observations of magnetopause surface waves at Mercury by the MESSENGER spacecraft during its third flyby on 29 September 2009. The panels show the three components and intensity of the magnetic field in Mercury solar orbital (MSO) coordinates, the Mercury equivalent of GSE coordinates. Note multiple crossings of the magnetopause (marked by arrows in the bottom panel) as MESSENGER moved from the magnetosheath into the dusk-flank magnetosphere (from Boardsen et al., 2010).

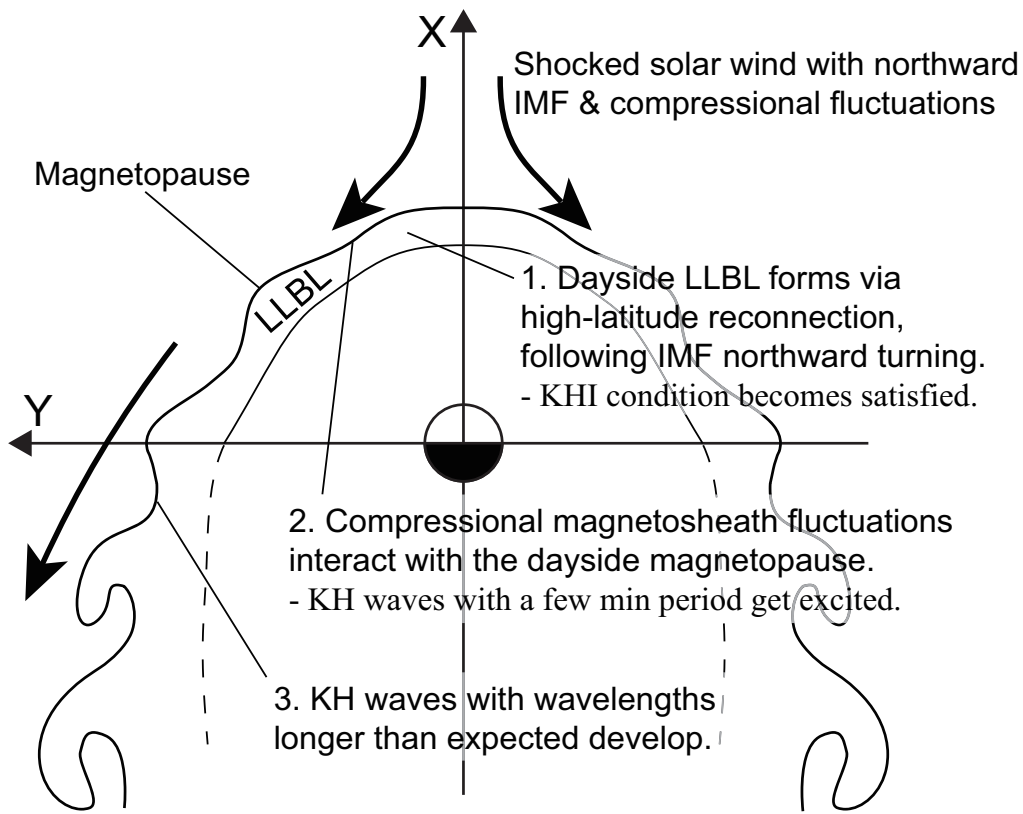

**Figure 26.** Schematic drawing of the equatorial magnetosphere (view from north), illustrating how KH waves with wavelengths longer than predicted by the linear theory of the KHI could be excited under northward IMF conditions (from Hasegawa et al., 2009b).

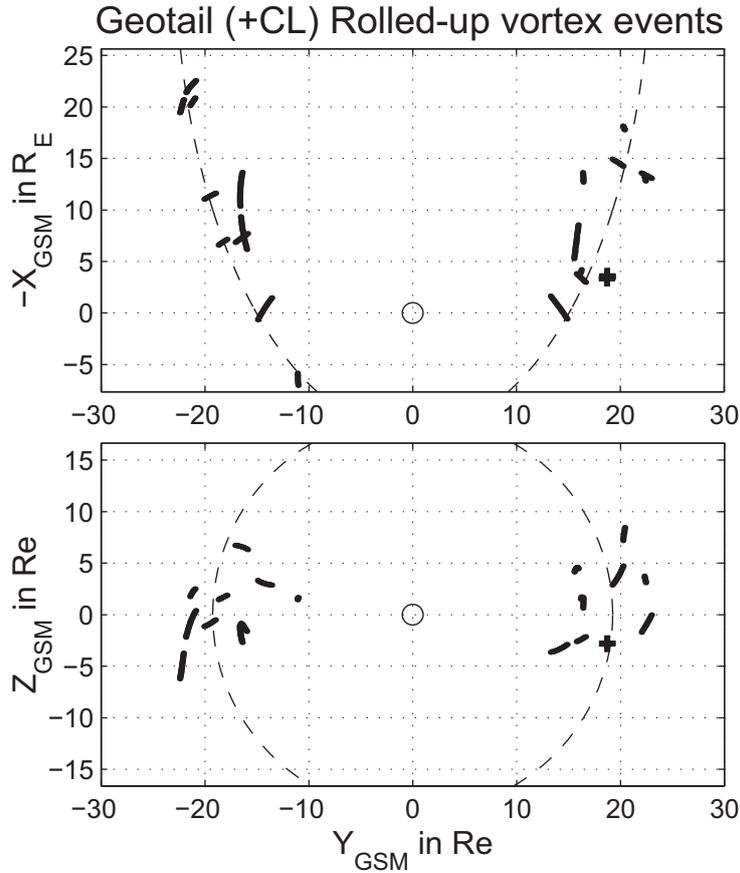

**Figure 27.** Locations in GSM coordinates of rolled-up vortices identified from Geotail data (solid curves) and Cluster (plus) at or around the magnetopause during northward IMF periods. The dashed curves in the top panel represent the average magnetopause position based on the Roelof and Sibeck (1993) model. The rolled-up vortices are found mostly behind the dawn-dusk terminator ($X_{GSM} = 0$), with a roughly equal occurrence probability for both sides of the magnetosphere (from Hasegawa et al., 2006b).

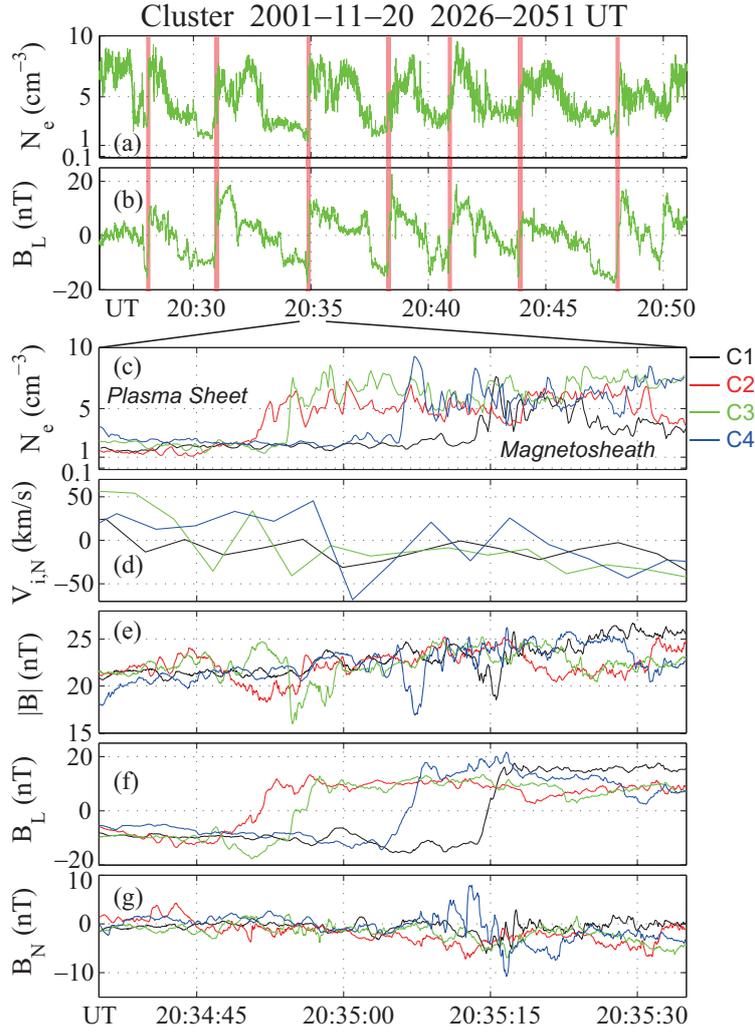

**Figure 28.** Cluster observations of thin current sheets developed at the trailing (sunward-facing) edge of rolled-up KH vortices. The event is the same as shown in Figures 7, 20, and 21. (a) Electron density estimated from the spacecraft potential measurement and (b) $L$ component of the magnetic field at C3 for the interval 2026–2051 UT. The juxtaposition of the current sheets and steep density gradients (magnetopause) is indicated by the red vertical bars. (c) Electron density, (d) $N$ component of the ion velocity in the boundary-rest frame, and (e-g) magnitude and the $L$ and $N$ components of the magnetic field for 2034:35–2035:35 UT. Here, $N$ is along the normal to the current sheet estimated by the four-spacecraft timing method (Schwartz, 1998), and $L$ is roughly along the maximum magnetic variance direction (Appendix A). In one of the current sheets, traversed by C3 at ~2034:55 UT (green lines in panels (e-g)), reconnection signatures as illustrated in Figure 29b were identified (from Hasegawa et al., 2009b).

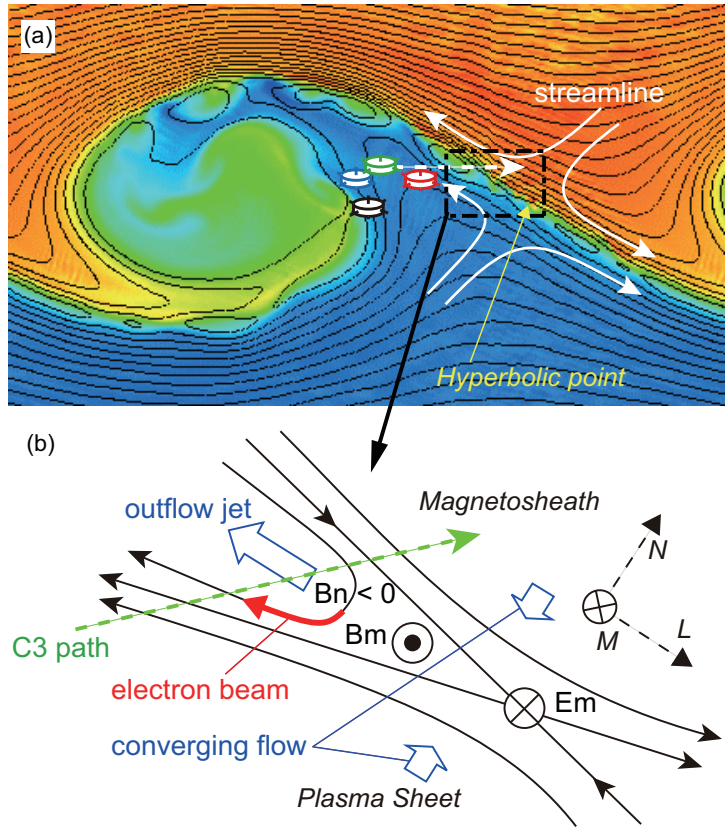

**Figure 29.** (a) Structure of an MHD-scale vortex developed in a 2D two-fluid simulation of the KHI (Nakamura et al., 2008), showing the plasma density in color (red, dense; blue, tenuous) with in-plane magnetic field lines overlaid (view from north). (b) Schematic drawing of reconnection signatures identified by C3 at the trailing edge of a rolled-up KH vortex (shown in Figure 28): (1) non-zero $B_N$ consistent with interconnection of the magnetosheath and plasma sheet field lines, (2) reconnection outflow jet, and (3) magnetic field-aligned electron beam (from Hasegawa et al., 2009b).

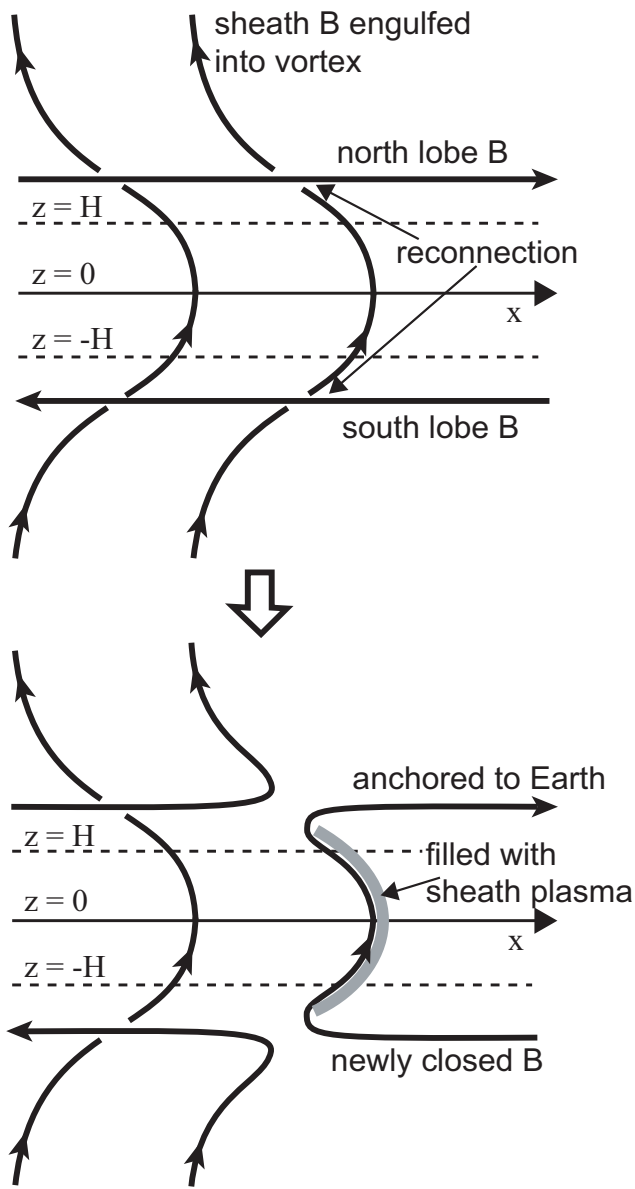

**Figure 30.** Schematic drawing of magnetic field lines around the dusk-flank magnetopause (viewed from dawnside), illustrating how magnetic reconnection could occur as a result of the KHI growth in the magnetotail flank geometry. The Earth is to the right and north is upward. Note that the magnetosheath plasma can be captured into closed flux tubes through this KHI-induced dual reconnection process (from Takagi et al., 2006).

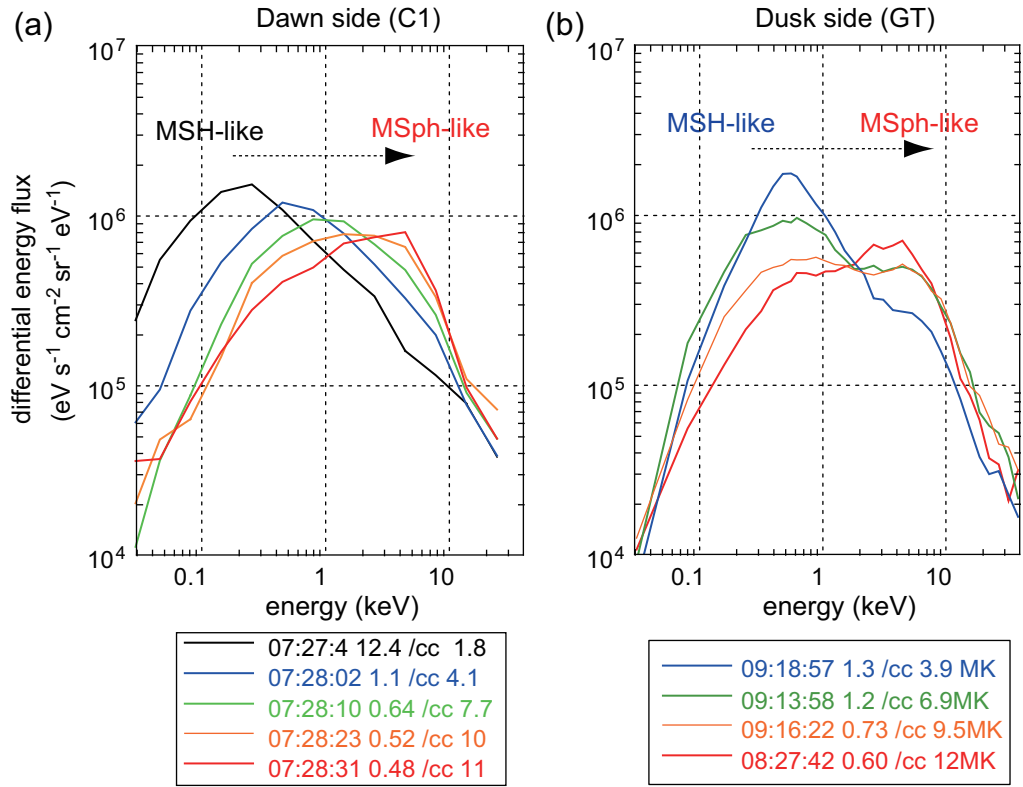

**Figure 31.** Dawn-dusk asymmetry in proton energy distributions observed around KH vortices detected in both tail flanks, implying dawn-dusk asymmetric transport of protons in energy space. The panels show evolution of the differential energy fluxes seen by Cluster (C1) on the dawn side (a) and by Geotail (GT) on the dusk side (b). On the dawn side, each energy distribution has a single flux peak, and the energy at the peak gradually increases as one moves from the dense magnetosheath-like to tenuous magnetosphere-like regions. On the other hand, typical distributions on the dusk side (e.g., green and yellow lines in panel b) are a superposition of low-energy magnetosheath-like (less than 1 keV) and high-energy magnetospheric (more than a few keV) populations, indicating that on the dusk side, proton transport in real space does occur but that in energy space is inefficient (from Nishino et al., 2011).